\documentclass[preprint,12pt]{elsarticle}
\usepackage[utf8]{inputenc}
\usepackage{amssymb}
\usepackage{amsmath}
\usepackage{amsthm}

\usepackage{subfiles}
\usepackage[section]{placeins}%so that figures appear on the right section
\usepackage{graphicx}
\usepackage{mathrsfs}
\usepackage{delarray}
\usepackage{stmaryrd}
\usepackage{ifthen}
\usepackage{paralist}
\usepackage[table]{xcolor}
\usepackage{slashbox}
\usepackage{xcolor}
\usepackage{color}%change color of single cell of table
\usepackage{multirow}
\usepackage{multicol}%multiple columns of text
\usepackage{makecell}%multiple rows in one cell
\usepackage{arydshln}
\usepackage{tikz}
\usetikzlibrary{arrows.meta}
\usepackage[colorlinks=true,
            linkcolor=blue]
           {hyperref}

\newtheorem{theorem}{Theorem}

\newtheorem{corollary}{Corollary}

\newtheorem{lemma}[theorem]{Lemma}

\newtheorem{definition}[theorem]{Definition}
\newtheorem{remark}{Remark}

\newcounter{finaln}
\newcommand{\integers}[2][2]{\ifthenelse{\equal{#1}{2}}
{\llbracket #2\rrbracket}
{\ifthenelse{\equal{#1}{0}}{
\ifthenelse{\equal{11}{\the\catcode`#2}}
{\{0,\ldots,#2 -1\}}
{\setcounter{finaln}{#2 -1}
\{0,\ldots,\thefinaln \}}}
{\{1,\ldots,#2\}}}}
\newcommand{\N}{\mathbb{N}}
\renewcommand{\int}[1]{\integers{#1}}
\newcommand{\lcm}{\text{lcm}}

\newcommand{\T}{\mathscr{T}} % trajectory
\renewcommand{\O}{\mathscr{O}} % orbit
\newcommand{\C}{\mathscr{C}} % limit dynamics
\newcommand{\G}{\mathscr{G}} % transition graph
\DeclareMathOperator{\deltaeq}{\;\overset{\Delta}{=}\;}
\newcommand{\p}{\textsc{par}}
\newcommand{\bip}{\textsc{bip}}
\newcommand{\seq}{\textsc{seq}}
\newcommand{\bs}{\textsc{bs}}
\newcommand{\bp}{\textsc{bp}}
\newcommand{\lc}{\textsc{lc}}

\newcommand{\PreserveBackslash}[1]{\let\temp=\\#1\let\\=\temp}
\newcolumntype{C}[1]{>{\PreserveBackslash\centering}p{#1}}
\newcommand{\Z}{\mathbb{Z}}

\newcommand{\B}{\mathbb{B}}
\newcommand{\Bn}{\mathbb{B}^n}

\renewcommand{\O}{\mathcal{O}}

\newcommand{\ie}{i.e.{}}

\journal{ArXiv}

\begin{document}

\begin{frontmatter}


%%comprender la influencia de los esquemas de actualización sobre los ACE a través de la complejidad asintótica.
%% understanding the influence of periodic update modes on ECA through their: 
%%impact of (a)synchronism on ECA: towards classification
\title{Impact of (a)synchronism on ECA: towards a new classification\footnote{Partial results of this work were presented at the international conference LATIN 2024 (Latin American Theoretical Informatics)~\cite{donoso2024}}}

\author[label1,label3]{Isabel Donoso-Leiva}
\author[label1]{Eric Goles}
\author[label1]{Mart{\'i}n R{\'i}os-Wilson }
\author[label2,label3]{Sylvain Sen{\'e} }
\affiliation[label1]{organization={Facultad de Ingeniería y Ciencias, Universidad Adolfo Ibáñez },country={Chile}}
\affiliation[label2]{organization={Université publique, Marseille},country={France}}
\affiliation[label3]{organization={Aix Marseille Univ, CNRS, LIS, Marseille},country={France}}

%% Abstract
\begin{abstract}
In this paper, we study the effect of (a)synchronism on the dynamics of elementary cellular automata. Within the framework of our study, we choose five distinct update schemes, selected from the family of periodic update modes: parallel, sequential, block sequential, block parallel, and local clocks. Our main measure of complexity is the maximum period of the limit cycles in the dynamics of each rule. In this context, we present a classification of the ECA rule landscape. We classified most elementary rules into three distinct regimes: constant, linear, and superpolynomial. Surprisingly, while some rules exhibit more complex behavior under a broader class of update schemes, others show similar behavior across all the considered update schemes. Although we are able to derive upper and lower bounds for the maximum period of the limit cycles in most cases, the analysis of some rules remains open. To complement the study of the 88 elementary rules, we introduce a numerical simulation framework based on two main measurements: the energy and density of the configurations. In this context, we observe that some rules exhibit significant variability depending on the update scheme, while others remain stable, confirming what was observed as a result of the classification obtained in the theoretical analysis.
\end{abstract}

\begin{highlights}
\item A new formal approach to analyze the impact on elementary cellular automata (ECA) of asynchronism/synchronism is presented.
\item Concepts of asyncronism/synchronism are considered through fundamental families of periodic update modes.
\item A new classification of ECA based on computational complexity is offered, measured through the length of their longest reached limit cycle is proposed.
\item Numerical experiments are analyzed for specific and well known chaotic and complex ECA, namely 54, 90, 110 and 150, through measures of density and energy.
\end{highlights}

%% Keywords

\end{frontmatter}

\section{Introduction}
\label{sec:intro}

Cellular automata are collections of discrete state entities (the cells) arranged over a grid that interact with each other according to a local rule over discrete time. They were first introduced by Ulam and Von Neumann in the 1940s~\cite{vonNeumann1966} following from automata networks which were defined by McCulloch and Pitts in the same decade~\cite{McCulloch1943}.
From there, fundamental results have been obtained such as 
the introduction of the retroaction cycle theorem~\cite{Robert1980}, 
the self organizing behavior~\cite{wolfram1983},
the Turing universality of the model itself~\cite{Smith1971,Goles1990}, 
the structure of the space of elementary cellular automata ~\cite{li1990structure},
the undecidability of all nontrivial properties of limit sets of cellular automata~\cite{Kari1994}, 
hierarchy of certain cellular automata~\cite{Concha2022}.
While CA are simple models, they are able to exhibit great complexity which ranges from biology and health sciences, as a representational model of disease spreading~\cite{White2007} and its impact\cite{Sirakoulis2000}, to 
social models~\cite{Sakoda1971,Schelling1971}, 
to parallel and distributed computation~\cite{Smith1972,Mazoyer1987}, to physics~\cite{Vichniac1984}.\smallskip

Despite significant theoretical contributions since the 1980s that have improved our understanding of these objects from both computational and behavioral perspectives~\cite{Robert1986,Goles1990}, the sensitivity of cellular automata (CA) to (a)synchronism remains an open question. Any progress in this area could have profound implications in computer science—particularly concerning synchronous versus asynchronous computation and processing~\cite{Chapiro1984,Charron1996}—as well as in systems biology, especially regarding the temporal organization of gene expression~\cite{Hubner2010,Fierz2019}.

\subsection*{Related Work}
Although synchronism, where all cells update simultaneously, can be an interesting case study (transforming the CA model into a massively parallel computing environment~\cite{cannataro1995parallel}) this assumption is often unrealistic for many applications. This issue arises from two key perspectives: first, from a mathematical modeling standpoint (based for example on the observation of  biological phenomena), perfectly simultaneous interactions between cells are rare~\cite{cornforth2003artificial}; second, from a computational perspective, a synchronous update scheme requires an internal clock to synchronize each processor computing cell states. In hardware design, this requirement increases complexity and reduces the efficiency of simulating the model.

Consequently, alternative approaches to the synchronous update scheme, collectively referred to as \textit{asynchronous update schemes}, have been proposed. Some of the most well-known types include:
\begin{enumerate}
	 \item Fully asynchronous update scheme \cite{fates2006fully}: at each time step, the local rule of the CA is applied to a single cell that is uniformly randomly chosen.
	\item $\alpha$-Asynchronous update scheme \cite{Fates2005}: at each time step, each cell has a probability $\alpha$ to update its state and a probability $1-\alpha$ to stay in the same state. The parameter $\alpha$ is called the \emph{synchronization rate}.
    \item Fixed random sweep~\cite{SCHONFISCH1999123,kitagawa1974cell}: A randomly chosen permutation of the cells is determined at initialization, and at each time step, cells are updated according to this fixed order.
    \item Random new sweep~\cite{SCHONFISCH1999123,kitagawa1974cell}: At each time step, a new random permutation of the cells is selected for updating.
    \item Non-random update schemes~\cite{Robert1986, goles2021complexity, goles2016pspace, rios2024c}: This category includes sequential update schemes, where cells are updated one by one in a predetermined order. When the context is clear, we refer to these simply as \textit{update schemes} or \textit{update modes}. Additionally, updates can occur in groups rather than individually, forming a \textit{block sequential update scheme}. Other periodic update schemes exist, all characterized by a fixed cycle in which the sequence of updated nodes (whether individually or in groups) repeats after a defined number of time steps.
\end{enumerate}

Beyond these variants, which differ based on deterministic or probabilistic updating schemes, some cellular automata rules exhibit intrinsically asynchronous dynamics, such as probabilistic CA. In these models, the local update rule itself is probabilistic. As noted by Fatès  \cite{fates2013guided}, this line of research has yielded significant recent developments concerning asynchronism.

Fatès identifies two primary approaches to studying asynchronism in CA. The first, an \textit{experimental approach}, relies on numerical simulations, while the second is based on mathematical analysis. In this work, we adopt both approaches. As explained in \cite{fates2013guided}, the experimental approach facilitates the detection and classification of changes induced by asynchornism in global dynamical properties, both qualitative and quantitative. In the same review, it is state that various observed properties for different elementary rules, including the Game of Life model (or more generally life-like rules)  can be capture by this approach (see for example \cite{ingerson1984, blok1999synchronous, Fates2005, miszczak2023rule, roy2024note}). Notably, as highlighted in \cite{fates2013guided}, while intuition might suggest that asynchronism increases dynamical complexity, certain cases reveal that it can instead simplify system behavior. Interestingly, our findings confirm this phenomenon, showing that highly complex rules can exhibit remarkably simple behavior under specific update schemes. 

An important question explored in \cite{fates2013guided}, is related to the definition of efficient and correct numerical simulation protocols to capture the changes induce changes induced by asynchronism in CA dynamics. In this context, the choice of relevant order parameters is a key factor. In this work, we propose an approach based on the maximum period of limit cycles as a measure of the complexity of the dynamics of an ECA rule .

On the other hand, the analytical approach employs mathematical techniques to study the effect of asynchronism in CA models. For probabilistic CA, a primary method involves analyzing these systems as Markov chains \cite{agapie2004markov,agapie2014probabilistic}. Additional approaches include computing asymptotic densities (for instance, exact density calculations in specific cases \cite{fuks2011orbits}), use of techniques aming to approximate the stationary measures of probabilistic CA such as block approximation \cite{cirillo2024block}, extensions of mean field approximations such as \cite{fuks2015local}  the study of the impact of asynchronism on dynamical properties such as reversibility \cite{das2012synthesis,sethi2014reversibility,roy2024reversibility}. A key advantage of our work is that, by focusing exclusively on the deterministic case (where both local rules and update schemes are deterministic), we primarily employ tools from discrete mathematics and combinatorics. While one might assume that analytical descriptions remain impractical even in deterministic settings, we show that in certain cases it is possible to completely characterize an elementary rule complexity from an analytic standpoint. Moreover, from our results is derived the fact that asynchronism can have widely varying effects in deterministic CA, implying that deterministic models are not necessarily simpler than their probabilistic counterparts.

\subsection*{Our Contribution}
Since the aim of this paper is to increase the knowledge on asynchronism sensitivity, elementary cellular automata were chosen because they are a family of restricted and  ``simple'' cellular automata, which has been well studied~\cite{Wolfram1984,Culik1988,Kurka1997}. By studying ECA we can put the focus on the periodic update modes (from the classical parallel update mode to a family of more general ones known as local clocks~\cite{Pauleve2022}) and their impact over the dynamics. Here, we approach the subject with ideas derived from~\cite{rios2024a,rios2024c} and pay attention to the influence of update modes on the resulting asymptotic dynamical behavior, in particular in terms of the maximal period of limit cycles. 
Additionally, for rules whose behavior is too complex to analyze directly, we have ran computational experiments that used measures of density and energy to study their asymptotic dynamical behavior. For these experiments we have implemented a number of update modes per family, for three different orders of magnitude of ring sizes.\smallskip

In this paper, we highlight formally that the choice of the update mode can have a deep influence on the dynamics of systems. In particular, two specific elementary cellular automata rules, namely rules $2$ and $184$ (known as the traffic rule) as defined by Wolfram's codification, are studied here. These rules were chosen because the results obtained require proofs which serve as example of how to obtain similar results for other rules that are mentioned in their respective section. In particular, the traffic rule also emphasizes sensitivity to (a)synchronism.

Note that both rules belong to the Wolfram's class II~\cite{Wolfram1984}, which means that, according to computational observations, these cellular automata evolve asymptotically towards a ``set of separated simple stable or periodic structures''. Since our (a)synchronism sensitivity measure consists in limit cycles maximal periods,   Wolfram's class II is naturally the most pertinent one in this context.

Secondly, we have run computational experiments over rules $90$ and $150$ (of Wolfram's class III) and rules $54$ and $110$ (of Wolfram's class IV) to discuss the influence of different update modes over rules whose dynamics is already known to be too chaotic (class III) or too complex (class IV) to show jumps in asymptotic complexity as a result of the influence of the different update modes. Instead, we show the changes (or lack thereof) in their dynamics by way of showing whether there are changes in the evolution of density and energy. \medskip

\subsection*{Structure of the Paper}
In Section~\ref{sec:def}, the main definitions and notations are formalized. 
The emphasizing of elementary cellular automata (a)synchronism sensitivity is presented in the first part of Section~\ref{sec:res} through demonstrations that set an upper-bound for the limit cycle periods of rules $2$ and $184$ depending on distinct families of periodic update modes. 
In the second part, experimental results for rules $90$, $150$, $54$ and $110$ under different update modes are presented, using measures of density and energy.
The paper ends with Section~\ref{sec:persp} in which we discuss some perspectives of this work.

\section{Preliminaries}
\label{sec:def}

\paragraph{General notations}

Let $\integers{n} = \integers[0]{n}$,
let $\B = \{0,1\}$, and 
let $x_i$ denote the $i$-th component of vector $x \in \B^n$.
Given a vector $x \in \B^n$, we can denote it classically as 
$(x_0, \dots, x_{n-1})$ or as the word $x_0 \dots x_{n-1}$ if it eases the reading.

\subsection{Boolean automata networks and elementary cellular automata}

Roughly speaking, a Boolean automata network (BAN)  applied over a grid of size $n$  is a 
collection of $n$ automata represented by the set $\integers{n}$, each having a 
state within $\B$, which interact with each other over discrete time.  
A \emph{configuration} $x$ is an element of $\Bn$, \ie a Boolean vector of 
dimension $n$. 
Formally, a \emph{BAN} is a function $f: \Bn \to \Bn$ defined by means of $n$ 
local functions $f_i: \Bn \to \B$, with $i \in \integers{n}$, such that $f_i$ is 
the $i$th component of $f$. 
Given an automaton $i \in \integers{n}$ and a configuration $x \in \Bn$, 
$f_i(x)$ defines the way that $i$ updates its states depending on the state of 
automata \emph{effectively} acting on it; automaton $j$ ``effectively'' acts on 
$i$ if and only if there exists a configuration $x$ in which the state of $i$ 
changes with respect to the change of the state of $j$; $j$ is then called a 
\emph{neighbor} of $i$.\smallskip
 
An \emph{elementary cellular automaton (ECA)} is a particular BAN dived into 
the cellular space $\Z$ so that \emph{(i)} the evolution of state $x_i$ of 
automaton $i$ (rather called cell $i$ in this context) over time only depends 
on that of cells $i-1$, $i$ itself, and $i+1$, and \emph{(ii)} all cells share 
the same and unique local function.
As a consequence, it is easy to derive that there exist $\smash{2^{2^3}} = 256$ 
distinct ECA, and it is well known that these ECA can be grouped into $88$ 
equivalence classes up to symmetry.\smallskip

In absolute terms, BANs as well as ECA can be studied as infinite models of 
computation, as it is classically done in particular with ECA.
In this paper, we choose to focus on finite ECA, which are ECA whose underlying 
structure can be viewed as a torus of dimension $1$ which leads naturally to work on 
$\Z/n\Z$, the ring of integers modulo $n$ so that the neighborhood of cell $0$ is 
$\{n-1, 0, 1\}$ and that of cell $n-1$ is $\{n-2, n-1, 0\}$.\smallskip

Now the mathematical objects at stake in this paper are statically defined, let 
us specify how they evolve over time, which requires defining when the cells 
state update, by executing the local functions. 

\begin{figure}[t!]
	\centerline{\includegraphics[width=\textwidth]{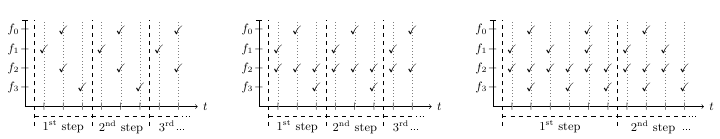}}
	\caption{Illustration of the execution over time of local transition 
		functions of any BAN $f$ of size $4$ according to 
		(left) $\mu_\bs = (\{0\}, \{2,3\}, \{1\})$,
		(center) $\mu_\bp = \{(1), (2,0,3)\}$, and
		(right) $\mu_\lc = ((1,3,2,2), (0,2,1,0))$.
		The $\checkmark$ symbols indicate the moments at which the automata update 
		their states; the vertical dashed lines separate periodical time steps 
		from each other.}
	\label{fig:um_exec}
\end{figure}

\subsection{Update modes}

To choose an organization of when cells update their state over time leads to 
define what is classically called an update mode (aka update schedule or 
scheme).
In order to increase our knowledge on (a)synchronism sensitivity, as evoked in 
the introduction, we pay attention in this article to deterministic and 
periodic update modes.
Generally speaking, given a BAN $f$  applied over a grid of size $n$ , a \emph{deterministic} (resp. 
\emph{periodic}) \emph{update mode} of $f$ is an infinite (resp. a finite) 
sequence $\mu = (B_k)_{k \in \N}$ (resp. $\mu = (B_0, \dots, B_{p-1})$), where 
$B_i$ is a subset of $\integers{n}$ for all $i \in \N$ (resp. for all $i \in 
\integers{p}$). 
Another way of seeing the update mode $\mu$ is to consider it as a function 
$\mu^\star: \N \to \mathcal{P}(\integers{n})$ which associates each time step 
with a subset of $\integers{n}$ so that $\mu^\star(t)$ gives the automata which 
update their state at step $t$; furthermore, when $\mu$ is periodic, there 
exists $p \in \N$ such that for all $t \in \N$, $\mu^\star(t+p) = 
\mu^\star(t)$.\smallskip

Three known update mode families are considered: the block-sequential 
ones ~\cite{Robert1986}, the block-parallel ones ~\cite{Demongeot2020,Perrot2024a,Perrot2024b} 
and the local clocks ones ~\cite{Rios2021-phd}.
Updates induced by each of them over time are depicted in 
Figure~\ref{fig:um_exec}.\smallskip

A \emph{block-sequential update mode} $\mu_\bs = (B_0, \dots, B_{p-1})$ is an 
ordered partition of $\integers{n}$, with $B_i$ a subset of $\integers{n}$ for 
all $i$ in $\integers{p}$. 
Informally, $\mu_\bs$ defines an update mode of period $p$ separating 
$\integers{n}$ into $p$ disjoint blocks so that all automata of a same block 
update their state in parallel while the blocks are iterated in series. 
The other way of considering $\mu_\bs$ is: 
$\forall t \in \N, \mu_\bs^\star(t) = B_{t \mod p}$.\smallskip

A \emph{block-parallel update mode} $\mu_\bp = \{S_0, \dots, S_{s-1}\}$ is a 
partitioned order of $\integers{n}$, with $S_j = (i_{j,k})_{0 \leq k \leq 
|S_j|-1}$ 
a sequence of $\integers{n}$ for all $j$ in $\integers{s}$. 
Informally, $\mu_\bp$ separates $\integers{n}$ into $s$ disjoint subsequences 
so that all automata of a same subsequence update their state in series while 
the subsequences are iterated in parallel.
Note that there exists a natural way to convert $\mu_\bp$ into a sequence of 
blocks of period $p = \lcm(|S_0|, \dots, |S_{s-1}|)$.
It suffices to define function $\varphi$ as: 
$\varphi(\mu_\bp) = (B_\ell)_{\ell \in \integers{p}}$ with $B_\ell = 
\{i_{j,\ell \mod |S_j|} \mid j \in \integers{s}\}$.  
The other way of considering $\mu_\bp$ is:
$\forall j \in \integers{s}, \forall k \in \integers{|S_j|}$, $i_{j,k} \in 
\mu_\bp^\star(t) \iff k = t \mod |S_j|$.\smallskip

A \emph{local clocks update mode} $\mu_{\lc} = (P,\Delta)$, with 
$P = (p_0, \dots, p_{n-1})$ and $\Delta = (\delta_0, \dots, \delta_{n-1})$, is 
an update mode such that each automaton $i$ of $\integers{n}$ is associated 
with a period $p_i \in \N^*$ and an initial shift $\delta_i \in \integers{p_i}$ 
such that $i \in \mu_{\lc}^\star(t) \iff t = \delta_i \mod p_i$, with 
$t \in \N$. \smallskip

\begin{figure}[t!]
\centering{
\includegraphics[]{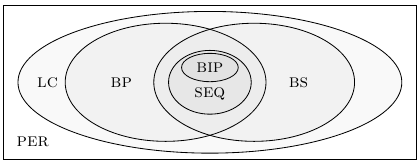}}
	\caption{Order of inclusion of the defined families of periodic update 
		modes, where \textsc{per} stands for ``periodic''.}
	\label{fig:um_inclusion}
\end{figure}

Let us now introduce three particular cases or subfamilies of these three 
latter update mode families. 
The \emph{parallel update mode} $\mu_\p = (\integers{n})$ makes every automaton 
update its state at each time step, such that $\forall t \in \N, 
\mu_\p^\star(t) = \integers{n}$.
A \emph{bipartite update mode} $\mu_\bip = (B_0, B_1)$ is a block-sequential 
update mode composed of two blocks such that the automata in a same block do 
not act on each other (notice that if the grid is finite and the boundary condition is periodic, then this definition induces that such update modes are necessarily associated with grids of even size, and that there are exactly two bipartite update modes, depending on if the even numbered cells are updated first or second.)
A \emph{sequential update mode} $\mu_\seq = (\phi(\integers{n}))$, where $
\phi(\integers{n}) = \{i_0\}, \dots, \{i_{n-1}\}$ is a permutation of 
$\integers{n}$, makes one and only one automaton update its state at each time 
step so that all automata have updated their state after $n$ time steps 
depending on the order induced by $\phi$.
All these update modes follow the order of inclusion pictured in 
Figure~\ref{fig:um_inclusion}.\smallskip

As we focus on periodic update modes, let us differentiate two kinds of time 
steps.
A \emph{substep} is a time step at which a subset of automata change their 
states.
A \emph{step} is the composition of substeps having occurred over a period.

\subsection{Dynamical systems}

An ECA $A$, with $f \in \integers{256}$, together with an update mode $\mu$, 
denoted by the pair $(f, \mu)$, define a \emph{discrete dynamical system}. 
$(f, M)$ denotes by extension any dynamical 
system related to $A$ under the considered update mode families, with $M \in \{\textsc{par}, \textsc{bip}, \textsc{seq}, \textsc{bp}, 
\textsc{bs}, \textsc{lc}, \textsc{per}\}$.\smallskip

Let $f$ be an ECA which is applied over a grid of size $n$ and let $\mu$ be a periodic update mode 
represented as a periodical sequence of subsets of $\integers{n}$ such that
$\mu = (B_0, ..., B_{p-1})$.
Let $F = (f, \mu)$ be a tuple composed of the global function $f$ that goes from $\B^n$ to itself defining the 
dynamical system related to ECA $f$ applied over a grid of size $n$ and update mode $\mu$. 
Let $x \in \B^n$ a configuration of $F$. \\
The \emph{trajectory} of $x$ is the infinite path 
$\T(x) \deltaeq x^0 = x \to x^1 = F(x) \to \dots \to x^t = F^t(x) \to \cdots$, 
where $t \in \N$, 
\begin{equation*}
	\begin{array}{lcl}
		F(x) & = & f_{B_{p-1}} \circ \dots \circ f_{B_0}\text{, with } \forall k 
			\in \integers{p}, \forall i \in \integers{n}, f_{B_k}(x)_i = 
			\begin{cases}
				f_i(x) & \text{if } i \in B_k\text{,}\\
				x_i & \text{otherwise,}
			\end{cases}\\		
			\text{and }\\			
F^t(x) & = & 
			\underbrace{F \circ \dots \circ F}_{t \text{ times}}(x)\text{,}
	\end{array}
\end{equation*}\smallskip

\begin{figure}[t!]
\includegraphics[width=\textwidth]{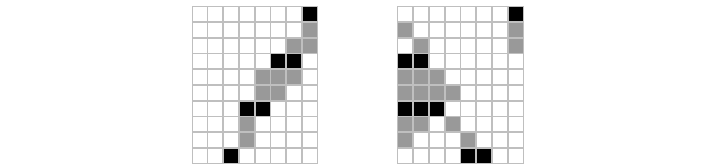}
	\caption{Space-time diagrams (time going downward) representing the $3$ first 
		(periodical) steps of the evolution of configuration 
		$x = (0,1,1,0,0,1,0,1)$ of dynamical systems 
		(left) $(156, \mu_\bs)$, and
		(right) $(178, \mu_\bp)$, %and (right) $(184, \mu_\lc)$, 
         where 
		$\mu_\bs = (\{1,3,4\}, \{0,2,6\}, \{5,7\})$, and
		$\mu_\bp = \{(1,3,4), (0,2,6), (5), (7)\}$, 
		The configurations obtained at each step are depicted by lines with cells 
		at state $1$ in black. 
		Lines with cells at state $1$ in light gray represent the configurations 
		obtained at substeps.
		Remark that $x$ belongs to a limit cycle of length $3$ (resp. $2$) in 
		$(156, \mu_\bs)$ (resp. $(178, \mu_\bp)$).}
	\label{fig:space-time-diag}
\end{figure}

In the context of ECA, it is convenient to represent trajectories by 
\emph{space-time diagrams} which give a visual aspect of the latter, as 
illustrated in Figure~\ref{fig:space-time-diag}. 
The \emph{orbit} of $x$ is the set $\O(x)$ composed of all the configurations 
which belongs to $\T(x)$.
Since $f$ is defined over a grid of finite size and the boundary condition is periodic, the temporal evolution of $x$ governed by the 
successive applications of $F$ leads it undoubtedly to enter into a \emph{limit 
phase}, \ie a cyclic subpath $\C(x)$ of $\T(x)$ such that $\forall y = 
F^k(x) \in \C(x), \exists t \in \N, F^t(y) = y$, with $k \in \N$.
$\T(x)$ is this separated into two phases, the limit phase and the 
\emph{transient phase} which corresponds to the finite subpath 
$x \to \dots \to x^\ell$ of length $\ell$ such that $\forall i \in 
\integers{\ell + 1}, \nexists t \in \N, x^{i + t} = x^{i}$. 
The \emph{limit set} of $x$ is the set of configurations belonging to 
$\C(x)$.\smallskip

From these definitions, we derive that $F$ can be represented as a graph 
$\G_F = (\B^n, T)$, where $(x,y) \in T \subseteq \B^n \times \B^n \iff 
y = F(x)$.
In this graph, which is classically called a \emph{transition graph}, 
the non-cyclic (resp. cyclic) paths represent the transient (resp. limit) 
phases of $F$. 
More precisely, the cycles of $\G_F$ are the \emph{limit cycles} of $F$. 
When a limit cycle is of length $1$, we call it a \emph{fixed point}. 
Furthermore, if the fixed point is such that all the cells of the configuration 
has the same state, then we call it an \emph{homogeneous fixed point.}\smallskip

%\emph{Notation for ECA rules}
Eventually, we make use of the following specific notations.
Let $x \in \B^n$ be a configuration and $[i,j] \subseteq \integers{n}$ be a 
subset of cells. 
We denote by $x_{[i,j]}$ the projection of $x$ on $[i,j]$. 
Since we work on ECA, such a projection defines a sub-configurations and can be 
of three kinds: 
\begin{itemize}
    \item $i < j$ and $x_{[i,j]} = (x_i, x_{i+1}, \dots, x_{j-1}, x_{j})$,
    \item or $i = j$ and $x_{[i,j]} = (x_i)$,
    \item or $i > j$ and $x_{[i,j]} = (x_i, x_{i+1}, \dots, x_n, 
x_0, \dots, x_{j-1}, x_j$).
\end{itemize}

Thus, given $x \in \B^n$ and $i \in \integers{n}$, an ECA $f$ can be rewritten as \\
$f(x) = (f(x_{[n-1,1]}), f(x_{[0,2]}), \dots, f(x_{[i,i+2]}), \dots, f(x_{[n-3,n-1]}),  
f(x_{[n-2,0]}).$\\
%''}
Abusing notations, the word $u \in \mathbb{B}^k$ is called a \emph{wall} for a 
dynamical system if for all $a, b \in \B$, $f(aub) = u$, and we assume in this work 
that walls are of size $2$, i.e. $k = 2$, unless otherwise stated. 
Such a word $u$ is an \emph{absolute wall} (resp. a \emph{relative wall}) for an ECA 
rule if it is a wall for any update mode (resp. strict subset of update modes).
We say that a rule $F$ can \emph{dynamically create new walls} if there is a time $t\in\N$ and an initial configuration $x^0\in\B^n$ such that $x^t(=F^t(x^0))$ has a higher number of walls than $x^0$. Finally, we say that a configuration $x$ is an \emph{isle} of $1's$ (resp. an isle of $0's$) if 
there exists an interval $I = [a,b] \subseteq \integers{n}$ such that $x_i = 1$ 
(resp. $x_i = 0$) for all $i \in I$ and $x_i = 0$ (resp. $x_i = 1$) otherwise.

\section{Results}
\label{sec:res}
In this section we will present the results of our research. 
In the first part we show the theoretical results, where we work towards a theoretical classification for ECA.
And in the second part we offer some experimental results where we compare the dynamics of energy and density under different update modes for some of the rules whose dynamics we have not yet been able to classify.

\subsection{Theoretical Results}
\label{subsec:thm}

\subsubsection{$\Theta(1)$}
\label{sec:constante}
There are 25 ECA that always reach limit cycles whose length does not depend on the length of the ring. These are divided into two groups:
\begin{itemize}
\item Rules that always reach fixed points, regardless of the length of the ring or the update mode under which they are applied. These rules are:  0, 4, 8, 12, 44, 72, 76, 78, 128, 132, 136, 140, 164, 200 and $204$.
\item Rules that can reach limit cycles of length $2$, $3$ or $6$. The length of the cycle does not depend on the length of the ring, and changing the update mode cannot increase the maximum length of the cycle. These rules are: $5$, $13$, $28$, $29$, $32$, $36$, $51$, $77$, $160$ and $232$.
\end{itemize}

\begin{table}[!ht]
\begin{center}
\begin{tabular}{|c|c|c|c|c|c|c|c|c|}
\hline
Rule&111&110&101&100&011&010&001&000\\\hline
128&1&0&0&0&0&0&0&0\\\hline
132&1&0&0&0&0&1&0&0\\\hline
136&1&0&0&0&1&0&0&0\\\hline
140&1&0&0&0&1&1&0&0\\\hline
\end{tabular}
\end{center}
\end{table}

\begin{theorem}
	\label{thm:128puntofijo}
	Rules $128$, $132$, $136$, $140$ always reach fixed points.
\end{theorem}

\begin{proof}
Let $x$ be an arbitrary initial configuration. Since we know that $x=0^n$ and $x=1^n$ are fixed points, let us assume that $x$ can be written as
$$x=0^{a_1}1^{b_1}\dots0^{a_k}1^{b_k}$$
such that $a_1+b_1+\dots+a_k+b_k=n$.\\
Since $f_{r}(001)=f_{r}(100)=0$, for $r\in\{128, 132, 136, 140\}$, groups of ones separated by two zeros cannot increase in size. And for these rules, groups of ones never increase in size, because $f_{128}(101)=f_{132}(101)=f_{136}(101)=f_{140}(101)=1$.\\
 neighbors updates first, then we have two consecutive $0$'s.\\
On the other hand, since $f_{r}(110)=0$ for $r\in\{128, 132, 136, 140\}$, groups of $1$'s always decrease in size at the right side by at least one cell per iteration. Additionally, since $f_{128}(011)=f_{132}(011)=0$, the size of the groups will decrease faster for these rules.\\
In every iteration, groups of ones will continue to decrease in size until only isolated $1$'s remain.\\
In the case of rules $128$ and $136$ those ones will disappear finally reaching $0^n$ which is a fixed point, because $f_{128}(010)=f_{136}(010)$.\\
For rules $132$ and $140$ the isolated $1$'s will be preserved, since $f_{132}(010)=f_{140}(010)==1$; with the fixed points being written as $x=0^{a_1}10^{a_2}\dots0^{a_{k-1}}0^{a_k}1$.\\
\end{proof}

\begin{table}[!ht]\label{table:0}
\begin{center}
\begin{tabular}{|c|c|c|c|c|c|c|c|c|}
\hline
Rule&111&110&101&100&011&010&001&000\\\hline
0&0&0&0&0&0&0&0&0\\\hline
4&0&0&0&0&0&1&0&0\\\hline
8&0&0&0&0&1&0&0&0\\\hline
12&0&0&0&0&1&1&0&0\\\hline
\end{tabular}
\end{center}
\end{table}

\begin{corollary}
	Rules $0$, $4$, $8$, $12$ always reach fixed points.
\end{corollary}
\begin{proof}[Sketch of proof]
First, note that $f_r(111)=0$, which forces all groups of ones of size at least three to be split, and then proceed with the exact same reasoning as the previous theorem.
\end{proof}

\begin{table}[!ht]\label{table:32}
\begin{center}
\begin{tabular}{|c|c|c|c|c|c|c|c|c|}
\hline
Rule&111&110&101&100&011&010&001&000\\\hline
32 &0&0&1&0&0&0&0&0\\\hline
36 &0&0&1&0&0&1&0&0\\\hline
44 &0&0&1&0&1&1&0&0\\\hline
160&1&0&1&0&0&0&0&0\\\hline
164&1&0&1&0&0&1&0&0\\\hline
\end{tabular}
\end{center}
\end{table}

\begin{theorem}
	Rules $32$ and $160$ can only reach cycles of length 2 under parallel update mode.
\end{theorem}
\begin{proof}[Sketch of proof.]
Note that for rules $32$ and $160$ there is a special case with parallel update schedule, in which for the configuration $x=(10)^{\frac{n}{2}}$ we have that $f_{32}((10)^{\frac{n}{2}})=f_{160}((10)^{\frac{n}{2}})=(01)^{\frac{n}{2}}$, which leads to cycles of length 2. Nevertheless, even if we start with the configuration $x$, if there is just one cell that updates asynchronously with respect to the others, the cycle is broken and we reach a fixed point. This is because when a cell is updated alone, it will produce a groups of three $1$'s or three $0$'s, and either is enough to lead to a fixed point.
\end{proof}

\begin{theorem}
	Rules $36,44$ and $164$ always reach fixed points regardless of the update mode.
\end{theorem}

\begin{proof}[Sketch of proof]
Follows the same reasoning as the proof of theorem \ref{thm:128puntofijo}.
\end{proof}

\begin{table}[!ht]
\begin{center}
\begin{tabular}{|c|c|c|c|c|c|c|c|c|}
\hline
Rule&111&110&101&100&011&010&001&000\\\hline
72  &0  &1  &0  &0  &1  &0  &0  &0\\\hline
76  &0  &1  &0  &0  &1  &1  &0  &0\\\hline
\end{tabular}
\end{center}
\end{table}

\begin{theorem}
	\label{thm:72puntofijo}
	Rules $72$ and $76$ always reach fixed points.
\end{theorem}

\begin{proof}
We know that $x=0^n$ is a fixed point, so let $x=0^{a_1}1^{b_1}\dots0^{a_k}1^{b_k}$, since  $f_r(111)=0$ for $r\in\{72,76\}$, meaning that the configuration $y=1^n$ will become $f(y)=0^{a_1}1^{b_1}\dots0^{a_k}1^{b_k}$.\\
Since $f_r(011)=f_r(110)=1$ and $f_r(001)=f_r(100)=0$, for all $r\in\{72,76\}$, groups composed of exactly two $1$'s are walls, meaning that once the word $0110$ appears it will not change.\\
This also means that groups of ones larger than two cannot increase in size.\\
In any other case, groups of $1$'s of size at least three will decrease in size in each iteration until:
\begin{itemize}
\item case 1: the group is left with exactly two ones, in which case it becomes a wall and remains unchanging.
\item case 2: for rule $72$, $f_{72}(010))=0$, isolated $1$'s will disappear, with which the resulting fixed points can be written as $0^{a_1'}110^{a_2'}\dots110^{a_{k'}'}11$.
\item case 3: for rule $76$, $f_{76}(010)=1$, isolated $1$'s will be preserved, which means that the resulting fixed points can be written as $0^{a_1'}1^{b_1'}0^{a_2'}1^{b_2'}\dots0^{a_{k'}'}1^{b_{k'}'}$, with $b_i\in\{1,2\},i\in\{1,\dots,k'\}$.

\end{itemize}
\end{proof}

\begin{table}[!ht]
\begin{center}
\begin{tabular}{|c|c|c|c|c|c|c|c|c|}
\hline
Rule&111&110&101&100&011&010&001&000\\\hline
200 &1  &1  &0  &0  &1  &0  &0  &0\\\hline
204 &1  &1  &0  &0  &1  &1  &0  &0\\\hline
\end{tabular}
\end{center}
\end{table}

\begin{theorem}
	\label{thm:200puntofijo}
	Rules $200$ and $204$ always reach fixed points.
\end{theorem}

\begin{proof}
Rule $204$ is the identity rule, meaning that all configurations are fixed points, and the only difference between rules $200$ and $204$ is $f_{204}(010)=1\neq0=f_{200}(010)$, which means that for rule $204$ all configurations are fixed points except for the ones that have isolated $1$'s, because said $1$'s will disappear as soon as the cell in which they are located is updated.
\end{proof}

\begin{table}[!ht]
\begin{center}
\begin{tabular}{|c|c|c|c|c|c|c|c|c|}
\hline
Rule&111&110&101&100&011&010&001&000\\\hline
232 &1  &1  &1  &0  &1  &0  &0  &0\\\hline
\end{tabular}
\end{center}
\end{table}

\begin{theorem}
	\label{thm:232puntofijo}
	Rule $232$ always reaches cycles of length at most 2.
\end{theorem}

\begin{proof}
We know that $x=0^n$ and $x=1^n$ are fixed points, so let $x=0^{a_1}1^{b_1}\dots0^{a_k}1^{b_k}$, with $a_1+b_1+\dots+a_k+b_k=n$.\\
Since $f_{232}(010)=0$ and $f_{232}(101)=1$, if there are isolated $0$'s or $1$'s, then they will disappear.\\
If $b_i>1$ with $i\in\{1,\dots,k\}$, those groups of ones will not decrease in size, since $f_{232}(111)=f_{232}(011)=f_{232}(110)=1$. Similarly, if $a_i>1$ with $i\in\{1,\dots,k\}$, those groups of zeros will not decrease in size, since $f_{232}(000)=f_{232}(001)=f_{232}(100)=0$. This leads to a configuration that can be written as $x'=0^{a_1'}1^{b_1'}\dots0^{a_k'}1^{b_k'}$, with $a_i,b_i>1$ for all $i\in\{1,\dots,k\}$.\\
There is a special case for the parallel update schedule, where if $x=(01)^{\frac{n}{2}}$ and $y=(10)^{\frac{n}{2}}$, then $f_{232}(x)=y$ and $f_{232}(y)=x$, which is when we find cycles of length 2.\\
However, even if the initial configuration is $x=(01)^{\frac{n}{2}}$, we only need one cell to be updated before or after its neighbours to give rise to a domino effect where we eventually reach a fixed point.\\
Indeed, if $x=(01)^{\frac{n}{2}-1}11$, this means that the wall $111$ has appeared, and note that one of the cells on either side of the wall must be equal to $0$, and once that cell is updated, the number of consecutive $1$'s will increase. So after less than $n$ iterations, the configuration will have reached the fixed point $1^n$. Analogous for $x=(01)^{\frac{n}{2}-1}00$, where the configuration reaches the fixed point $0^n$.\\
If there is more than one cell that updates before or after its neighbours, then the configuration $x=(01)^{\frac{n}{2}}$ is destroyed faster and it reaches fixed points that can be written as $x'=0^{a_1'}1^{b_1'}\dots0^{a_k'}1^{b_k'}$, same as before.
\end{proof}

\begin{table}[!ht]
\begin{center}
\begin{tabular}{|c|c|c|c|c|c|c|c|c|}
\hline
Rule&111&110&101&100&011&010&001&000\\\hline
51 &0  &0  &1  &1  &0  &0  &1  &1\\\hline
\end{tabular}
\end{center}
\end{table}

\begin{theorem}
	\label{thm:51puntofijo}
	Rule $51$ always reaches cycles of length at most 2.
\end{theorem}

\begin{proof}
This rule is the negation function. Each time a cell is updated its value will change to the opposite than the one it had. This means that a single cell will have a cycle of period 1 if it is updated an even number of times in each step, and of period 2 if the number of times it updates in each step is odd.

Obviously, the cycle of the entire configuration will have a period of length 2 is there is a cell that is updated an odd number of times, and it will be a fixed point if all cells are updated an even number of times.
\end{proof}

\subsubsection{$\mathcal{O}(n)$}
\label{sec:linear}
We have found $33$ ECA such that the longest cycle we can reach is directly proportional to the length of the ring, regardless of the update mode. These rules are: $2$, $3$, $6$, $7$, $10$, $11$, $14$, $15$, $18$, $19$, $23$, $24$, $26$, $27$, $33$, $34$, $35$, $38$, $40$, $42$, $43$, $46$, $50$, $94$, $104$, $130$, $134$, $138$, $142$, $162$, $168$, $170$ and $172$.

There are three rules that behave differently to the rest. These rules are $40$, $168$ and $172$, and they are only able to reach limit cycles of length $\mathcal{O}(n)$ when all cells are updated simultaneously, that is, with the parallel update mode. If there is just one cell that is updated before or after the rest, or that is updated more than once per iteration, the rule is no longer able to reach these limit cycles and instead reaches fixed points for all update modes. In simpler terms, rules $40$, $168$ and $172$ always reach fixed points for all update modes \textit{except} the parallel one.

\begin{table}[!ht]
\begin{center}
\begin{tabular}{|c|c|c|c|c|c|c|c|c|}
\hline
Rule&111&110&101&100&011&010&001&000\\\hline
2   &0  &0  &0  &0  &0  &0  &1  &0\\\hline
6   &0&0&0&0&0&1&1&0\\\hline
10  &0&0&0&0&1&0&1&0\\\hline
14  &0&0&0&0&1&1&1&0\\\hline
\end{tabular}
\end{center}
\end{table}

\begin{theorem}\label{thm:2}
$(2,\bs)$ of size $n$ always reach limit cycles of size $\mathcal{O}(n)$.
\end{theorem}
\begin{proof}
Since the update schedule is sequential, that means that two consecutive cells always update one after the other, which means that $\mu_i\leq\mu_{i+1}$ or $\mu_i\leq\mu_{i+1}$. Let $L=\{\ell_j\}_{j=1}^N=\{i\in\{0,\dots,n-1\}|\mu_{\ell_i}\leq\mu_{\ell_i+1}\}$ the set of cells that are updated before or at the same time as their right-hand neighbor.

Let $x=1^k0^{n-k}$ be an initial configuration that has an isle of $1$'s and all the state of all other cells is equal to $0$. 

We can find four cases:\\
\underline{Case 1:} The first cell of the isle updates \textbf{after} the cell outside the isle, and last cell updates \textbf{before} the one outside, denoted $\mu_{p-1}\leq\mu_{p}$ and $\mu_{q}\leq\mu_{q+1}$. This means that $p-1,q\in L$. Let $p-1=\ell_j$ and $q=\ell_k$, with $k=j+s$ (the reasoning is the same if $k=j+1$).

\begin{center}
\begin{tabular}{ccc|ccc|ccc|ccc|ccc|ccc}
 & &$\ell_{j-1}$ & & &$\ell_j$&$p$&$\dots$&$\ell_{k-1}$&&$\dots$&$\ell_{k}$&&&$\ell_{k+1}$&&\\\hline
0&$\dots$&0&0&$\dots$&0&1    &$\dots$&1       &1&$\dots$&1&0&$\dots$&0&0&$\dots$&0\\\hline
\end{tabular}
\end{center}

Since all cells between $\ell_{j-1}$ and $\ell_{j}$ must be updated from right to left, and $f_{2}(001)=1$, we will gain $1$'s to the left until we reach $\ell_{j-1}$. Analogously, because $f_{2}(110)=0$ and $f_2(111)=0$, on the right side of the isle we will lose $1$'s until we reach $\ell_{j}$.

\begin{center}
\begin{tabular}{ccc|ccc|ccc|ccc|ccc|ccc}
 & &$\ell_{j-1}$ & & &$\ell_j$&$p$&$\dots$&$\ell_{k-1}$&&$\dots$&$\ell_{k}$&&&$\ell_{k+1}$&&\\\hline
0&$\dots$&0&0&$\dots$&0&1    &$\dots$&1       &1&$\dots$&1&0&$\dots$&0&0&$\dots$&0\\\hline
0&$\dots$&0&1&$\dots$&1&0    &$\dots$&0       &0&$\dots$&0&0&$\dots$&0&0&$\dots$&0\\\hline
\end{tabular}
\end{center}

After the first iteration is completed, the isle is again delimited by cells that belong to $L$, so the process repeats. It is easy to see that after $N<n$ iterations the isle will return to its position from $\ell_{j-1}$ to $\ell_j$.

\underline{Case 2:} First and last cell of the isle update \textbf{after} the cells outside the isle, denoted $\mu_{p-1}\leq\mu_{p}$ and $\mu_{q}>\mu_{q+1}$. This means that $p-1\in L$.

\begin{center}
\begin{tabular}{ccc|ccc|cccccc|ccc}
 & &$\ell_{j-1}$ & & &$\ell_j$&$p$&$\dots$&$q$&$q+1$&$\dots$&$\ell_{j+1}$&&\\\hline
0&$\dots$&0&0&$\dots$&0&1    &$\dots$&1       &0&$\dots$&0&0&$\dots$&0\\\hline
\end{tabular}
\end{center}

Since cell $l_j$ must update before $p$ does, and $f_{2}(001)=1$,then we gains ones to the left, and since all cells between $\ell_{j-1}$ and $\ell_j$ update from right to left, we continue to gain ones until we reach $\ell_{j-1}$. On the other hand, since all cells between a $\ell_{j}$ and $q$ also update from right to left, we will lose ones at the right side of the isle, until we reach $\ell_{j}$. In effect, the isle moves.

\begin{center}
\begin{tabular}{ccc|ccc|cccccc|ccc}
 & &$\ell_{j-1}$ & & &$\ell_j$&$p$&$\dots$&$q$&$q+1$&$\dots$&$\ell_{j+1}$&&\\\hline
0&$\dots$&0&0&$\dots$&0&1    &$\dots$&1       &0&$\dots$&0&0&$\dots$&0\\\hline
0&$\dots$&0&1&$\dots$&1&0    &$\dots$&0       &0&$\dots$&0&0&$\dots$&0\\\hline
\end{tabular}
\end{center}

Note that if there are one or more cells that belong to $L$ inside the aisle, then we will only lose ones starting from $q$ until the first cell in $L$ that it reaches.

From here it develops following the analysis from Case 1.

\underline{Case 3:} First and last cells of the isle update \textbf{before} the ones outside the isle, denoted $\mu_{p-1}>\mu_{p}$ and $\mu_{q}<\mu_{q+1}$. This means that $q\in L$, let $q=\ell_k$.

\begin{center}
\begin{tabular}{ccc|ccc|cccc|ccc|ccc}
 & & & & &$p-1$&$p$&$p+1$&$\dots$&$\ell_{k-1}$& &$\dots$&$\ell_{k}$&$q+1$&\\\hline
0&$\dots$&0&0&$\dots$&0&1 &1   &$\dots$&1       &1&$\dots$&1&0&$\dots$&0\\\hline
\end{tabular}
\end{center}

We know from Case 2 that after the first iteration we will lose $1$'s from $l_k$ to $l_{k-1}$. On the other hand, since there could be a cell between $p$ and $\ell_k$ that is in $L$, that cell would update before its left side neighbour, meaning that it would change its state. But we know that the first cell of the isle of ones has to update before the one outside, so the neighborhood will be $011$ and $f_{2}(011)=0$, so we also lose $1$'s to both sides of the isle.

\begin{center}
\begin{tabular}{ccc|ccc|cccc|ccc|ccc}
 & & & & &$p-1$&$p$&$p+1$&$\dots$&$\ell_{k-1}$& &$\dots$&$\ell_{k}$&$q+1$&\\\hline
0&$\dots$&0&0&$\dots$&0&1 &1   &$\dots$&1       &1&$\dots$&1&0&$\dots$&0\\\hline
0&$\dots$&0&0&$\dots$&0&0 &0   &$\dots$&0       &0&$\dots$&0&0&$\dots$&0\\\hline
\end{tabular}
\end{center}

This case always leads to a fixed point.

\underline{Case 4:}  First cell of the isle updates \textbf{before} the one outside and last cell of the isle updates \textbf{after} the one outside of it, denoted $\mu_{p-1}>\mu_{p}$ and $\mu_{q}>\mu_{q+1}$. This means that $p,q\not\in L$.

\begin{center}
\begin{tabular}{cccccc|ccc|cccccc}
 &$p-1$&$p$&$p+1$&$\dots$&$\ell_j$&$\dots$&&$\ell_{k}$& &$\dots$&$q$&$q+1$&\\\hline
$\dots$&0&1&1&$\dots$&1 &1   &$\dots$&1       &1&$\dots$&1&0&$\dots$&0\\\hline
\end{tabular}
\end{center}

We know from Case 2 that we will $1$'s to the left because $\mu_{p-1}>\mu_{p}$, which means that this case always leads to Case 3 (which leads to fixed points):

\begin{center}
\begin{tabular}{cccccc|ccc|cccccc}
 &$p-1$&$p$&$p+1$&$\dots$&$\ell_j$&$\dots$&&$\ell_{k}$& &$\dots$&$q$&$q+1$&\\\hline
$\dots$&0&1&1&$\dots$&1 &1   &$\dots$&1       &1&$\dots$&1&0&$\dots$&0\\\hline
$\dots$&0&0&0&$\dots$&0 &0   &$\dots$&0       &0&$\dots$&0&0&$\dots$&0\\\hline
\end{tabular}
\end{center}

Note that if there are more than one isle of 1's in the initial configuration, and since isles always move from right to left (going from $\ell_j$ to $\ell_{j-1}$), they cannot interact with each other under Rule 2, from where we have the result.
\end{proof}

\begin{theorem}\label{thm:6y10y14}
Rules $(6,\bs),(10,\bs)$ and $(14,\bs)$ have cycles of length $\mathcal{O}(n)$.
\end{theorem}

\begin{proof}[Sketch of proof.]
Following a similar analysis as the previous theorem, we know that when the initial configuration is $x=1^k0^{n-k}$, we have that cases 1 and 2 give us the same result.

But since $f_6(010)=1$, cases 3 and 4 are able to reach cycles of length $\mathcal{O}(n)$.

Similarly, since $f_{10}(011)=1$, cases 3 and 4 are also able to reach cycles of length $\mathcal{O}(n)$ for this rule.

And finally, because $f_{14}(010)=f_{14}(011)=1$, we can repeat the previous analysis.
\end{proof}

\begin{table}[!ht]
\begin{center}
\begin{tabular}{|c|c|c|c|c|c|c|c|c|}
\hline
Rule&111&110&101&100&011&010&001&000\\\hline
130 &1  &0  &0  &0  &0  &0  &1  &0\\\hline
134 &1  &0  &0  &0&0&1&1&0\\\hline
138 &1  &0  &0  &0&1&0&1&0\\\hline
142 &1  &0  &0  &0&1&1&1&0\\\hline
\end{tabular}
\end{center}
\end{table}

\begin{corollary}
Rules $(130,\bs),(134,\bs),(138,\bs)$ and $(142,\bs)$ have cycles of length $\mathcal{O}(n)$
\end{corollary}
\begin{proof}[Sketch of proof.]
The only difference between these rules and the previous ones is that $f(111)=1$, which prevents isles of ones from being separated. The rest of the analysis is analogous.
\end{proof}

\begin{table}[!ht]
\begin{center}
\begin{tabular}{|c|c|c|c|c|c|c|c|c|}
\hline
Rule&111&110&101&100&011&010&001&000\\\hline
34  &0  &0  &1  &0  &0  &0  &1  &0\\\hline
38  &0  &0  &1  &0&0&1&1&0\\\hline
42  &0  &0  &1  &0&1&0&1&0\\\hline
46  &0  &0  &1  &0&1&1&1&0\\\hline
\end{tabular}
\end{center}
\end{table}

\begin{corollary}\label{thm:34}
Rules $(34,\bs),(38,\bs),(42,\bs)$ and $(46,\bs)$ have cycles of length $\mathcal{O}(n)$
\end{corollary}
\begin{proof}[Sketch of proof.]
The only difference between these rules and the ones from theorems \ref{thm:2} and \ref{thm:6y10y14} is that $f(101)=1$, which could allow isles to fuse together. The rest of the analysis is analogous.
\end{proof}

\begin{table}[!ht]
\begin{center}
\begin{tabular}{|c|c|c|c|c|c|c|c|c|}
\hline
Rule&111&110&101&100&011&010&001&000\\\hline
162 &1  &0  &1  &0  &0  &0  &1  &0\\\hline
166 &1  &0  &1  &0  &0  &1  &1  &0\\\hline
170 &1  &0  &1  &0  &1  &0  &1  &0\\\hline
174 &1  &0  &1  &0  &1  &1  &1  &0\\\hline
\end{tabular}
\end{center}
\end{table}

Note that Rule 174 is equivalent to rule 138, while rule 166 is equivalent to rule 154. Furthermore, the result is weaker for rule 154 than the ones we have seen so far, because when two neighbouring cells can be updated simultaneously the interaction between isles of $1$'s generates a longer dynamic.

\begin{corollary}\label{thm:162}
Rules $(154,\seq),(162,\bs)$ and $(170,\bs)$ have cycles of length $\mathcal{O}(n)$
\end{corollary}
\begin{proof}[Sketch of proof.]
For $(154,\seq)$ we must consider $L=\{\ell_j\}_{j=1}^N=\{i\in\{0,\dots,n-1\}|\mu_{\ell_i} < \mu_{\ell_i+1}\}$ and the rest of the analysis proceeds analogously.

The reasoning for the other rules is identical to the rest.
\end{proof}

\begin{table}[!ht]
\begin{center}
\begin{tabular}{|c|c|c|c|c|c|c|c|c|}
\hline
Rule&111&110&101&100&011&010&001&000\\\hline
138&1&0&0&0&1&0&1&0\\\hline
170&1&0&1&0&1&0&1&0\\\hline
\end{tabular}
\end{center}
\end{table}

\begin{theorem}\label{thm:170lc}
$(138,\lc)$ and $(170,\lc)$ of size $n$ have largest limit cycles of size $\mathcal{O}(n)$.
\end{theorem}
\begin{proof}
Let $x$ be a configuration in some limit cycle for $(170,\lc)$ (analogous for $(138,\lc)$\\
By definition $f_{170}(001)=f_{138}(001)=1$ and $f_{170}(110)=f_{138}(110)=0$, so we know that the last time the cell to the left of the isle was updated must have been \textit{before or at the same time as} the first cell of the isle was updated (otherwise it would have also become 1 during the iteration). Similarly, the last cell of the isle must have been updated for the last time \textit{before or simultaneously} to the cell at its right.

This means that by the time one iteration is completed, the isle of ones must have moved at least one cell to the left, because both the first and last cells of the isle must have been updated at least once.

Now, let $L=\{\ell_j\}_{j=0}^N$ the set of cells that under the update schedule $\mu$ are updated for the last time before or after the cell to its right.

It is easy to see that the only locations in which the isle of ones can stop moving are cells in $L$, the cycle cannot be larger than $N\leq n$.

Since $x$ belongs to a cycle, its position is delimited by cells in $L$, which means that the isles cannot interact with each other. 
\end{proof}

\begin{corollary}
$(138,\p),(138,\seq),(138,\bs),(138,\bp)$ and $(170,\p)$,\\$(170,\seq)$,$(170,\bs)$,$(170,\bp)$ of size $n$ have largest limit cycles of size $\mathcal{O}(n)$.
\end{corollary}

\begin{table}[!ht]
\begin{center}
\begin{tabular}{|c|c|c|c|c|c|c|c|c|}
\hline
Rule&111&110&101&100&011&010&001&000\\\hline
40 &0&0&1&0&1&0&0&0\\\hline
168&1&0&1&0&1&0&0&0\\\hline
172&1&0&1&0&1&1&0&0\\\hline
\end{tabular}
\end{center}
\end{table}

\begin{table}[!ht]\label{table:168}
\begin{center}\renewcommand{\arraystretch}{2}
\begin{tabular}{|C{1.5cm}|C{1.5cm}C{2cm}C{2cm}C{2cm}C{2cm}|}
\hline
Rule&Parallel&Bipartite&$\bs\setminus\p$&$\bp\setminus\p$&$\lc\setminus\p$\\\hline
40 &$\mathcal{O}(n)$&$1$&$1$&$1$&$1$\\\hline
168&$\mathcal{O}(n)$&$1$&$1$&$1$&$1$\\\hline
172&$\mathcal{O}(n)$&$1$&$1$&$1$&$1$\\\hline
\end{tabular}
\end{center}
\caption{Longest limit cycle for Rules 40, 168 and 172, according to update schedule.}
\end{table}

\begin{theorem}
	\label{thm:168puntofijo}
	Rules $168$ and $172$ always reach fixed points under update schedules different than the parallel.
\end{theorem}

\begin{proof}
Let $x$ be an arbitrary initial configuration. Since we know that $x=0^n$ and $x=1^n$ are fixed points, let us assume that $x$ can be written as
$$x=0^{a_1}1^{b_1}\dots0^{a_k}1^{b_k}$$
such that $a_1+b_1+\dots+a_k+b_k=n$.\\
Since $f_{r}(001)=f_{r}(100)=0$, for $r\in\{168,172\}$, groups of ones separated by two zeros cannot increase in size.\\
When there exists a $j\in\{1,\dots,k\}$ such that $a_j=1$, those $0$'s will disappear for rules $168$ and $172$ ($f_{168}(101)=f_{172}(101)=1$) if said $0$ is updated \textbf{before} its neighbours. However, if either of its neighbours updates \textbf{first}, then we have two consecutive $0$'s.\\
On the other hand, since $f_{r}(110)=0$ for $r\in\{168,172\}$, groups of $1$'s always decrease in size at the right side by at least one cell per iteration.\\
In every iteration, groups of ones will continue to decrease in size until only isolated $1$'s remain.\\
In the case of rule $168$ those ones will disappear(because $f_{168}(010)=0$) finally reaching $0^n$ which is a fixed point.\\
And for rule $172$ the isolated $1$'s will be preserved, since $f_{172}(010)=1$; with the fixed points being written as $x=0^{a_1}10^{a_2}\dots0^{a_{k-1}}0^{a_k}1$.\\
\end{proof}

\begin{theorem}
\label{thm:168paralelo}
	Rules $(40,\p)$, $(168,\p)$ and $(172,\p)$ can reach cycles $\mathcal{O}(n)$.
\end{theorem}
\begin{proof}
For rules $40$ and $168$\\
Starting with an initial configuration $x=(10)^{\frac{n}{2}-1}11$, $f(x)=(01)^{\frac{n}{2}-1}11$, $f^2(x)=(10)^{\frac{n}{2}-2}11(10)$.
\begin{equation*}
\begin{split}
f^3(x)&=(01)^{\frac{n}{2}-2}11(01)\\
f^4(x)&=(10)^{\frac{n}{2}-3}11(10)^2\\
\vdots&\\
f^{t}(x)&=(10)^{\frac{n}{2}-\frac{t}{2}-1}11(10)^{\frac{t}{2}}\text{  if t is even}\\
f^{t}(x)&=(01)^{\frac{n}{2}-\frac{t-1}{2}-1}11(01)^{\frac{t-1}{2}}\text{  if t is odd}\\
\vdots&\\
f^{n-4}(x)&=(10)^{\frac{n}{2}-\frac{n-4}{2}-1}11(10)^{\frac{n-4}{2}}=(10)11(10)^{\frac{n-4}{2}}\\
f^{n-3}(x)&=(01)^{\frac{n}{2}-\frac{n-4}{2}-1}11(01)^{\frac{n-4}{2}}=(01)11(01)^{\frac{n-4}{2}}\\
f^{n-2}(x)&=(10)^{\frac{n}{2}-\frac{n-2}{2}-1}11(10)^{\frac{n-2}{2}}=11(10)^{\frac{n-4}{2}}\\
f^{n-1}(x)&=(01)^{\frac{n}{2}-\frac{n-2}{2}-1}11(01)^{\frac{n-4}{2}}=11(01)^{\frac{n-4}{2}}\\
\end{split}
\end{equation*}
And finally, $f^n(x)=(10)^{\frac{n}{2}-1}11$. Note that if $n$ is an odd number, then we can start from $x=(10)^{\frac{n-1}{2}}1$.

For rule $172$, starting with an initial configuration $x=1^{n-1}0$, we have $f(x)=1^{n-2}01$, $f^2(x)=1^{n-3}01^2$
\begin{equation*}
\begin{split}
f^3(x)&=1^{n-4}01^3\\
\vdots&\\
f^t(x)&=1^{n-(t+1)}01^t\\
\vdots&\\
f^{n-1}(x)&=1^{n-(n-1+1)}01^{n-1}=01^{n-1}\\
\end{split}
\end{equation*}
And finally, $f^n(x)=1^{n-1}0$.

In all three cases cases, if there is just one cell that does not update at the same time as all the others all configurations reach a fixed point.
\end{proof}

\clearpage
\newpage
\subsection{$\Omega\left(2^{(\sqrt{n\log(n)})}\right)$}
In this section we will show rules for which we have found update modes such that cycles of superpolynomial length can be reached. \textcolor{blue}{There are 9 rules in this category (so far).}
\begin{table}[!ht]\label{table:increasing}
\begin{center}\renewcommand{\arraystretch}{2}
\begin{tabular}{|C{0.85cm}|C{0.85cm}C{2.3cm}C{2.3cm}C{2.3cm}C{2.3cm}|}
\hline
Rule&Parallel&Sequential&\makecell{Block\\Sequential}&\makecell{Block\\Parallel}&\makecell{Local\\Clocks}\\\hline
1&$\Theta(1)$&$\mathcal{O}(n)$&$\mathcal{O}(n)$&$\Omega\left(2^{(\sqrt{n\log(n)})}\right)$&$\Omega\left(2^{(\sqrt{n\log(n)})}\right)$\\\hline
9&$\mathcal{O}(n)$&$\mathcal{O}(n)$&$\mathcal{O}(n)$&$\Omega\left(2^{(\sqrt{n\log(n)})}\right)$&$\Omega\left(2^{(\sqrt{n\log(n)})}\right)$\\\hline
56&$\mathcal{O}(n)$&$1$&$\mathcal{O}(n)$&$\Omega\left(2^{(\sqrt{n\log(n)})}\right)$&$\Omega\left(2^{(\sqrt{n\log(n)})}\right)$\\\hline
73&?&$\Theta\left(2^{(\sqrt{n\log(n)})}\right)$&$\Omega\left(2^{(\sqrt{n\log(n)})}\right)$&$\Omega\left(2^{(\sqrt{n\log(n)})}\right)$&$\Omega\left(2^{(\sqrt{n\log(n)})}\right)$\\\hline
108&$\Theta(1)$&$\Theta\left(2^{(\sqrt{n\log(n)})}\right)$&$\Omega\left(2^{(\sqrt{n\log(n)})}\right)$&$\Omega\left(2^{(\sqrt{n\log(n)})}\right)$&$\Omega\left(2^{(\sqrt{n\log(n)})}\right)$\\\hline
152&$\mathcal{O}(n)$&$1$&$\mathcal{O}(n)$&$\Omega\left(2^{(\sqrt{n\log(n)})}\right)$&$\Omega\left(2^{(\sqrt{n\log(n)})}\right)$\\\hline
156&$\Theta(1)$&$\Theta\left(2^{(\sqrt{n\log(n)})}\right)$&$\Omega\left(2^{(\sqrt{n\log(n)})}\right)$&$\Omega\left(2^{(\sqrt{n\log(n)})}\right)$&$\Omega\left(2^{(\sqrt{n\log(n)})}\right)$\\\hline
178&$\Theta(1)$&$\mathcal{O}(n)$&$\mathcal{O}(n)$&$\Omega\left(2^{(\sqrt{n\log(n)})}\right)$&$\Omega\left(2^{(\sqrt{n\log(n)})}\right)$\\\hline
184&$\mathcal{O}(n)$&$1$&$\mathcal{O}(n)$&$\Omega\left(2^{(\sqrt{n\log(n)})}\right)$&$\Omega\left(2^{(\sqrt{n\log(n)})}\right)$\\\hline
\end{tabular}
\end{center}
\caption{Longest cycle for each rule and update mode}
\end{table}

\begin{table}[!ht]
\begin{center}
\begin{tabular}{|c|c|c|c|c|c|c|c|c|}
\hline
Rule&111&110&101&100&011&010&001&000\\\hline
156&1&0&0&1&1&1&0&0\\\hline
\end{tabular}
\end{center}
\end{table}

\begin{table}[!ht]\label{table:156}
\begin{center}\renewcommand{\arraystretch}{2}
\begin{tabular}{|C{1cm}|C{1.2cm}C{2.3cm}C{2.3cm}C{2.3cm}C{2.3cm}|}
\hline
Rule&Parallel&Bipartite&\makecell{Block\\Sequential}&\makecell{Block\\Parallel}&\makecell{Local\\Clocks}\\\hline
156&$\Theta(1)$&$\Theta\left(2^{(\sqrt{n\log(n)})}\right)$&$\Omega\left(2^{(\sqrt{n\log(n)})}\right)$&$\Omega\left(2^{(\sqrt{n\log(n)})}\right)$&$\Omega\left(2^{(\sqrt{n\log(n)})}\right)$\\\hline
\end{tabular}
\end{center}
\caption{Longest limit cycle for Rule 156, according to update schedule.}
\end{table}

\begin{figure}[!ht]
	\centerline{\includegraphics{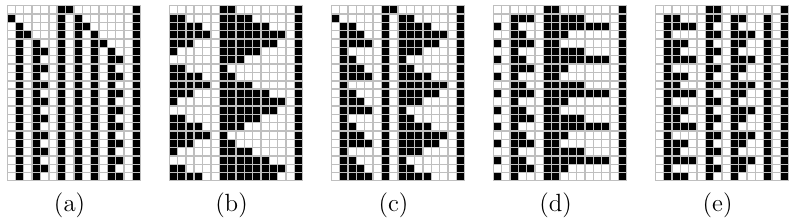}}
	\caption{Space-time diagrams (time going downward) of configuration 
		$0000001100000001$ following rule $156$ depending on: 
		(a) the parallel update mode $\mu_\p = (\integers{16})$, 
		(b) the bipartite update mode $\mu_\bip = (\{i \in \integers{16} \mid i \equiv
			0 \mod 2\}, \{i \in \integers{16} \mid i \equiv 1 \mod 2\})$, 
		(c) the block-sequential update mode $\mu_\bs = 
			(\{10,15\},\{0,1,5,7,8,12\}, \{4,6,9,11,14\},\{3,13\},\{2\})$, 
		(d) the block-parallel update mode $\mu_\bp = \{(0,1),(2,3,4),(5),(6,8,7),(11,10,9),(14,13,12),(15)\}$, 
		(e) the local clocks update mode $\mu_\lc = 
		(P = (2,2,2,2,4,4,4,4,3,3,3,3,1,4,1,1), 
		\Delta = (1,1,1,0,3,3,3,2,1,1,1,0,0,3,0,0))$.
		}
	\label{fig:rule156}
\end{figure}

\begin{lemma}
	\label{lem:156wall01}
	ECA rule $156$ admits only one wall, namely the word $w = 01$.
\end{lemma}

\begin{proof}
	For all $a, b \in \B$, $f_{156}(a01) = 0$ and $f_{156}(01b) = 1$. 
	By definition of the rule, no other word of length $2$ gets this property. 
	Thus $w$ is the unique wall for ECA rule $156$.\qed
\end{proof}

\begin{lemma}
	\label{lem:156wallcreation}
	ECA rule $156$ can dynamically create new walls only if the underlying update mode makes two consecutive cells update their state simultaneously. 
\end{lemma}

\begin{proof}
	Remark that to create walls in the trajectory of a configuration $x 
	\in \B^n$, $x$ need to have at least one wall. 
	Indeed, the only configurations with no walls are $0^n$ and $1^n$, and they are 
	fixed points.
	Thus, with no loss of generality, let us focus on a subconfiguration $y \in \B^k$, 
	with $k \leq n-2$, composed of free state cells surrounded by two walls.
	Notice that since the configurations are toric, the wall ``at the left'' and ``at 
	the right'' of $y$ can be represented by the same two cells.\smallskip
	
	Configuration $y$ is necessarily of the form $y = (1)^\ell(0)^r$, with $k = \ell + 
	r$. 
	Furthermore, since $f_{156}(000) = 0$ and $f_{156}(100) = 1$, the only cells whose 
	states can change are those where $1$ meets $0$, i.e. $y_{\ell-1}$ and $y_\ell$.
	Such state changes depends on the schedule of updates between these two cells.
	Let us proceed with case disjunction:
	\begin{enumerate}
	\item Case of $\ell \geq 1$ and $r \geq 1$:
		\begin{itemize}
		\item if $y_\ell$ is updated strictly before $y_{\ell-1}$, then $y^1 = 
			(1)^{\ell+1}(0)^{r-1}$, and the number of $1$s (resp. $0$s) increases 
			(resp. decreases);
		\item if $y_\ell$ is updated strictly after $y_{\ell-1}$, then $y^1 = 
			(1)^{\ell-1}(0)^{r+1}$, and the number of $1$s (resp. $0$s) decreases 
			(resp. increases);
		\item if $y_\ell$ and $y_{\ell-1}$ are updated simultaneously, then $y^1 = 
			(1)^{\ell-1}(01)(0)^{r-1}$, and a wall is created.
		\end{itemize}
	\item Case of $\ell = 0$ (or of $y = 0^k$): nothing happens until $y_0$ is updated. 
		Let us admit that this first state change has been done for the sake of 
		clarity and focus on $y^1 = (1)^{\ell = 1}(0)^{r = k-1}$, which falls into Case 
		1.
	\item Case of $r = 0$ (or of $y = 1^k$): symmetrically to Case 2, nothing happens 
		until $y_{k-1}$ is updated. 
		Let us admit that this first state change has been done for the sake of 
		clarity and focus on $y^1 = (1)^{\ell = k-1}(0)^{1}$, which falls into Case 
		1.
	\end{enumerate}
	As a consequence, updating two consecutive cells simultaneously is a necessary 
	condition for creating new walls in the dynamics of ECA rule $156$.
\end{proof}

\begin{theorem}
	\label{thm:156parallel}
	$(156, \p{})$ has only fixed points and limit cycles of length two.
\end{theorem}

\begin{proof}
	We base the proof on Lemmas~\ref{lem:156wall01} and~\ref{lem:156wallcreation}.
	So, let us analyze the possible behaviors between two walls, since by definition, 
	what happens between two walls is independent of what happens between two other 
	walls.
	
	Let us prove the results by considering the three possible distinct cases for a 
	configuration between two walls $y \in \B^{k+4}$:
	\begin{enumerate}
	\item Consider $y = (01)(0)^k(01)$ the configuration with only $0$s between the two 
		walls. 
		Applying the rule twice, we obtain $y^2 = (01)^2(0)^{k-2}(01)$.
		So, a new wall appears every two iterations so that, for all $t < \frac{k}{2}$,  
		$y^{2t} = (01)^{t+1}(0)^{k-2t}(01)$, until a step is reached where there is no 
		room for more walls. 
		This step is reached after $k$ (resp $k-1$) iterations when $k$ is even (resp. 
		odd) and is such that there is only walls if $k$ is even (which implies that 
		the dynamics has converged to a fixed point), and only walls except one cell 
		otherwise.
		In this case, considering that $t = \lfloor \frac{k}{2} \rfloor$, because 
		$f_{156}(100) = 1$ and $f_{156}(110) = 0$, we have that 
		$y^{2t} = (01)^{t+1}(0)(01) \to y^{2t+1} = (01)^{t+1}(1)(01) \to 
		y^{2t} = (01)^{t+1}(0)(01)$, which leads to a limit cycle of length 
		$2$.
	\item Consider now that $y = (01)(1)^k(01)$. 
		Symmetrically, for all $t < \frac{k}{2}$, $y^{2t} = (01)(0)^{k-2t}(01)^{t+1}$. 
		The same reasoning applies to conclude that the length of the largest limit 
		cycle is $2$. 
	\item Consider finally configuration $y$ where $k = \ell + r$ such that 
		$y = (01)(1)^\ell(0)^r(01)$. 
		Applying $f_{156}$ on it leads to $y^1 = (01)(1)^{\ell-1}(01)(0)^{r-1}(01)$, 
		which falls into the two previous cases.
	\end{enumerate}
\end{proof}

\begin{theorem}
	\label{thm:156BIP}
	$(156, \bip{})$ applied over a grid of size $n$ has largest limit cycles of length 
	$\Theta(2^{\sqrt{n\log(n)}})$.
\end{theorem}

\begin{proof}
	First, by definition of a bipartite update mode and by Lemmas~\ref{lem:156wall01} 
	and~\ref{lem:156wallcreation}, the only walls appearing in the dynamics are the 
	ones present in the initial configuration. 
	Let us prove that, given two walls $u$ and $v$, distanced by $k$ cells, the largest
	limit cycles of the dynamics between $u$ and $v$ are of length $k+1$. 
	We proceed by case disjunction depending on the nature of the subconfiguration $y 
	\in \B^{k+4}$, with the bipartite update mode $\mu = (\{i \in \integers{k+4} \mid i 
	\equiv 0 \mod 2\}, \{i \in \integers{k+4} \mid i \equiv 0 \mod 2\})$:\smallskip
	
	\begin{enumerate}
	\item Case of $y = (01)(0)^k(01)$: we prove that configuration $(01)(1)^k(01)$ is 
		reached after $\lceil \frac{k}{2} + 1 \rceil$ steps:\\
		\begin{minipage}{.4\textwidth}
			\begin{enumerate}
			\item If $k$ is odd, we have:\\[-3mm]
				\begin{equation*}
					\begin{array}{ccc}
						y & = & (01)(0)^k(01)\\
						y^1 & = & (01)(1)(0)^{k-1}(01)\\
						y^2 & = & (01)111(0)^{k-3}(01)\\
						& \vdots & \\
						y^{\lfloor\frac{k}{2}\rfloor} & = & (01)(1)^{k-2}(00)(01)\\
						y^{\lceil\frac{k}{2}\rceil} & = & (01)(1)^{k}(01)\\
					\end{array}
				\end{equation*}
			\end{enumerate}
		\end{minipage}
		\quad
		\begin{minipage}{.4\textwidth}
			\begin{enumerate}
			\item If $k$ is even, we have:\\[-3mm]
				\begin{equation*}
					\begin{array}{ccc}
						y & = & (01)(0)^k(01)\\
						y^1 & = & (01)(1)(0)^{k-1}(01)\\
						y^2 & = & (01)(111)(0)^{k-2}(01)\\
						& \vdots & \\
			            y^{\frac{k}{2}} & = & (01)(1)^{k-1}(0)(01)\\
						y^{\frac{k}{2}+1} & = & (01)(1)^{k}(01)\\
					\end{array}
				\end{equation*}
			\end{enumerate}			
		\end{minipage}\smallskip
		
	\item Case of $y = (01)(1)^k(01)$: we prove that configuration $(01)(0)^k(01)$ is 
		reached after $\lceil \frac{k}{2} + 1 \rceil$ steps:\\
		\begin{minipage}{.4\textwidth}
			\begin{enumerate}
			\item If $k$ is odd, we have:\\[-3mm]
				\begin{equation*}
					\begin{array}{ccc}
						y & = & (01)(1)^k(01)\\
						y^1 & = & (01)(1)^{k-1}(0)(01)\\
						y^2 & = & (01)(1)^{k-3}(000)(01)\\
						& \vdots & \\
						y^{\lfloor\frac{k}{2}\rfloor} & = & (01)(11)(0)^{k-2}(01)\\
						y^{\lceil\frac{k}{2}\rceil} & = & (01)(0)^{k}(01)\\
					\end{array}
				\end{equation*}
			\end{enumerate}
		\end{minipage}
		\quad
		\begin{minipage}{.4\textwidth}
			\begin{enumerate}
			\item If $k$ is even, we have:\\[-3mm]
				\begin{equation*}
					\begin{array}{ccc}
						y & = & (01)(1)^k(01)\\
						y^1 & = & (01)(1)^{k-2}(00)(01)\\
						y^2 & = & (01)(1)^{k-4}(0000)(01)\\
						& \vdots & \\
						y^{\frac{k}{2}} & = & (01)(11)(0)^{k-2}(01)\\
						y^{\frac{k}{2}+1} & = & (01)(0)^{k}(01)\\
					\end{array}
				\end{equation*}
			\end{enumerate}
		\end{minipage}\smallskip
		
	\item Case of $y = (01)(1)^\ell(0)^r(01)$: this case is included in Cases 1 and 2.
	 \end{enumerate}\smallskip

	As a consequence, the dynamics of any $y$ leads indeed to a limit cycle of length 
	$k+1$. 
	Remark that if we had chosen the other bipartite update mode as reference, 
	the dynamics of any $y$ would have been symmetric and led to the same limit 
	cycle.\smallskip 
	
	Finally, since the dynamics between two pairs of distinct walls of independent 
	of each other, the asymptotic dynamics of a global configuration $x$ such that 
	$x = (01)(0)^{k_1}(01)(0)^{k_2}(01)\dots(01)(0)^{k_m}$ is a limit cycle whose 
	length equals to the least common multiple of the lengths of all limit cycles 
	of the subconfigurations $(01)(0)^{k_1}(01), (01)(0)^{k_2}(01), \dots, (01)
	(0)^{k_m}(01)$.
	We derive that the largest limit cycles are obtained when the $(k_i+1)$s are 
	distinct primes whose sum is equal to $n - 2m$, with $m$ is constant.
	As a consequence, the length of the largest limit cycle is lower- and upper-bounded  
	by the primorial of $n$ (that is, the maximal product of distinct primes whose sum is 
	$\leq n$.), denoted by function $h(n)$.
	In~\cite{Deleglise2015}, it is shown  in Theorem~$18$ that when $n$ tends to infinity, $\log h(n) \sim \sqrt{n \log n}$.
	Hence,  we deduce that the length of the largest limit cycles of $(156, \bip{})$ of 
	size $n$ is $\Theta(2^{\sqrt{n\log(n)}})$. 
	\end{proof}

\begin{corollary}
	\label{cor:156other}
	The families $(156, \bs{})$, $(156, \bp{})$ and $(156, \lc{})$ of size 
	$n$ have largest limit cycles of length $\Omega(2^{\sqrt{n\log(n)}})$.
\end{corollary}

\begin{proof}
	Since the bipartite update modes are specific block-sequential and block-parallel 
	update modes, and since both block-sequential and block-parallel update modes are 
	parts of local-clocks update modes, all of them inherit the property stating that 
	the lengths of the largest limit cycles are lower-bounded by 
	$2^{\sqrt{n\log(n)}}$. 
\end{proof}

\begin{table}[!ht]
\begin{center}
\begin{tabular}{|c|c|c|c|c|c|c|c|c|}
\hline
Rule&111&110&101&100&011&010&001&000\\\hline
56&0&0&1&1&1&0&0&0\\\hline
162&1&0&0&1&1&0&0&0\\\hline
184&1&0&1&1&1&0&0&0\\\hline
\end{tabular}
\end{center}
\end{table}

\begin{table}[!ht]\label{table:184y152}
\begin{center}\renewcommand{\arraystretch}{2}
\begin{tabular}{|C{1.5cm}|C{1.5cm}C{2cm}C{2cm}C{2.5cm}C{2.5cm}|}
\hline
Rule&Parallel&Sequential&\makecell{Block\\Sequential}&\makecell{Block\\Parallel}&\makecell{Local\\Clocks}\\\hline
56&$\mathcal{O}(n)$&$1$&$\mathcal{O}(n)$&$\Omega\left(2^{(\sqrt{n\log(n)})}\right)$&$\Omega\left(2^{(\sqrt{n\log(n)})}\right)$\\\hline
152&$\mathcal{O}(n)$&$1$&$\mathcal{O}(n)$&$\Omega\left(2^{(\sqrt{n\log(n)})}\right)$&$\Omega\left(2^{(\sqrt{n\log(n)})}\right)$\\\hline
184&$\mathcal{O}(n)$&$1$&$\mathcal{O}(n)$&$\Omega\left(2^{(\sqrt{n\log(n)})}\right)$&$\Omega\left(2^{(\sqrt{n\log(n)})}\right)$\\\hline
\end{tabular}
\end{center}
\caption{Longest cycle for rules 152 and 184, for each update schedule.}
\end{table}

\begin{figure}[!ht]
	\centerline{
	% (184,PAR)	
	\begin{minipage}{.16\textwidth}
	\centerline{
	\scalebox{.45}{%
}}
	\end{minipage}
	}\smallskip
	
	\centerline{
	\begin{minipage}{.16\textwidth}
	\centerline{(a)}
	\end{minipage}
	\quad
	\begin{minipage}{.16\textwidth}
	\centerline{(b)}
	\end{minipage}
	\quad
	\begin{minipage}{.16\textwidth}
	\centerline{(c)}
	\end{minipage}
	\quad
	\begin{minipage}{.16\textwidth}
	\centerline{(d)}
	\end{minipage}
	\quad
	\begin{minipage}{.16\textwidth}
	\centerline{(e)}
	\end{minipage}
	}
	\caption{Space-time diagrams (time going downward) of configuration 
		$0011000000011000$ following rule $184$ depending on: 
		(a) the parallel update mode $\mu_\p = (\integers{16})$, 
		(b) the bipartite update mode \\$\mu_\bip = (\{i \in \integers{16} \mid i \equiv
			0 \mod 2\}, \{i \in \integers{16} \mid i \equiv 1 \mod 2\})$, 
		(c) the block-sequential update mode $\mu_\bs = 
			(\{0,1,2,5,6,7,10,11,12,15\},\{3,4,8,9,13,14\})$, 
		(d) the block-parallel update mode $\mu_\bp = \{(0,3),(4,5,6,7),(1,8,9,10),(2,14,13,15),(11,12)\}$, 
		(e) the local clocks update mode $\mu_\lc = 
		(P = (2,4,4,2,4,4,4,4,4,2,4,4,2,4,4,4), 
		\Delta = (0,0,0,1,0,1,2,1,3,0,0,0,1,0,1,3))$.
		}
	\label{fig:rule184}
\end{figure}

\begin{theorem}\label{thm:184seq}
$(184,\seq)$ can only reach homogeneous fixed points.
\end{theorem}
\begin{proof}
By definition, with sequential update mode, it is not possible for two consecutive cells to be updated simultaneously. 
Let us consider an initial configuration such that it consist of an isle of ones surrounded by zeros: $y=1^{k}0^{n-k}$. Since $f_{184}(000)=0$ and $f_{184}(111)=1$, we can focus on what happens where zeros and ones meet. Without loss of generality, let us assume that the first one is on cell $0$ and the last one is on cell $k-1$.
We will proceed with a case-by-case analysis.
\begin{enumerate}
\item Case 1. The cells inside the isle update before the ones outside: $\mu_{0}>\mu_{1}$ and $\mu_{k-1}<\mu_k$: since $f_{184}(011)=1$ and $f_{184}(110)=0$ we have that $y^1=(1)^{k-\ell}(0)^{n-k+\ell}$, where $\ell$ will depend on the update mode. It will be given by the first cell going from right to left that updates \textit{before} the one on the right, meaning that $\mu_{k-\ell}<\mu_{k-\ell+1}$.\\ It is easy to see that on the second iteration the isle is still on the same case, from where we can conclude that the process repeats and the number of ones decreases until we reach the homogeneous fixed point $0^n$.
\item Case 2. The cells inside the isle update before the ones outside: $\mu_{0}<\mu_{1}$ and $\mu_{k-1}>\mu_k$. Since $f_{184}(001)=0$, the state of the cells to the left of the isle remain unchanged. But, since $f_{184}(100)=1$ the number of ones can increase to the right: $y^1=1^{k+r}0^{n-k-r}$, where $r$ will be given by the first cell going from left to right that updates \textit{after} the one to its right, meaning $\mu_{k+r}>\mu_{k+r+1}$. Its easy to see that for the next iteration the ends of the isle follow the same order of updating as it had to begin with, from where we can conclude that the isle will continue to increase its size until it reaches the homogeneous fixed point $1^n$.
\item Case 3. $\mu_0<\mu_1$ and $\mu_{k-1}<\mu_k$. Follows the same analysis as case 1. As it has been shown, there is no increasing or decreasing the number of ones at the left side of the isle, because $f_{184}(011)=1$ and $f_{184}(001)=0$. Which means that it will lose ones until it reaches the homogeneous fixed point $0^n$. 
\item Case 4. $\mu_{0}>\mu_1$ and $\mu_{k-1}>\mu_{k}$. The analysis is identical to case 2, because the left side of the isle doesn't change its behavior regardless of if the cell inside the isle updates before or after the one outside. Meaning that it will gain ones until it reaches the homogeneous fixed point $1^n$.
\end{enumerate}
\begin{table}[!ht]
\begin{center}
\begin{tabular}{cccccccccccc}
0&1&$\dots$& $k-\ell$&$k-\ell+1$&$\dots$ &$k-2$&$k-1$&$k$&$k+1$&$\dots$&n-1\\\hline
1&1&$\dots$&1        &1         &$\dots$ &1  &1  &0&0  &$\dots$&0\\\hline
1&1&$\dots$&1        &0         &$\dots$ &0  &0  &0&0  &$\dots$&0\\\hline
$\vdots$\\\hline
0&0&$\dots$&0        &0         &$\dots$ &0  &0  &0&0  &$\dots$&0\\\hline
\end{tabular}
\end{center}
\caption{Representation of Cases 1 and 3: $\mu_{0}>\mu_{1}$ or $\mu_{0}<\mu_{1}$ and $\mu_{k-1}<\mu_k$.}
\begin{center}
\begin{tabular}{cccccccccccc}
0&1&$\dots$&$k$&$k+1$&$\dots$ &$k+r$&$k+r+1$&$\dots$&n-1\\\hline
1&1&$\dots$&1  &1    &$\dots$ &0    &0      &$\dots$&0\\\hline
1&1&$\dots$&1  &1    &$\dots$ &1    &0      &$\dots$&0\\\hline
$\vdots$\\\hline
1&1&$\dots$&1  &1    &$\dots$ &1    &1      &$\dots$&1\\\hline
\end{tabular}
\end{center}
\caption{Representation of Cases 2 and 4: $\mu_{0}>\mu_{1}$ or $\mu_{0}<\mu_{1}$ and $\mu_{k-1}>\mu_k$.}
\end{table}
Now, if we generalize this by considering any initial configuration as isles of ones separated by zeros, we can see that every isle has to follow one of the four previous cases, from where we can conclude that isles that follow cases 1 and 3 will disappear, while the ones that follow cases 2 and 4 will grow in size.

Note that if an isle of type 2 reaches one of type 1 or 3 as it grows they will combine which will result in an isle of case 3. Similarly, if an isle of case 4 reaches one of types 1 or 3, the combination of them results  in an isle of case 1. 

From the previous analysis we can conclude that sequential update modes always lead to homogeneous fixed points.
\end{proof}

\begin{theorem}\label{thm:184bs}
$(184,\bs)$ of size $n$ has largest limit cycles of size $\mathcal{O}(n)$.
\end{theorem}
\begin{proof}
Doing a similar analysis as the one done for theorem \ref{thm:184seq}, is easy to see that $(184,\bs)$ can only reach (homogeneous) fixed points if there exist only one group of two consecutive cells that update on the same substep.

Let $\mu$ be a $\bs$ update mode such that there is one group of (at least) three consecutive cells that are updated simultaneously.\\
Let us consider an initial configuration $x=10^{n-1}$, such that the only one is on one of the three cells that update on the same substep.\\
We will denote by $e_1,e_2$ the cells that update at the same substep than their right hand neighbor ($\mu_{e_{1}}=\mu_{e_{2}}=\mu_{e_{2}+1}$), the set $\{r_{i}\}_{i=1}^j$ will be the cells that are updated \textit{after} their right hand neighbor ($\mu_{r_{i}}>\mu_{r_{i}+1}$) and the set $\{\ell_i\}_{i=1}^k$ will be the cells that are updated \textit{before} their right hand neighbor ($\mu_{\ell_{k}}<\mu_{\ell_{k}+1}$).

We will proceed by cases, considering the order in which cells are updated outside our group.

First case: $r_je_1e_2\ell_k$.\\
\underline{$t=1$} the only $1$ moves once cell to the right because $f_{184}(100)=1$ and $f_{184}(010)=0$.\\
\underline{$t=2$} the $1$ advances again, but since the next cell to the right is updated \textit{before} the one on the right, we gain ones until the \textit{first} cell such that $\mu_i>\mu_{i+1}$, which means that it is a cell in $\{r_{i}\}_{i=1}^j$. Let us call that cell $r_1$.\\
\underline{$t=3$} We continue to gain ones until the next cell to the right that belongs to $\{r_{i}\}_{i=1}^j$, which will be $r_2$. In this instance, we do not lose ones to the left, because $f_184(001)=0$ and $f_{184}(011)=1$.

\begin{center}
\begin{tabular}{ccc|ccc|ccc|ccc}
 &   &$r_j$&$e_1$&$e_2$&$\ell_k$&$\dots$&&$r_1$&&&$r_2$\\\hline
0&$\dots$&0&1    &0    &0       &0&$\dots$&0&0&$\dots$&0\\\hline
0&$\dots$&0&0    &1    &0       &0&$\dots$&0&0&$\dots$&0\\\hline
0&$\dots$&0&0    &0    &1       &1&$\dots$&1&0&$\dots$&0\\\hline
0&$\dots$&0&0    &0    &1       &1&$\dots$&1&1&$\dots$&1\\\hline
$\vdots$\\\hline
1&$\dots$&1&0    &0    &1       &1&$\dots$&1&1&$\dots$&1\\\hline
1&$\dots$&1&1    &0    &1       &1&$\dots$&1&1&$\dots$&1\\\hline
\end{tabular}
\end{center}

\underline{$t\in\{4,\dots,j\}$} Similar as to what occurs on $t=3$, and because we are on this first case, the cell to the left of $e_1$ will have to be $r_j$.\\
\underline{$t=j+1$} $s_1$ changes its state to $1$, and it is the only cell whose state can change.\\
Now, let us see what happens with one 0 surrounded by ones.\\

\underline{$t=1$} the only $0$ moves once cell to the left.\\
\underline{$t=2$} the $0$ advances again, but this time we gain $0$'s to the left until just \textit{before} the \textit{first} cell marked with an $\ell$. Note that the cell marked with the $\ell$ has not changed yet.\\
\underline{$t=\{3,\dots,k\}$} Analogously, the number of zeros keeps increasing  until it finally reaches $\ell_k$, which leaves us with two surviving $1$'s.

\begin{center}
\begin{tabular}{ccc|ccc|ccc|ccc}
 &&$\ell_1$&$\dots$& &$r_j$&$s_1$&$s_2$&$\ell_k$\\\hline
1&$\dots$&1&1&$\dots$&1&1    &1    &0       &1&$\dots$&1\\\hline
1&$\dots$&1&1&$\dots$&1&1    &0    &1       &1&$\dots$&1\\\hline
1&$\dots$&1&0&$\dots$&0&0    &1    &1       &1&$\dots$&1\\\hline
$\vdots$\\\hline
0&$\dots$&0&0&$\dots$&0&0    &1    &1       &0&$\dots$&0\\\hline
0&$\dots$&0&0&$\dots$&0&0    &1    &0       &0&$\dots$&0\\\hline
\end{tabular}
\end{center}

\underline{$t=k+1$} $\ell_k$ becomes $0$, since $f_{110}=0$, with which the only surviving $1$ is on cell $s_2$, and we already know from the previous analysis what happens next.

It takes $j$ steps to go from one $1$ surrounded by zeros on the position $s_2$ to one $0$ surrounded by ones, on that same position, and it takes $j$ steps to return to the configuration we had on $t=1$ on the first analysis, and because $j+k<n$, we have found cycles $\mathcal{O}(n)$.

Using the same reasoning, we can prove that a second case with $\ell_1s_1s_2r_1$ follows the same behavior and it produces cycles of length $(j+1)+(k+1)\leq n$. Similarly, cases 3 ($r_js_1s_2r_1$) and 4 ($\ell_1s_1s_2\ell_k$) lead to cycles of length $(j+1)+k<n$ and $j+(k+1)<n$, resp.
\end{proof}
\begin{remark}
Note that cycles $\mathcal{O}(n)$ can also be found with block sequential configurations with (at least) two groups of size two or more consecutive cells that update on the same substep.
\end{remark}
\begin{lemma}\label{lemma:184walls}
$w=0011$ is a relative wall for ECA rule $184$.
\end{lemma}
\begin{proof}We proceed case by case.

\begin{table}[!ht]
\begin{center}
\begin{tabular}{c|cccc|c c c|cccc|c}
\cline{1-6}\cline{8-13}
0&\cellcolor{gray}0&0&1&\cellcolor{gray}1&0& &0&\cellcolor{gray}0&0&1&\cellcolor{gray}1&1\\\cline{1-6}\cline{8-13}
0&\cellcolor{gray}0&0&1&\cellcolor{gray}0&0& &0&\cellcolor{gray}0&0&1&\cellcolor{gray}1&1\\\cline{1-6}\cline{8-13}
0&0&\cellcolor{gray}0&\cellcolor{gray}1&1&0& &0&0&\cellcolor{gray}0&\cellcolor{gray}1&1&1\\\cline{1-6}\cline{8-13}
0&0&0&1&1&0& &0&0&0&1&1&1\\\cline{1-6}\cline{8-13}
\end{tabular}\hspace{10pt}
\begin{tabular}{c|cccc|c c c|cccc|c}
\cline{1-6}\cline{8-13}
1&\cellcolor{gray}0&0&1&\cellcolor{gray}1&0& &1&\cellcolor{gray}0&0&1&\cellcolor{gray}1&1\\\cline{1-6}\cline{8-13}
1&\cellcolor{gray}1&0&1&\cellcolor{gray}0&0& &1&\cellcolor{gray}1&0&1&\cellcolor{gray}1&1\\\cline{1-6}\cline{8-13}
1&0&\cellcolor{gray}0&\cellcolor{gray}1&1&0& &1&0&\cellcolor{gray}0&\cellcolor{gray}1&1&1\\\cline{1-6}\cline{8-13}
1&0&0&1&1&0& &1&0&0&1&1&1\\\cline{1-6}\cline{8-13}
\end{tabular}
\end{center}
\end{table}

As long as the cells at the border of the $w$ update twice before the ones around them are updated, the update mode preserves the word.
\end{proof}
\begin{theorem}\label{thm:184bp}
$(184,\bp)$ of size $n$ has largest limit cycles of size $\Omega\left(2^{\sqrt{n\log n}}\right)$.
\end{theorem}
\begin{proof}[Sketch of proof]
We know that the homogeneous configurations $1^n$ and $0^n$ are fixed points, so we will study configurations that have ones and zeros. \\
From lemma \ref{lemma:184walls} we know that we can find in the family of block parallel update modes examples of iteration schedules such that the relative walls are preserved.\\
Similar to what we have done before, we will focus on what happens between walls, starting by choosing update modes within $\bp$ that preserve walls for a given initial configuration.\\
Then, we prove that the behavior between walls that are at a distance $k$ has cycles of length $\mathcal{O}(k)$.\\
And because of the independence of the dynamics between two pairs of walls, the asymptotic dynamics of a global configuration $x$ is a limit cycle whose length is given by the least common multiple of the lengths of all limit cycles of the subconfigurations and, with the same argument as was used for $(156,\bip)$, we can conclude that the largest limit cycle of the family $(184,\bp)$ applied over a ring of size $n$ is  $\Omega\left(2^{\sqrt{n\log n}}\right)$.\\
\end{proof}
\begin{remark}
Note that the previous theorem requires the ring to be of a larger size than was the case for $(156,\bip)$ and $(178,\bp)$, because the relative walls being used in this proof are composed of four cells instead of 2, as was the case for the other two theorems.
\end{remark}

\begin{corollary}\label{cor:rule152}
The previous theorems are also valid for $(152,\seq)$, $(152,\bs)$ and $(152,\bp)$, respectively.
\end{corollary}
\begin{proof}
The only difference between Rules $184$ and $152$ is $f_{184}(101)=1\neq0=f_{152}(101)$, and that part of the definition is not used during the previous proofs.
\end{proof}

\begin{corollary}\label{cor:rule56}
The previous theorems are also valid for $(56,\seq)$ and $(56,\bs)$ respectively.
\end{corollary}
\begin{proof}
The only difference between Rules $184$ and $56$ is $f_{184}(111)=1\neq0=f_{56}(111)$, and that part of the definition is only used during the previous proofs to establish that $1^n$ is a fixed point; which means we can follow the same reasoning to prove the same properties for rule $56$.
\end{proof}
\clearpage
\begin{table}[!ht]
\begin{center}
\begin{tabular}{|c|c|c|c|c|c|c|c|c|}
\hline
Rule&111&110&101&100&011&010&001&000\\\hline
108&0&1&1&0&1&1&0&0\\\hline
\end{tabular}
\end{center}
\end{table}

\begin{table}[!ht]\label{table:rule108}
\begin{center}\renewcommand{\arraystretch}{2}
\begin{tabular}{|C{1cm}|C{1.5cm}C{2.3cm}C{2.3cm}C{2.3cm}C{2.3cm}|}
\hline
Rule&Parallel&Bipartite&\makecell{Block\\Sequential}&\makecell{Block\\Parallel}&\makecell{Local\\Clocks}\\\hline
108&$\Theta(1)$&$\Omega\left(2^{(\sqrt{n\log(n)})}\right)$&$\Omega\left(2^{(\sqrt{n\log(n)})}\right)$&$\Omega\left(2^{(\sqrt{n\log(n)})}\right)$&$\Omega\left(2^{(\sqrt{n\log(n)})}\right)$\\\hline
\end{tabular}
\end{center}
\caption{Longest limit cycle for Rule 108, according to update schedule.}
\end{table}

\begin{figure}[!h]
	\centerline{\includegraphics{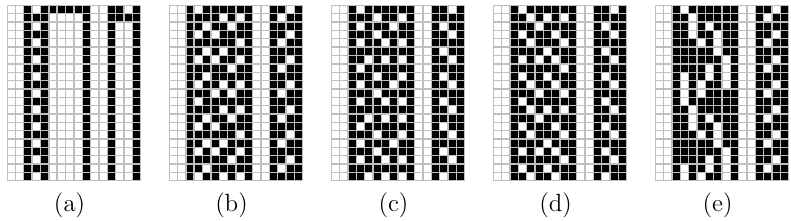}}
	\caption{Space-time diagrams (time going downward) of configuration 
		$0010111111001101$ following rule $108$ depending on: 
		(a) the parallel update mode $\mu_\p = (\integers{16})$, 
		(b) the bipartite update mode $\mu_\bip = (\{i \in \integers{16} \mid i \equiv
			0 \mod 2\}, \{i \in \integers{16} \mid i \equiv 1 \mod 2\})$, 
		(c) the block-sequential update mode $\mu_\bs = (\{0,4,8,12\},\{1,3,5,7,9,11,13,15\},\{2,6,10,14\})$, 
		(d) the block-parallel update mode $\mu_\bp =\{(0,1),(10,11),(2,3,4,5),(6,7,8,9),(12,13,14,15)\}$, 
		(e) the local clocks update mode $\mu_\lc = 
			(P = (3,3,2,2,2,2,2,2,2,3,3,3,2,3,3,2), 
			\Delta = (0,1,0,1,0,1,0,1,0,1,0,1,0,1,2,1))$.}
	\label{fig:rule108}
\end{figure}

\begin{lemma}
	\label{lem:108walls}
	ECA rule $108$ admits two walls, $001$ and $100$.
\end{lemma}

\begin{proof}
	By definition of the rule, $f_{108}(a00)=0$, $f_{108}(001)=0$ and $f_{108}(01b)=1$, for all $a,b\in\{0,1\}$, which means that $\omega_1=001$ is a wall. Also by definition, $f_{108}(a10)=1$, $f_{108}(100)=0$ and $f_{108}(00b)=0$, meaning that $\omega_2=100$ is also a wall.
\end{proof}
We will now show that there exists one sequential update mode for which there are cycles of length $\Omega\left(2^{(\sqrt{n\log(n)})}\right)$
\begin{theorem}
	\label{thm:108seq}
	Rule $(108, \seq)$  applied over a grid of size $n$  can reach largest limit cycles of length $\Omega\left(2^{(\sqrt{n\log(n)})}\right)$
\end{theorem}
\begin{proof}
Let $\mu_{\seq}=(0,1,2,3,\dots,n)$ a sequential update schedule.
Let $x$ be an initial configuration such that $x=\omega_1 1^k \omega_2=(001) 1^k (100)$, with $k$ an even number. We need to calculate the state for the configuration $f(x)$.\\
Since $f_{108}(111)=0$, and $f_{108}(011)=1$, we have that
\begin{equation*}
\begin{matrix}
x     &= &001& 1^k                      & 100\\
f(x)  &= &001& (01)^{\frac{k}{2}}       & 100\\
f^2(x)&= &001& 1^{k-2}(10)              & 100\\
f^3(x)&= &001& (01)^{\frac{k-2}{2}}(1)^2& 100\\
f^4(x)&= &001& 1^{k-4}(10)^2            & 100\\
\end{matrix}
\end{equation*}
It is easy to see that on odd iterations there is a number of words $(01)$ that appears in the configuration, but that amount goes down as the numbers of $1$'s to the right side increases. And for the even iterations, we can see the number of ones going down to the left, while the number of words $(10)$ increases. This can be written as
\begin{equation*}
\begin{matrix}
f^t(x)    &= &001& (01)^{\frac{k-(t-1)}{2}}(1)^{t-1}& 100\\
f^{t+1}(x)&= &001& (1)^{k-t}(10)^{\frac{t}{2}}      & 100\\
\end{matrix}
\end{equation*}
when $t$ is odd. Which means that the $(k-1)$-th iteration is
\begin{equation*}
\begin{matrix}
f^{k-1}(x)&= &001& (01)^{\frac{k-((k-1)-1)}{2}}(1)^{(k-1)-1}& 100\\
		  &= &001& (01)^{1}(1)^{k-2}                        & 100\\
\end{matrix}
\end{equation*}
And then $k-$th
\begin{equation*}
\begin{matrix}
f^{k}(x)&= &001& (1)^{k-k}(10)^{\frac{k}{2}}& 100\\
        &= &001& (10)^{\frac{k}{2}}         & 100\\
\end{matrix}
\end{equation*}
And finally, the $(k+1)$-th iteration:
\begin{equation*}
\begin{matrix}
f^{k+1}(x)&= &001& 1^k                      & 100\\
\end{matrix}
\end{equation*}
Which brings us back to the original configuration, meaning that between two walls we can find cycles of length $k$ when $k$ is even.\\

If $k$ is odd, $x=\omega_1 1^k\omega_2$ results in cycles of length 2. We will instead focus on an initial configuration $x$ such that $x=\omega_1 1^{k-1}0\omega_2=(001)(1)^{k-1}0(100)$.
\begin{equation*}
\begin{matrix}
f(x)  &=& 001& (01)^{\frac{k-1}{2}}1^2    &100\\
f^2(x)&=& 001& (1)^{k-2}01                &100\\
f^3(x)&=& 001& (01)^{\frac{k-3}{2}}1(10)  &100\\
f^4(x)&=& 001& (1)^{k-4}0(1)^3            &100\\
f^5(x)&=& 001& (01)^{\frac{k-5}{2}}1(10)^2&100\\
f^6(x)&=& 001& (1)^{k-6}0(1)^5            &100\\
\end{matrix}
\end{equation*}
From where we can deduce that after $t$ iterations (with $t$ odd) we will have
\begin{equation*}
\begin{matrix}
f^t(x)    &=& 001& (01)^{\frac{k-t}{2}}1(10)^{\frac{t-1}{2}}&100\\
f^{t+1}(x)&=& 001& (1)^{k-t}0(1)^{t-1}                      &100\\
\end{matrix}
\end{equation*}
Which means that after $k-1$ iterations we will have
\begin{equation*}
\begin{matrix}
f^{k-1}(x)&=& 001& (1)0(1)^{(k-1)-1}    &100\\
f^{k}(x)  &=& 001& 1(10)^{\frac{k-1}{2}}&100\\
f^{k+1}(x)&=& 001& 0(1)^{k-1}           &100\\
f^{k+2}(x)&=& 001& (10)^{\frac{k-1}{2}}1&100\\
f^{k+3}(x)&=& 001& (1)^{k-1}0           &100\\
\end{matrix}
\end{equation*}
Which brings us back to the starting configuration, meaning that between two walls we can find cycles of length $k+2$ when $k$ is odd.\\
Using the same argument used for Theorem \ref{thm:156BIP}, we conclude that the largest limit cycles of the family $(108,\seq)$ when applied over a ring of length $n$ is $\Omega\left(2^{\sqrt{n\log n}}\right)$.
\end{proof}

\begin{corollary}\label{thm:108bloques}
    $(108,\bs),(108,\bp),(108,\lc)$ have largest limit cycles of length $\Omega\left(2^{\sqrt{n\log n}}\right)$.
\end{corollary}
\begin{proof}
    The family of sequential update schedules are a subset of block parallel and block sequential, which themselves are a subset of local clocks, so they inherit the result.
\end{proof}

\begin{table}[!ht]
\begin{center}
\begin{tabular}{|c|c|c|c|c|c|c|c|c|}
\hline
Rule&111&110&101&100&011&010&001&000\\\hline
1&0&0&0&0&0&0&0&1\\\hline
9&0&0&0&0&1&0&0&1\\\hline
\end{tabular}
\end{center}
\end{table}

\begin{table}[!ht]\label{table:rule1}
\begin{center}\renewcommand{\arraystretch}{2}
\begin{tabular}{|C{1.5cm}|C{1.5cm}C{2cm}C{2cm}C{2.5cm}C{2.5cm}|}
\hline
Rule&Parallel&Bipartite&\makecell{Block\\Sequential}&\makecell{Block\\Parallel}&\makecell{Local\\Clocks}\\\hline
1&$2$&$\Omega\left(n\right)$&$\Omega\left(n\right)$&$\Omega\left(2^{(\sqrt{n\log(n)})}\right)$&$\Omega\left(2^{(\sqrt{n\log(n)})}\right)$\\\hline
9&$2$&$\Omega\left(n\right)$&$\Omega\left(n\right)$&$\Omega\left(2^{(\sqrt{n\log(n)})}\right)$&$\Omega\left(2^{(\sqrt{n\log(n)})}\right)$\\\hline
\end{tabular}
\end{center}
\caption{Longest limit cycle for Rules 1 and 9, according to update schedule.}
\end{table}

\begin{figure}[!h]
	\centerline{\includegraphics{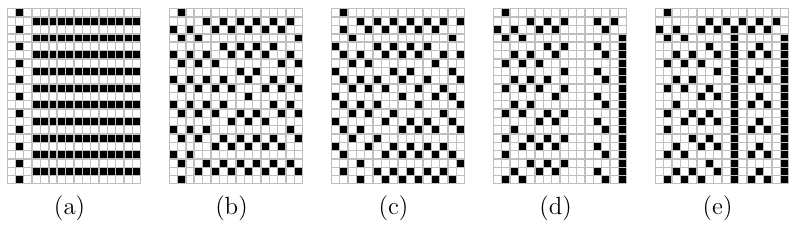}}
	\caption{Space-time diagrams (time going downward) of configuration 
		$0100000000000000$ following rule $1$ depending on: 
		(a) the parallel update mode $\mu_\p = (\integers{16})$, 
		(b) the bipartite update mode $\mu_\bip = (\{i \in \integers{16} \mid i \equiv
			0 \mod 2\}, \{i \in \integers{16} \mid i \equiv 1 \mod 2\})$, 
		(c) the block-sequential update mode $\mu_\bs = (\{0,3,5,7,9,11,13\},\{1,4,6,8,10,12,14\},\{2,15\})$, 
		(d) the block-parallel update mode $\mu_\bp =\{(0,1),(2,3),(4,5),(6,7),(8,9),(12,11),(14,13),(10),(15)\}$, 
		(e) the local clocks update mode $\mu_\lc = 
			(P = (2,2,2,2,2,2,2,2,2,1,2,2,2,2,2,1), 
			\Delta = (1,0,1,0,1,0,1,0,1,0,1,0,1,0,1,0))$.}
	\label{fig:1}
\end{figure}

\begin{theorem}\label{thm:1bip}
$(1,\bip)$ of size $n$ can reach cycles of length at least $\Theta(n)$
\end{theorem}
\begin{proof}
Let $x$ be an initial configuration such that $x=0^k01$.
Then, for $\mu_{\bip}$ we have 
\begin{equation*}
\begin{matrix} 
x     &= & 0^k    01                 \\
f(x)  &= & 00(10)^{\frac{k}{2}-1}   00    \\
f^2(x)&= & (10)(01)^{\frac{k}{2}-2} 00(10)\\
\end{matrix}
\end{equation*}
\begin{equation*}
\begin{matrix} 
f^3(x)&= & (01)0^{k-4}  (01)^2                 \\
f^4(x)  &= & 0^4(10)^{\frac{k}{2}-3}   0^4    \\
f^5(x)&= & (10)^2(01)^{\frac{k}{2}-4} 00(10)^2\\
\end{matrix}
\end{equation*}
\begin{equation*}
\begin{matrix} 
f^6(x)&= & (01)^20^{k-8}  (01)^3                 \\
f^7(x)&= & 0^6(10)^{\frac{k}{2}-5}   0^6    \\
f^8(x)&= & (10)^3(01)^{\frac{k}{2}-6} 00(10)^3\\
\end{matrix}
\end{equation*} 
Since the shape of the configuration depends on the multiplicity of the number of step with respect to 3, let $T=\left\lfloor\frac{t}{3}\right\rfloor$.
\begin{equation*}
\begin{matrix} 
f^T(x)&= & (01)^T 0^{k-4T}  (01)^{T+1}                 \\
f^{T+1}(x)&= & (00)^{T+1}(10)^{\frac{k}{2}-(2T+1)}(00)^{T+1}    \\
f^{T+2}(x)&= & (10)^{T+1}(01)^{\frac{k}{2}-2(T+1)} 00(10)^{T+1}\\
\end{matrix}
\end{equation*} 
From where we can deduce that there are two times on which we reach a cycle, depending on the length of $k$.\\
If $k\mod 4=0$, then for $T^*=\frac{k-8}{4}$ we have
\begin{equation*}
\begin{matrix} 
f^{T^*}(x)&= & (01)^{T^*} 0^{k-4T^*}  (01)^{T^*+1}                 \\
          &= & (01)^{\frac{k-8}{4}} 0^{k-4(\frac{k-8}{4})}  (01)^{\frac{k-8}{4}+1}                 \\
          &= & (01)^{\frac{k-8}{4}} 0^{8}  (01)^{\frac{k-8}{4}+1}                 \\
f^{T^*+1}(x)&= & (00)^{T^*+1}(10)^{\frac{k}{2}-(2T^*+1)}(00)^{T^*+1}    \\
          &= & (00)^{\frac{k-8}{4}+1}(10)^{\frac{k}{2}-(2(\frac{k-8}{4})+1)}(00)^{\frac{k-8}{4}+1}    \\
          &= & (00)^{\frac{k-8}{4}+1}(10)^{\frac{k}{2}-((\frac{k-8}{2})+1)}(00)^{\frac{k-8}{4}+1}    \\
          &= & (00)^{\frac{k-8}{4}+1}(10)^{3}(00)^{\frac{k-8}{4}+1}    \\
f^{T^*+2}(x)&= & (10)^{T^*+1}(01)^{\frac{k}{2}-2(T^*+1)} 00(10)^{T^*+1}\\
            &= & (10)^{\frac{k-8}{4}+1}(01)^{\frac{k}{2}-(\frac{k-8}{2}+2)} 00(10)^{\frac{k-8}{4}+1}\\
            &= & (10)^{\frac{k-8}{4}+1}(01)^{2} 00(10)^{\frac{k-8}{4}+1}\\
f^{T^*+3}(x)&= & (01)^{T^*+1} 0^{k-4(T^*+1)}  (01)^{T^*+2}                 \\
          &= & (01)^{\frac{k-8}{4}+1} 0^{4}  (01)^{\frac{k-8}{4}+2}                 \\
f^{T^*+4}(x)&= & (00)^{T^*+2}(10)^{\frac{k}{2}-(2T^*+2)}(00)^{T^*+2}    \\
          &= & (00)^{\frac{k-8}{4}+2}(10)^{\frac{k}{2}-(2(\frac{k-8}{4})+2)}(00)^{\frac{k-8}{4}+2}    \\
          &= & (00)^{\frac{k}{4}}(10)(00)^{\frac{k}{4}}    \\
\end{matrix}
\end{equation*} 
It is easy to see that $f^{T^*+4}(x)$ is similar to $x$ up to translation, and if we repeat the previous analysis, we obtain that the overall cycle is of length $p=2(3(\frac{k-8}{4})+4)=\frac{3k}{2}-4$.

If instead $k\mod 4=2$, then for $T^*=\frac{k-6}{4}$ we have
\begin{equation*}
\begin{matrix} 
f^{T^*}(x)&= & (01)^{T^*} 0^{k-4T^*}  (01)^{T^*+1}                 \\
          &= & (01)^{\frac{k-6}{4}} 0^{6}  (01)^{\frac{k-6}{4}+1}                 \\
f^{T^*+1}(x)&= & (00)^{T^*+1}(10)^{\frac{k}{2}-(2T^*+1)}(00)^{T^*+1}    \\
          &= & (00)^{\frac{k-6}{4}+1}(10)^{2}(00)^{\frac{k-6}{4}+1}    \\
f^{T^*+2}(x)&= & (10)^{T^*+1}(01)^{\frac{k}{2}-2(T^*+1)} 00(10)^{T^*+1}\\
            &= & (10)^{\frac{k-6}{4}+1}(01)^{1} 00(10)^{\frac{k-6}{4}+1}\\
f^{T^*+3}(x)&= & (01)^{T^*+1} 0^{k-4(T^*+1)}  (01)^{T^*+2}                 \\
          &= & (01)^{\frac{k-6}{4}+1} 0^{2}  (01)^{\frac{k-6}{4}+2}                 \\
f^{T^*+4}(x)&= & (00)^{\frac{k-2}{4}}(01)(00)^{\frac{k+2}{4}}    \\
\end{matrix}
\end{equation*} 
From where we can follow the same reasoning as before, which gives us overall cycles of length $p=2(3(\frac{k-6}{4})+4)=\frac{3k}{2}-5$
\end{proof}

\begin{lemma}
	\label{lem:1walls}
	ECA rule $1$ admits two relative walls, $010$ and $000$.
\end{lemma}

\begin{proof}We can analyse them side by side, with $a,b\in\{0,1\}$.

\begin{table}[!ht]
\begin{center}
\begin{tabular}{c|ccc|c c c|ccc|c}
\cline{1-5}\cline{7-11}
a&\cellcolor{gray}0&\cellcolor{gray}1&\cellcolor{gray}0&b& &a&0&\cellcolor{gray}0&0&b\\\cline{1-5}\cline{7-11}
a&0&\cellcolor{gray}0&0&b& &a&\cellcolor{gray}0&\cellcolor{gray}1&\cellcolor{gray}0&b\\\cline{1-5}\cline{7-11}
a&0&1&0&b& &a&0&0&0&b\\\cline{1-5}\cline{7-11}
\end{tabular}
\end{center}
\end{table}
\end{proof}

We will now show that there exists (at least) one initial configuration for which a block-parallel update mode can reach cycles of length $\Omega\left(2^{(\sqrt{n\log(n)})}\right)$
\begin{theorem}
	\label{thm:1bp}
	Rule $(1, \bp)$  applied over a grid of size $n$  can reach largest limit cycles of length $\Omega\left(2^{(\sqrt{n\log(n)})}\right)$
\end{theorem}
\begin{proof}
Without loss of generality, let $\mu_{\bip}=(\{i|i\mod 2\equiv 0\},\{i|i\mod 2\equiv 1\})$ a bipartite update schedule.\\
Let $x$ be an initial configuration such that $x=\omega_1 0^k 1 \omega_0=(010) 0^k 01 (000)$, with $k$ an even number. We need to calculate the state for the configuration $f(x)$.\\
Since $f_{1}(000)=1$, and $f_{1}(101)=0$, we have that
\begin{equation*}
\begin{matrix} 
&x     &= &010& (10) 0^k                  & 000\\
&f(x)  &= &010& (00)(01)^{\frac{k}{2}}    & 000\\
&f^2(x)&= &010& (01)00(10)^{\frac{k}{2}-1}& 000\\
\end{matrix}
\end{equation*}
\begin{equation*}
\begin{matrix} 
&f^3(x)&= &010& (10)^2 0^{k-2}                & 000\\
&f^4(x)&= &010& (00)^2(01)^{\frac{k}{2}-1}    & 000\\
&f^5(x)&= &010& (01)^2 00(10)^{\frac{k}{2}-2} & 000\\
\end{matrix}
\end{equation*}
\begin{equation*}
\begin{matrix} 
&f^6(x)&= &010& (10)^3 0^{k-6}                & 000\\
&f^7(x)&= &010& (00)^3(01)^{\frac{k}{2}-2}    & 000\\
&f^8(x)&= &010& (01)^3 00(10)^{\frac{k}{2}-3} & 000\\
\end{matrix}
\end{equation*}
Let $T=\left\lfloor\frac{t}{3}\right\rfloor$
\begin{equation*}
\begin{matrix}
f^T(x)    &= &010& (10)^T 0^{k-2T}                   & 000\\
f^{T+1}(x)&= &010& (00)^T(01)^{\frac{k}{2}-T}        & 000\\
f^{T+2}(x)&= &010& (01)^T 00(10)^{\frac{k}{2}-(T+1)} & 000\\
\end{matrix}
\end{equation*}
Now, let $T^*=\frac{k-2}{2}$
\begin{equation*}
\begin{matrix} 
f^{T^*}(x)  &= &010& (10)^{\frac{k-2}{2}} 0^{k-2(\frac{k-2}{2})}         & 000\\
            &= &010& (10)^{\frac{k-2}{2}} 0^{2}                          & 000\\
f^{T^*+1}(x)&= &010& (00)^{\frac{k-2}{2}}(01)^{\frac{k}{2}-\frac{k-2}{2}}& 000\\
            &= &010& (00)^{\frac{k-2}{2}}(01)^{1}                        & 000\\
\end{matrix}
\end{equation*}
We have that $f^{T^*+1}(x)$ is $x'= \omega_1 0^k 1 \omega_0$, and through an analysis symmetrical to what we just did, we obtain that $f^{T^*+1}(x')=x$, meaning that when $k$ is even, the length of the cycle between walls can reach at least a length of $k$ (with the distance between walls equal to $k+2$).\\
Analogously, if $k$ is odd we have that:\\
\begin{equation*}
\begin{matrix} 
x     &= &010& (10)0^k                   & 010\\
f(x)  &= &010& 000(10)^{\frac{k-1}{2}}   & 010\\
f^2(x)&= &010& (01) 0(01)^{\frac{k-1}{2}}& 010\\
\end{matrix}
\end{equation*}
\begin{equation*}
\begin{matrix} 
f^3(x)&= &010& (10)^2 0^{k-2}                 & 010\\
f^4(x)&= &010& (00)^2 0(10)^{\frac{k-1}{2}-1} & 010\\
f^5(x)&= &010& (01)^2 0(01)^{\frac{k-1}{2}-1} & 010\\
\end{matrix}
\end{equation*}
\begin{equation*}
\begin{matrix} 
f^{T}(x)  &= &010& (10)^{T+1} 0^{k-2T}            & 010\\
f^{T+1}(x)&= &010& (00)^{T+1} 0(10)^{\frac{k-1}{2}-T} & 010\\
f^{T+2}(x)&= &010& (01)^{T+1} 0(01)^{\frac{k-1}{2}-T} & 010\\
\end{matrix}
\end{equation*}
\begin{equation*}
\begin{matrix} 
f^{T^*}(x)  &= &010& (10)^{T^*+1} 0^{k-2T^*}            & 010\\
            &= &010& (10)^{\frac{k-1}{2}+1} 0^{k-2(\frac{k-1}{2})}& 010\\
            &= &010& (10)^{\frac{k-1}{2}+1} 0^{1}            & 010\\
f^{T^*+1}(x)&= &010& (00)^{T^*+1} 1 & 010\\
            &= &010& (00)^{\frac{k-1}{2}+1} 1 & 010\\
\end{matrix}
\end{equation*}
with $T^*=\frac{k-1}{2}$.\\
Using the same argument used for \ref{thm:156BIP}, we conclude that the overall limit cycle of the family $(1,\bp)$ when applied over a ring of length $n$ can reach $\Omega\left(2^{\sqrt{n\log n}}\right)$.
\end{proof}

\begin{corollary}\label{cor:1lc}
    $(1,\lc)$ can reach limit cycles of length $\Omega\left(2^{\sqrt{n\log n}}\right)$.
\end{corollary}
\begin{proof}
    The family of sequential update schedules are a subset of block parallel and block sequential, which themselves are a subset of local clocks, so they inherit the result.
\end{proof}

\begin{corollary}\label{cor:rule9}
The previous results hold for $(9,\bip)$,$(9,\bp)$,$(9,\lc)$
\end{corollary}
\begin{proof}
Using the exact same reasoning as for theorems \ref{thm:1bip} and \ref{thm:1bp}, we can prove that the results are hold for $(9,\bip)$ and $(9,\bp)$. Thus, for the same reasons as corollary  \ref{cor:1lc} for $(9,\lc)$.
\end{proof}

Interestingly, the exact same update schedule can be used to prove a result for $(110,\bp)$, however, the same cannot be said for $(110,\bip)$ nor $(110,\p)$.
\begin{corollary}\label{thm:110}
    $(110,\bp)$ and $(110,\lc)$ can reach limit cycles of length $\Omega\left(2^{\sqrt{n\log n}}\right)$.
\end{corollary}
\begin{proof}
Using the exact same reasoning as for theorem \ref{thm:1bp} and subsequently for corollary \ref{cor:1lc}, we can prove that the results are hold for rule 110.
\end{proof}
\begin{figure}[!h]
	\centerline{\includegraphics{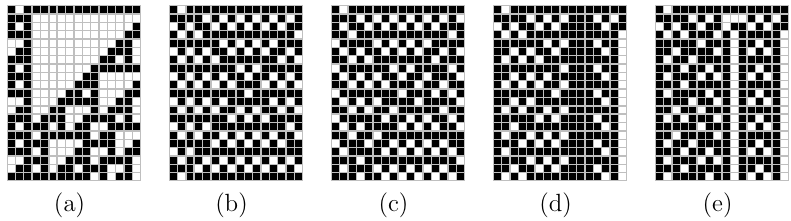}}
	\caption{Space-time diagrams (time going downward) of configuration 
		$1011111111111111$ following rule $110$ depending on: 
		(a) the parallel update mode $\mu_\p = (\integers{16})$, 
		(b) the bipartite update mode $\mu_\bip = (\{i \in \integers{16} \mid i \equiv
			0 \mod 2\}, \{i \in \integers{16} \mid i \equiv 1 \mod 2\})$, 
		(c) the block-sequential update mode $\mu_\bs = (\{0,3,5,7,9,11,13\},\{1,4,6,8,10,12,14\},\{2,15\})$, 
		(d) the block-parallel update mode $\mu_\bp =\{(0,1),(2,3),(4,5),(6,7),(8,9),(12,11),(14,13),(10),(15)\}$, 
		(e) the local clocks update mode $\mu_\lc = 
			(P = (2,2,2,2,2,2,2,2,2,1,2,2,2,2,2,1), 
			\Delta = (1,0,1,0,1,0,1,0,1,0,1,0,1,0,1,0))$.}
	\label{fig:110}
\end{figure}

\begin{table}[!ht]
\begin{center}
\begin{tabular}{|c|c|c|c|c|c|c|c|c|}
\hline
Rule&111&110&101&100&011&010&001&000\\\hline
73&0&1&0&0&1&0&0&1\\\hline
\end{tabular}
\end{center}
\end{table}

\begin{table}[!ht]\label{table:rule73}
\begin{center}\renewcommand{\arraystretch}{2}
\begin{tabular}{|C{1.5cm}|C{1.5cm}C{2cm}C{2cm}C{2.5cm}C{2.5cm}|}
\hline
Rule&Parallel&Bipartite&\makecell{Block\\Sequential}&\makecell{Block\\Parallel}&\makecell{Local\\Clocks}\\\hline
73&&$\Omega\left(2^{(\sqrt{n\log(n)})}\right)$&$\Omega\left(2^{(\sqrt{n\log(n)})}\right)$&$\Omega\left(2^{(\sqrt{n\log(n)})}\right)$&$\Omega\left(2^{(\sqrt{n\log(n)})}\right)$\\\hline
\end{tabular}
\end{center}
\caption{Longest limit cycle for Rule 73, according to update schedule.}
\end{table}

\begin{figure}[!h]
	\centerline{\includegraphics{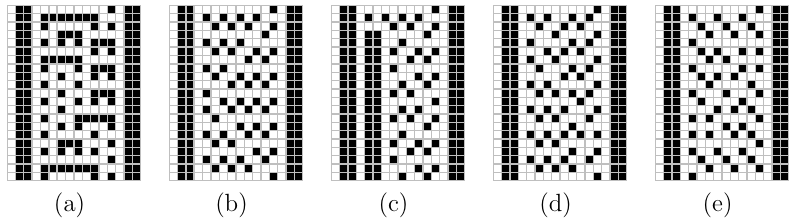}}
	\caption{Space-time diagrams (time going downward) of configuration 
		$011000000000000001011$ following rule $73$ depending on: 
		(a) the parallel update mode $\mu_\p = (\integers{16})$, 
		(b) the bipartite update mode $\mu_\bip = (\{i \in \integers{16} \mid i \equiv
			0 \mod 2\}, \{i \in \integers{16} \mid i \equiv 1 \mod 2\})$, 
		(c) the block-sequential update mode $\mu_\bs = (\{0,1,2,3,6,8,10,12,14,15\},\{4,5,7,9,11,13\})$, 
		(d) the block-parallel update mode $\mu_\bp =\{(0,1)(2,3,4,5),(9,8,6,7),(13,12,11,10),(15,14)\}$, 
		(e) the local clocks update mode $\mu_\lc = 
			(P = (3,3,3,3,3,3,3,3,3,3,3,3,3,3,3,3), 
			\Delta = (0,1,2,0,1,2,0,1,2,0,1,2,0,1,2,0,1,2,1))$.}
	\label{fig:rule73}
\end{figure}

\begin{lemma}
	\label{lem:73walls}
	ECA rule $73$ admits one wall, $0110$.
\end{lemma}

\begin{proof}
	By definition of the rule, $f_{73}(a01)=0$, with $a\in\{0,1\}$, $f_{73}(011)=f_{73}(110)=1$ and $f_{73}(10b)=0$, with $b\in\{0,1\}$. From where we know that $f_{73}(a0110b)=0110$, regardless of the update mode.
\end{proof}

We will now show that there exists (at least) one initial configuration for which a bipartite update mode can reach cycles of length $\Omega\left(2^{(\sqrt{n\log(n)})}\right)$
\begin{theorem}
	\label{thm:73bip}
	Rule $(73, \bip)$  applied over a grid of size $n$  can reach largest limit cycles of length $\Omega\left(2^{(\sqrt{n\log(n)})}\right)$
\end{theorem}
\begin{proof}
Without loss of generality, let $\mu_{\bip}=(\{i|i\mod 2\equiv 0\},\{i|i\mod 2\equiv 1\})$ a bipartite update schedule.\\
Let $x$ be an initial configuration such that $x=\omega 0^k 1 \omega=(0110) 0^k 01 (0110)$, with $k$ an even number. We need to calculate the state for the configuration $f(x)$.\\
Since $f_{73}(000)=1$, and $f_{73}(101)=0$, we have that
\begin{equation*}
\begin{matrix} 
x     &= &0110& 0^k    01                 & 0110\\
f(x)  &= &0110& (10)^{\frac{k}{2}}    00  & 0110\\
f^2(x)&= &0110& (01)^{\frac{k}{2}-1} 0010 & 0110\\
\\
f^3(x)&= &0110& 0^{k-2}  (01)^2                 & 0110\\
f^4(x)&= &0110& (10)^{\frac{k}{2}-1} 0000     & 0110\\
f^5(x)&= &0110& (01)^{\frac{k}{2}-2}(00)(10)^2& 0110\\
\end{matrix}
\end{equation*}
We can see that a pattern emerges, depending on the number of iteration $\mod 3$. Let $T=\lfloor\frac{t}{3}\rfloor$, with $t\in\N$, then we can write
\begin{equation*}
\begin{matrix} 
f^T(x)    &= &0110& 0^{k-2T}   (01)^{T+1}                 & 0110\\
f^{T+1}(x)&= &0110& (10)^{\frac{k}{2}-T}(00)^{T+1}        & 0110\\
f^{T+2}(x)&= &0110& (01)^{\frac{k}{2}-(T+1)}(00)(10)^{T+1}& 0110\\
\end{matrix}
\end{equation*}
Now, let $T^*=\frac{k-2}{2}$
\begin{equation*}
\begin{matrix} 
f^{T^*}(x)  &= &0110& 0^{k-2T^*}   (01)^{T^*+1}                    & 0110\\
            &= &0110& 0^{k-2(\frac{k-2}{2})}(01)^{\frac{k-2}{2}+1} & 0110\\
            &= &0110& 0^{2}   (01)^{\frac{k}{2}}                   & 0110\\
f^{T^*+1}(x)&= &0110& (10)^{\frac{k}{2}-T^*}(00)^{T^*+1}           & 0110\\
            &= &0110& (10)^{1}(00)^{\frac{k}{2}}                   & 0110\\
\end{matrix}
\end{equation*}
We have that $f^{T^*+1}(x)$ is $x'= \omega 1 0^k \omega$, and through an analysis symmetrical to what we just did, we obtain that $f^{T^*+1}(x')=x$, meaning that when $k$ is even, the length of the cycle between walls can reach at least a length of $k$ (with the distance between walls equal to $k+2$).\\
Analogously, if $k$ is odd we have that:\\
\begin{equation*}
\begin{matrix} 
x     &= &0110& 0^k    01               & 0110\\
f(x)  &= &0110& (10)^{\frac{k+1}{2}} 0  & 0110\\
f^2(x)&= &0110& (01)^{\frac{k-1}{2}} 001& 0110\\
\\
f^3(x)&= &0110& 0^{k-1}  (10)                 & 0110\\
f^4(x)&= &0110& (10)^{\frac{k+1}{2}-1} 000    & 0110\\
f^5(x)&= &0110& (01)^{\frac{k-1}{2}-1} 0(01)^2& 0110\\
\vdots\\
f^T(x)    &= &0110& 0^{k-T}  (10)^T                   & 0110\\
f^{T+1}(x)&= &0110& (10)^{\frac{k+1}{2}-T} 0(00)^T    & 0110\\
f^{T+2}(x)&= &0110& (01)^{\frac{k-1}{2}-T} 0(01)^{T+1}& 0110\\
\vdots\\
f^{T^*}(x)  &= &0110& 0^{k-T^*}  (10)^{T^*}               & 0110\\
            &= &0110& 0^{3}  (10)^{\frac{k-1}{2}}         & 0110\\
f^{T^*+1}(x)&= &0110& (10)^{\frac{k+1}{2}-T^*} 0(00)^{T^*}& 0110\\
            &= &0110& (10)^{1} 0(00)^{\frac{k-1}{2}}      & 0110\\
\end{matrix}
\end{equation*}
with $T^*=\frac{k-1}{2}$.\\
Using the same argument used for \ref{thm:156BIP}, we conclude that the largest limit cycles of the family $(73,\bip)$ when applied over a ring of length $n$ is $\Omega\left(2^{\sqrt{n\log n}}\right)$.
\end{proof}

\begin{corollary}\label{thm:73bloques}
    $(73,\bs),(73,\bp),(73,\lc)$ have largest limit cycles of length $\Omega\left(2^{\sqrt{n\log n}}\right)$.
\end{corollary}
\begin{proof}
    The family of sequential update schedules are a subset of block parallel and block sequential, which themselves are a subset of local clocks, so they inherit the result.
\end{proof}

\begin{table}[!ht]
\begin{center}
\begin{tabular}{|c|c|c|c|c|c|c|c|c|}
\hline
Rule&111&110&101&100&011&010&001&000\\\hline
178&1&0&1&1&0&0&1&0\\\hline
\end{tabular}
\end{center}
\end{table}

\begin{table}[!ht]\label{table:rule178}
\begin{center}\renewcommand{\arraystretch}{2}
\begin{tabular}{|C{1.5cm}|C{1.5cm}C{2cm}C{2cm}C{2.5cm}C{2.5cm}|}
\hline
Rule&Parallel&Bipartite&\makecell{Block\\Sequential}&\makecell{Block\\Parallel}&\makecell{Local\\Clocks}\\\hline
178&$\Theta(1)$&$\mathcal{O}(n)$&$\mathcal{O}(n)$&$\Omega\left(2^{(\sqrt{n\log(n)})}\right)$&$\Omega\left(2^{(\sqrt{n\log(n)})}\right)$\\\hline
\end{tabular}
\end{center}
\caption{Longest limit cycle for Rule 178, according to update schedule.}
\end{table}

\begin{figure}[!h]
	\centerline{\includegraphics{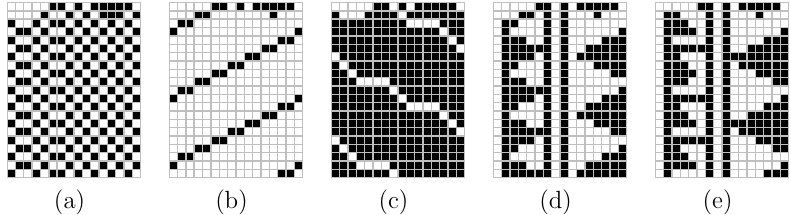}}
	\caption{Space-time diagrams (time going downward) of configuration 
		$0000011010111110$ following rule $178$ depending on: 
		(a) the parallel update mode $\mu_\p = (\integers{16})$, 
		(b) the bipartite update mode $\mu_\bip = (\{i \in \integers{16} \mid i \equiv
			0 \mod 2\}, \{i \in \integers{16} \mid i \equiv 1 \mod 2\})$, 
		(c) the block-sequential update mode $\mu_\bs = (\{3,9,15\}, \{2,4,8,10,14\}, 
			\{11,5,7,11,13\},\{0,6,12\})$, 
		(d) the block-parallel update mode $\mu_\bp =\{(0), (1), (2,3), (4,5), (6),
			(7), (8), (9), (10, 11), (12, 13), (14, 15)\}$, 
		(e) the local clocks update mode $\mu_\lc = 
			(P = (1,1,2,2,2,2,1,1,1,4,4,4,4,4,4,4), 
			\Delta = (0,0,1,0,1,0,0,0,0,0,1,0,1,0,1,0))$.}
	\label{fig:rule178}
\end{figure}
\begin{lemma}
	\label{lem:178walls}
	ECA rule $178$ admits two walls, $01$ and $10$, which are relative walls.
\end{lemma}

\begin{proof}
	Let $a, b \in \B$. %such that $a \neq b$. 
	Notice that by definition of the rule: 
	$f(a00) \neq f(b00)$ and $f(00a) \neq f(00b)$, and 
	$f(a11) \neq f(b11)$ and  $f(11a) \neq f(11b)$ if $a \neq b$; 
	$f(a01) = 1$ and $f(01a) = 0$; and
	$f(a10) = 0$ and $((10a) = 1$.
	Thus, neither $00$ nor $01$ nor $10$ nor $11$ are absolute walls. 
	
	From what precedes, notice that the properties of $00$ and $11$ prevent 
	them to be relative walls. 
	Consider now the two words $u = 01$ and $v = 10$ and let us show that they 
	constitute relative walls. 
	Regardless the states of the cells surrounding $u$, every time both 
	cells of $u$ are updated simultaneously, $u$ changes to $v$ and similarly, $v$ will change to $u$ independently of the states of the cells that surround it, as long as both its cells are updated together.
	Thus, $u$ and $v$ are relative walls.
\end{proof}
We will show certains $\bp$ and $\lc$ update modes that are able to produce these relative walls.
\begin{theorem}
	\label{thm:178BIP}
	Each representative of $(178, \bip{})$  applied over a grid of size $n$  has largest limit cycles of 
	length $\Theta(n)$.
\end{theorem}
\begin{proof}[Sketch of proof.]
From lemma \ref{lem:178walls} we know that the representatives of $(178,\bip)$ do not have walls, because by definition of $\bip$ two consecutive cells cannot be updated simultaneously.\\
Thus we can start by first considering configurations $x$ composed by \textit{one} group of contiguous $1$s and proving that if the number of $1$s is odd then the dynamics lead to homogeneous fixed points. Meanwhile, if the number of $1$s is even, then such configurations lead to limit cycles of length $\frac{n}{2}$.\\
Secondly, we consider configurations with \textit{several} groups of contiguous $1$s, which can be analyzed by studying the behaviour with just two groups of $1$s. These kinds of configurations are divided into $10$ cases, depending on the parity of the size of the groups, and the local bipartite update mode that each of the groups follows. We prove that each case converges either to a homogeneous fixed point or to limit cycles of length $\frac{n}{2}$.
\end{proof}

\begin{theorem}\label{thm:178bp}
    The family $(178,\bp)$ of size $n$ has largest limit cycles of length $\Omega\left(2^{\sqrt{n\log n}}\right)$
\end{theorem}

\begin{proof}[Sketch of proof]
First we notice that since $1^n$ and $0^n$ are fixed points, the configurations of interest have at least one relative wall. By Lemma $\ref{lem:178walls}$, we know that $w_1$ and $w_2$ are relative walls for the family $(178,\bp)$. Thus, we consider an initial configuration with  at least one wall. Similar to the proof of theorem \ref{thm:156BIP}, the idea is to focus on what can happen between two walls: $y=w_{\ell}y_1y_2\dots y_ky_{k+1}w_r$, given that the dynamics of subconfigurations delimited by two pairs of walls are independent.\\
We analyse different cases depending on if $w_{\ell}=w_r=01$ (similarly, $w_{\ell}=w_r=10$), or $w_{\ell}=01$ and $w_r=10$ (similarly, $w_{\ell}=10$ and $w_r=01$) and we prove that each of the cases converges either to a fixed point or to limit cycles of length $k+1$.\\
And because of the independence of the dynamics between two pairs of walls, the asymptotic dynamics of a global configuration $x$ is a limit cycle whose length equals the least common multiple of the lengths of all limit cycles of the subconfigurations, and with the same argument as was used in the proof of theorem \ref{thm:156BIP}, we conclude that the largest limit cycles of the family $(178,\bp)$  applied over a grid of size $n$  is $\Omega\left(2^{\sqrt{n\log n}}\right)$.
\end{proof}

\begin{theorem}
    The family $(178,\bp)$ of size $n$ has largest limit cycles of length $\Omega\left(2^{\sqrt{n\log n}}\right)$
\end{theorem}
\begin{proof}
    Notice that the configurations of interest here are those having at least one relative wall since $(0)^n$ and $(1)^n$ are fixed points. By Lemma \ref{lem:178walls} (and its proof) $w_1$ and $w_2$ are relative walls for the family $(178,\bp)$ since there exist block-parallel updates modes guarantying that the two contiguous cells carrying $w_1$ (resp. $w_2$) are updated simultaneously and an even number of substeps over the period. Thus, let us consider in this proof an initial configuration $x$ with at least one wall. The idea is to focus on what can happen between two walls because the dynamics of two subconfigurations delimited by two distinct pairs of walls are independent from each other.\\
    Now, let $y$ (resp. $y'$) be a subconfiguration of size $k+4$ such that $y = (w_{\ell},y_2,\dots,y_{k+1},w_r)$, with $w_{\ell}=y_0y_1,w_r=y_{k+2}y_{k+3}$ two relative walls in $W = \{w_1,w_2\}$ and such that for all $i\in\{2,\dots,k+1\},y_i=0$ (resp. $y_i'=1$) and the block-parallel mode
    \begin{equation*}
        \begin{split}
            \mu_{\bp} &= \{(y_0),(y_1),(y_2,y_3),\dots,(y_k,y_{k+1}),(y_{k+2}),(y_{k+3})\\
            &\equiv (\{y_0,y_1,y_2,y_4,\dots,y_{k},y_{k+2},y_{k+3}\},\{y_0,y_1,y_3,y_5,\dots,y_{k+1},y_{k+2},y_{k+3}\}),
        \end{split}
    \end{equation*}
    if $k$ is even, and:
    \begin{equation*}
        \begin{split}
            \mu_{\bp} &= \{(y_0),(y_1),(y_2,y_3),\dots,(y_k,y_{k+1}),(y_{k+2}),(y_{k+3})\\
            &\equiv (\{y_0,y_1,y_2,y_4,\dots,y_{k+1},y_{k+2},y_{k+3}\},\{y_0,y_1,y_3,y_5,\dots,y_{k},y_{k+2},y_{k+3}\}),
        \end{split}
    \end{equation*}    
    if $k$ is odd. The dynamics of $y$ follows four cases:
    \begin{enumerate}
        \item $w_{\ell}=w_r=01$, given $\tau\in\Z$, denoting the subconfigurations obtained at a substep by $y^{\frac{\tau}{2}}$:\\
        \overfullrule=0pt
        \begin{minipage}{.48\textwidth}
        -- If $k$ is even, we have
            \begin{align*}
    y&=&(01)(0)^k(01)\\
    y^{\left(\frac{1}{2}\right)}&=&(10)(1)(0)^{k-1}(10)\\
    y^1&=&(01)(1)^2(0)^{k-3}(1)(01)\\
    y^{\left(\frac{3}{2}\right)}&=&(10)(1)^3(0)^{k-5}(1)^2(10)\\
    y^2&=&(01)(1)^4(0)^{k-7}(1)^3(01)\\
    &\vdots&\\    
    y^i&=&(01)(1)^{2i}(0)^{k-4i+1}(1)^{2i-1}(01)\\
    &\vdots&\\
    y^{\frac{k-1}{2}}&=&(01)(1)^k(01)\\
    y^{\frac{k}{2}-1}&=&(10)(1)^k(10)\\
    y^{\frac{k}{2}}&=&(01)(1)^k(01)\\                
            \end{align*}
        \end{minipage}        
        \begin{minipage}{.48\textwidth}
        -- If $k$ is odd, we have
            \begin{align*}
    y&=&(01)(0)^k(01)\\
    y^{\left(\frac{1}{2}\right)}&=&(10)(1)(0)^{k-1}(10)\\
    y^1&=&(01)(1)^2(0)^{k-2}(01)\\
    &\vdots&\\
    y^i&=&(01)(1)^{2i}(0)^{k-2i}(01)\\
    &\vdots&\\    
    y^{\frac{k+1}{2}}&=&(01)(1)^k(01)\\
    y^{\frac{k+2}{2}}&=&(10)(1)^{k-1}(0)(10)\\
    y^{\frac{k+3}{2}}&=&(01)(1)^{k-2}(0)^2(01)\\
    &\vdots&\\
    y^{k+1}&=&(01)(0)^k(01)\\                
            \end{align*}
        \end{minipage}
    Thus, subconfiguration $y$ converges towards fixed point $(01)(1)^k(01)$ when $k$ is even and leads to a limit cycle of length $k+1$ when $k$ is odd.
    \item $w_{\ell}=w_r=10$: taking $y'$ as the initial subconfiguration, and applying the same reasoning, we can show that this case is analogous to the previous one up to a symmetry, which allows us to conclude that $y'$ converges towards fixed point $(10)(0)^k(10)$ when $k$ is even and leads to a limit cycle of length $k+1$ when $k$ is odd.
    \item $w_{\ell}=01$ and $w_r=10$:\\
    \begin{minipage}{.48\textwidth}
    --- If $k$ is even, we have
        \begin{align*}
    y&=&(01)(0)^k(10)\\
    y^{\left(\frac{1}{2}\right)}&=&(10)(1)(0)^{k-1}(01)\\
    y^1&=&(01)(1)^2(0)^{k-2}(10)\\
    &\vdots&\\
    y^i&=&(01)(1)^{2i}(0)^{k-2i}(10)\\
    &\vdots&\\
    y^{\frac{k}{2}}&=&(01)(1)^k(10)\\
    y^{\left(\frac{k+1}{2}\right)}&=&(10)(1)^k(01)&\\
    y^{\left(\frac{k+2}{2}\right)}&=&(01)(1)^{k-1}(0)(10)\\
    &\vdots&\\
    y^{k+1}&=&(01)(0)^k(10)\\            
        \end{align*}
    \end{minipage}
    \begin{minipage}{.48\textwidth}
    --- If $k$ is odd, we have
        \begin{align*}
    y&=&(01)(0)^k(10)\\
    y^{\left(\frac{1}{2}\right)}&=&(10)(1)(0)^{k-2}(01)\\
    y^1&=&(01)(1)^2(0)^{k-4}(1)^2(10)\\
    y^{\left(\frac{3}{2}\right)}&=&(10)(1)^3(0)^{k-6}(1)^3(01)\\
    y^2&=&(01)(1)^{4}(0)^{k-8}(1)^4(01)\\
    &\vdots&\\
    y^i&=&(01)(1)^{2i}(0)^{k-4i}(1)^{2i}(10)\\
    &\vdots&\\
    y^{\left(\frac{k-1}{2}\right)}&=&(01)(1)^k(10)\\
    y^{\frac{k}{2}}&=&(10)(1)^k(01)\\
    y^{\frac{k+1}{2}}&=&(01)(1)^k(10)\\            
        \end{align*}
    \end{minipage}
 Thus, subconfiguration $y$ leads to a limit cycle of length $k+1$ when $k$ is even and converges towards fixed point $(01)(1)^k(10)$ when $k$ is odd.
 \item $w_{\ell}=10$ and $w_r=10$: taking $y'$ as the initial subconfiguration, and applying the same reasoning we can show that this case is analogous to the previous one up to a symmetry, which allows us to conclude that $y'$ leads to a limit cycle of length $k+1$ when $k$ is even and converges towards fixed point $(10)(0)^k(01)$ when $k$ is odd.
 \end{enumerate} Finally, since the dynamics between two pairs of distinct walls is independent of each other, the asymptotic dynamics of a global configuration $x$ is a limit cycle whose length equals the least common multiple of the lengths of all limit cycles of the subconfigurations embedded into pairs of walls. With the same argument as the one used in the proof of theorem \ref{thm:156BIP}, we derive that the length of the largest limit cycles of the family $(178,\bp)$ of size $n$ is $\Omega\left(2^{\sqrt{n\log n}}\right)$.\qed
\end{proof}

\begin{theorem}\label{thm:178SEQ}
    $(178,\seq)$ of size $n$ has largest limit cycles of length  $\mathcal{O}(n)$
\end{theorem}
\begin{proof}[Sketch of proof]
    We analyse the dynamics that start from an isle of $1's$ surrounded by $0's$, dividing it on four cases, depending on if the first and last cells of the isle are updated before or after their neighboring $0$. The first two cases result in fixed points, a third in which the isle of $1's$ shifts to the right, and a fourth in which the isle of $1's$ shifts to the left.\\
    Because of the condition necessary for the shift to occur, we know that an isle shifting left-to-right (or right-to-left) will complete a cycle in less than $n$ iterations.\\
    We consider an initial configuration $x$ written as isles of $1's$ separated by $0's$. Each of the isles must fall into one of the four cases analysed on the previous point.\\
    Isles that correspond to the first two cases either disappear without ever interacting with another isle or they fuse with the nearest isle, which will result in a new isle which case will be of the same type as the isle the original one fused with.\\
    We prove that an isle traveling left-to-right cannot interact with another one going in the same direction, but will fuse with one shifting right-to-left into a new isle which will correspond to the case where it will eventually disappearing with no possibility of ever interacting with another isle.\\
    We conclude that if in the initial configuration there is an equal number of isles that correspond to the cases where the limit cycle will be a fixed point, but if there is more of one of those cases than the other, the limit cycle will be of length less than n.
\end{proof}
\begin{theorem}\label{thm:178BS}
    $(178,\bs)$  applied over a grid of size $n$  has largest limit cycles of length $\mathcal{O}(n)$
\end{theorem}
\begin{proof}[Sketch of proof]
    First, we use the fact that if two neighboring cells cannot be updated simultaneously, then the update mode is equivalent to a sequential update mode, whose limit cycles we have already proven to be strictly less than n.\\
    Then, we prove that if there are two consecutive cells that update simultaneously then limit cycle is of length at most 2. In order to prove this, we start by analysing the case with just one pair of cells that update on the same sub-step $\{s,s+1\}$, and just one isle of ones that stars on one of the neighboring cells of interest. This analysis is then divided in four cases, similar to the previous proof.\\
    We determine that every case produces two isles of ones, which travel in opposite directions, and we know that they will have to meet on the other side, once their combined movement circumnavigates the ring, whereupon they will fuse, become one isle that will have to disappear.\\
    Then, we note that the previous analysis holds with an isle that starts in a different place of the configuration and eventually arrives to the cells $s$ and $s+1$, and finally, it holds when there are multiple isles and/or multiple pairs of neighboring cells that update simultaneously.
\end{proof}
\begin{theorem}\label{thm:178PAR}
    $(178,PAR)$ has largest limit cycles of length 2.
\end{theorem}
\begin{proof}
    Direct from the proof of theorem \ref{thm:178BS}.
\end{proof}

\subsection{Experimental Results}
\label{subsec:exp}

There are around twenty rules that are too complex or too chaotic to obtain clear mathematical proofs. In order to study them we devised experiments where we calculated the dynamics of the density and energy. This allows us to obtain an idea about what happens when we change the update modes from $\p$ to $\seq$, $\bs$, $\bp$ and $\lc$. Indeed, we would like to understand how these properties are affected by the different update modes. With this goal in mind we have chosen rules 90, 150, 54 and 110. These rules are well known and famous for their complexity, with the first two belonging to class III~\cite{wolfram1984computation} and the other two to class IV of Wolfram's complexity classification~\cite{li1990structure}. 

\begin{definition}[Density]
The \emph{density} is defined as the average number of ones in a configuration $x\in\Bn$ (introduced as magnetization in~\cite{Zabolitzky1988}), that is: 
\begin{equation*}
d(x)=\frac{1}{n}\sum_{i=0}^{n-1} x_i,
\end{equation*}
with $x_i$ the state of cell $i$ for all $i\in\{0,\dots,n-1\}$. 
\end{definition}

\begin{definition}[Energy]
The \emph{energy} of a configuration is defined as 
\begin{equation*}
e(x) = \sum_{i=0}^{n-1}\frac{1-2x_i}{2}\left((2x_{i-1}-1)+(2x_{i+1}-1)\right),
\end{equation*}
with $x_i$ the state of cell $i$ for all $i\in\{0,\dots,n-1\}$. 
\end{definition}
Informally, the energy of a configuration $x$ is a measure of the total number of cells in a configuration that have a different state to that of their neighbors. Note that the energy can have values between $-n$ and $n$.

 For example, a homogeneous configuration will have maximum energy \ie $-n$, and in a configuration with alternating states $1010\dots$ the energy will be maximum,~\ie~ $n$.
This concept was used in the context of the study of discrete dynamical systems for threshold networks~\cite{Hopfield1982,Goles1985} and icing models~\cite{Mccoy2012}.
In order to compare different ring sizes, we can define \emph{normalized energy} as
\begin{equation*}
\bar{e}(x) = \frac{1}{n} \sum_{i=0}^{n-1}\frac{1-2x_i}{2}\left((2x_{i-1}-1)+(2x_{i+1}-1)\right).
\end{equation*}

\subsubsection*{Protocol}

The programs used, as well as .csv files containing the numerical results of the experiments are available on the following repository:
\begin{center}
\url{https://framagit.org/Isabel-Camila/Impact-of-Asynchronism}
\end{center}

We considered three parameters to start: the size of the ring ($n$), the size of the sample of configurations ($s$) and the size of the sample of update modes ($m$). 
We decided to perform the experiments through a number of samples of configurations, instead of calculating for all possible initial configurations for rings of size 20 and higher because the computational time grew too large which would not allow to experiment with different update modes.\smallskip

We chose rings of sizes $8$, $38$ and $138$, with sample sizes of $32$ and $128$ configurations. The configuration samples were created such that each cell had the same probability to hold state $0$ or $1$, thus each sample has average density equal to $0.5$ and average energy equal to $0$. Furthermore, we ran both configuration sample sizes for each ring size under 32 different update mode from each family. We calculated the dynamics over 1000 time steps of the average density and energy (averaging over both the configuration sample sizes and the update mode sample size). Table~\ref{table:summary_SampleSizes1} shows a summary of the considered sample configurations and update modes.
\begin{table}[t!]
\begin{center}
\begin{tabular}{|c|c|c|c|c|c|}
\hline
$n$&$s$&$m$ $\seq$& $m$ $\bs$& $m$ $\bp$& $m$ $\lc$\\\hline
8&32,128&32&32&32&32\\
38&32,128&32&32&32&32\\
138&32,128&32&32&32&32\\\hline
\end{tabular}
\end{center}
\caption{Summary of configuration ($s$) and update mode ($m$) sample sizes for rings of sizes $n=8$, $n=38$ and $n=138$.}
\label{table:summary_SampleSizes1}
\end{table}

After performing these simulations, we realized that higher orders of magnitude for the size of the ring were not needed because the experiments hinted that the overall behavior for each rule with its respective update modes resulted in similar graphs for rings of sizes $n=8$, $n=38$ and $n=138$. This meant that we could experiment on smaller rings in which every possible initial configuration could be tested, which lets us pay particular attention to update modes.

We chose a ring of size $16$, which allowed us to perform $1000$ time steps of the rule under different update modes for all possible configurations in a reasonable amount of time. As a result, we were able to perform the experiments with different number of blocks for $\bs$ and different period length for $\lc$. Table~\ref{table:summary_SampleSizes2} shows a summary of the considered samples sizes of update modes.
\begin{table}[t!]
\begin{center}
\begin{tabular}{|c|c|c|c|}
\hline
$m$ $\seq$&$m$ $\bs$&$m$ $\bp$&$m$ $\lc$\\\hline
$32$&$96$&$32$&$96$\\\hline
\end{tabular}
\end{center}
\caption{Update mode ($m$) sample sizes for $n = 16$}
\label{table:summary_SampleSizes2}
\end{table}

\FloatBarrier
\subsubsection{Different ring and sample sizes}

In this section we will show the behaviors of density and energy for rule 110 (rules 54, 90 and 150 will be discussed in the appendix). We can see that said measures are very similar when comparing rings of sizes $8$, $38$ and $138$, which gives us examples in three different orders of magnitude. Note that on the graphs we are only showing the first 200 iterations, since that is enough to see that the behavior stabilizes.

Examples of the update modes for $n=8$ are defined on Table~\ref{table:um8}.

\begin{table}[t!]
\centering
\begin{tabular}{|c|c|}
\hline
\makecell{Update\\Mode}&Definition\\\hline
Sequential&$(2, 3, 4, 5, 7, 6, 1, 0)$\\\hline
\makecell{Block\\Sequential}& $(\{1\},\{2, 7\},\{6, 4\}, \{5, 0, 3\})$\\\hline
\makecell{Block\\Parallel}& $\{(1, 2), (5, 3), (7), (6), (4, 0)\}$\\\hline
\makecell{Local\\Clocks}& $\{P= (2, 3, 3, 1, 1, 2, 3, 3),\Delta = (1, 0, 2, 0, 0, 0, 1, 0)\}$\\\hline
\end{tabular}
\caption{Example of the update modes used for $n=8$.}
\label{table:um8}
\end{table}

\FloatBarrier
\subsubsection*{Rule 110}
\begin{figure}[ht]
\includegraphics[width=\textwidth]{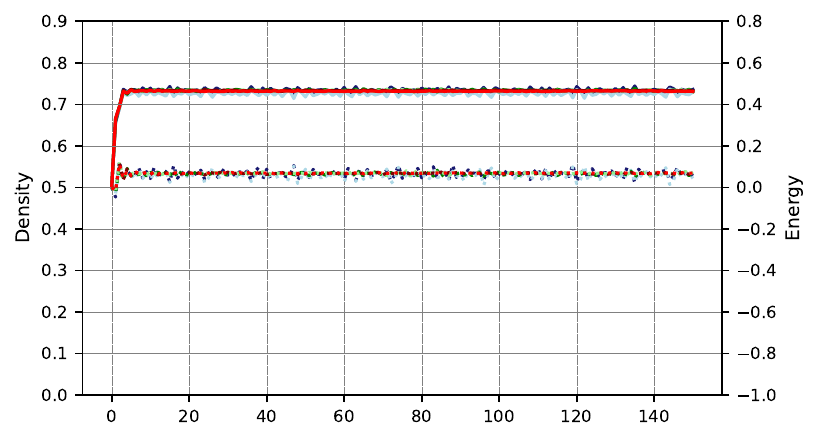}
\includegraphics[width=\textwidth]{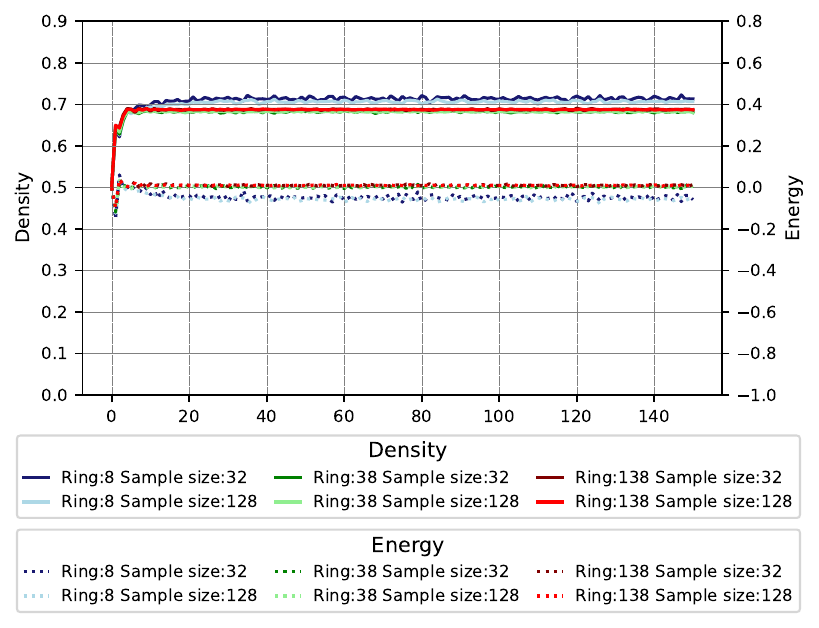}
\caption{Density (left y-axis) and normalized energy (right y-axis) for a ring of sizes $n=8$ (blue), $n=38$ (green) and $n=138$ (red) under rule 110 with sequential (top) and block-sequential (bottom) update mode with sample sizes of $s = 32$ and $s=128$ initial configurations, over 150 time steps.}
\label{fig:110 seq 38}
\end{figure}

As seen on Fig.~\ref{fig:110 seq 38}, we have that for all three ring sizes with sequential update modes the average density increases until it stabilizes at around $.74$, while there is a slight increase of energy for all three ring sizes. Similarly, for block-sequential update modes, on Fig.~\ref{fig:110 seq 38} we see that kind of behavior for the density, while the energy averages close to $0$ for all three ring sizes.

\begin{figure}[ht]
\includegraphics[width=\textwidth]{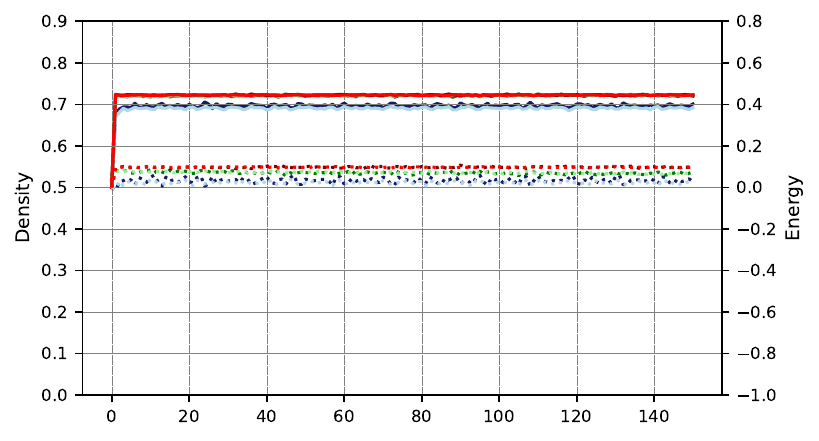}
\includegraphics[width=\textwidth]{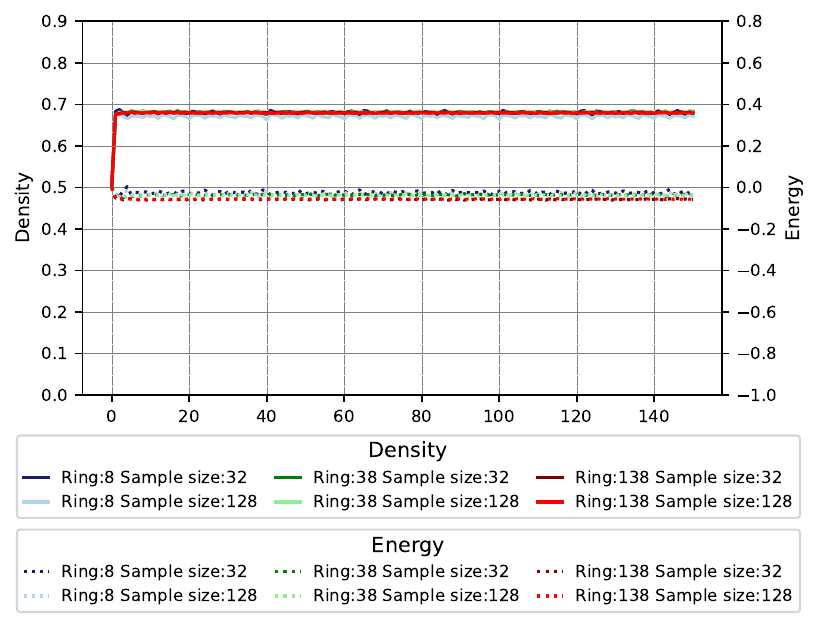}
\caption{Density (left y-axis) and normalized energy (right y-axis) for a ring of sizes $n=8$ (blue), $n=38$ (green) and $n=138$ (red) under rule 110 with block-parallel (top) and local clocks (bottom) update mode with sample sizes of $s = 32$ and $s=128$ initial configurations, over 150 time steps.}
\label{fig:110 lc 38}
\end{figure}

For the block-parallel update modes we can see on Fig.~\ref{fig:110 lc 38} that while the density always increases to very similar amounts, the energy does show different behaviors for each ring size, while for the local clocks update modes we have that the energy always decreases (Fig.~\ref{fig:110 lc 38}). We attribute the more irregular behavior of these two families to the fact that they allow for single cells to be updated more than once per time step, which gives a greater variety in their dynamics.

\subsubsection*{Rule 54}
\begin{figure}[ht]
\includegraphics[width=\textwidth]{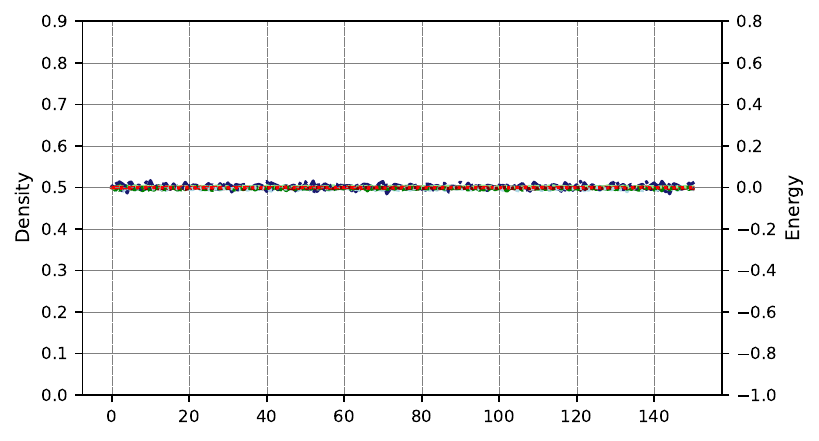}
\includegraphics[width=\textwidth]{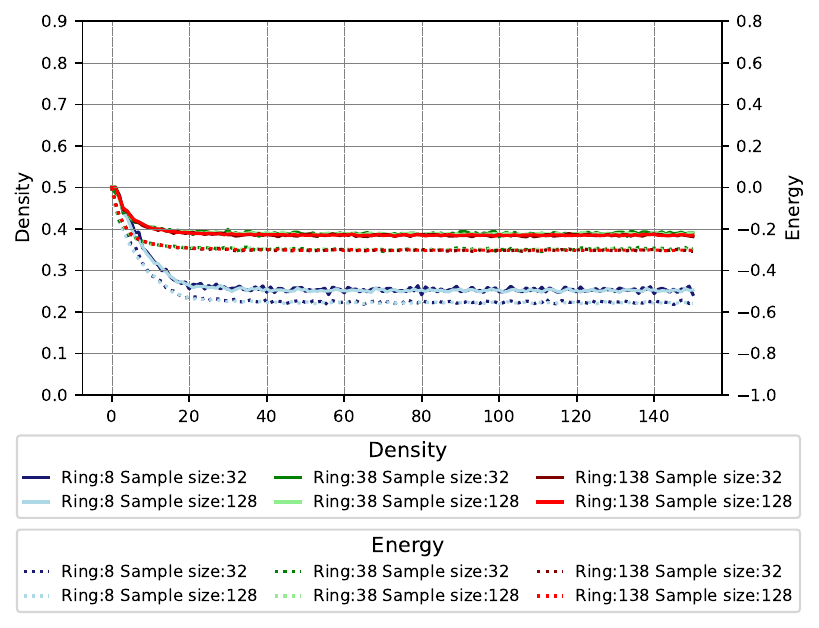}
\caption{Density (left y-axis) and normalized energy (right y-axis) for a ring of sizes $n=8$ (blue), $n=38$ (green) and $n=138$ (red) under rule 54 with sequential (top) and block-sequential (bottom) update mode with sample sizes of $s = 32$ and $s=128$ initial configurations, over 150 time steps.}
\label{fig:54 seq 38}
\end{figure}

\begin{figure}[ht]
\includegraphics[width=\textwidth]{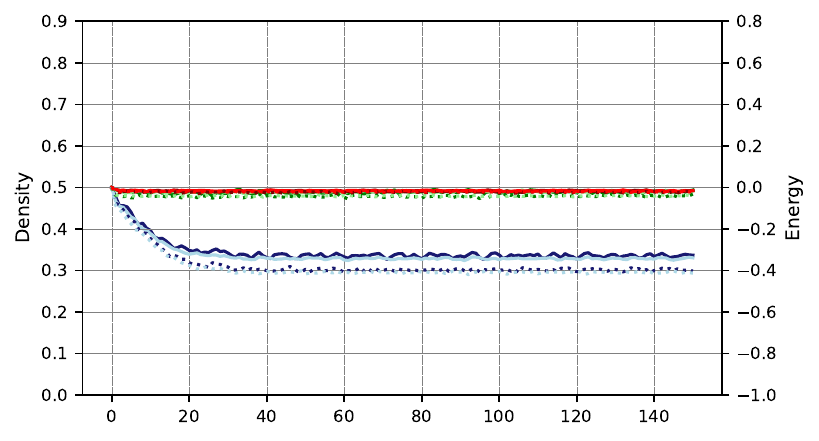}
\includegraphics[width=\textwidth]{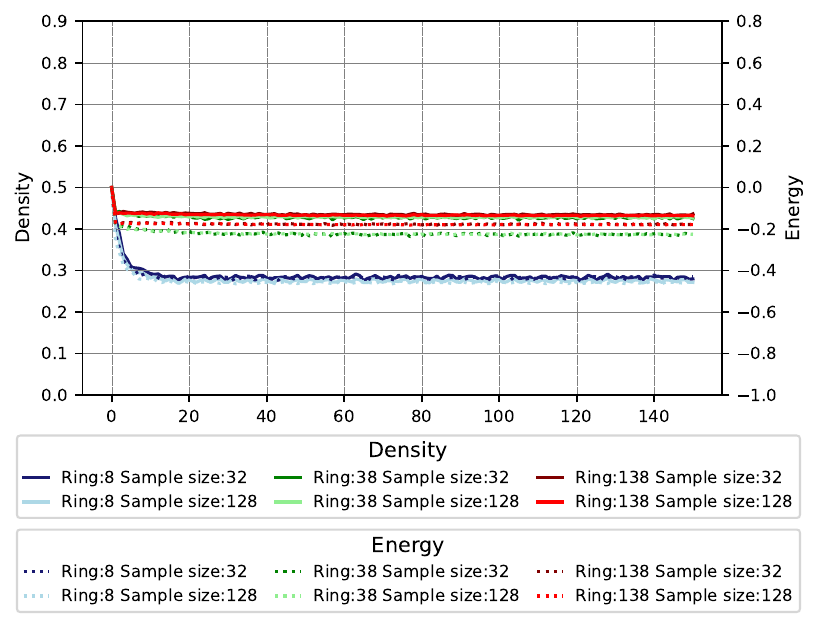}
\caption{Density (left y-axis) and normalized energy (right y-axis) for a ring of sizes $n=8$ (blue), $n=38$ (green) and $n=138$ (red) under rule 54 with block-parallel (top) and local clocks (bottom) update mode with sample sizes of $s = 32$ and $s=128$ initial configurations, over 150 time steps.}
\label{fig:54 lc 38}
\end{figure}

We can see on Fig.~\ref{fig:54 seq 38} (top) that with bigger sample size both density and energy stabilize at around $.5$ and $0$ resp. for sequential update modes, unlike with block-sequential, block-parallel and local clock where the average value of the density and energy decrease, as seen on Figs.~\ref{fig:54 seq 38} (bottom) and~\ref{fig:54 lc 38}. Note that for each update modes both density and energy reach similar values, especially for rings of size $38$ and $138$, which agrees with our assessment that the size of the ring is not as important.

\subsubsection*{Rule 90}
In the case of Rule 90, it appears that the density and energy are not affected, regardless of the update mode, with the density averaging at around .5 and the energy around 0. 

\begin{figure}[ht]
\includegraphics[width=\textwidth]{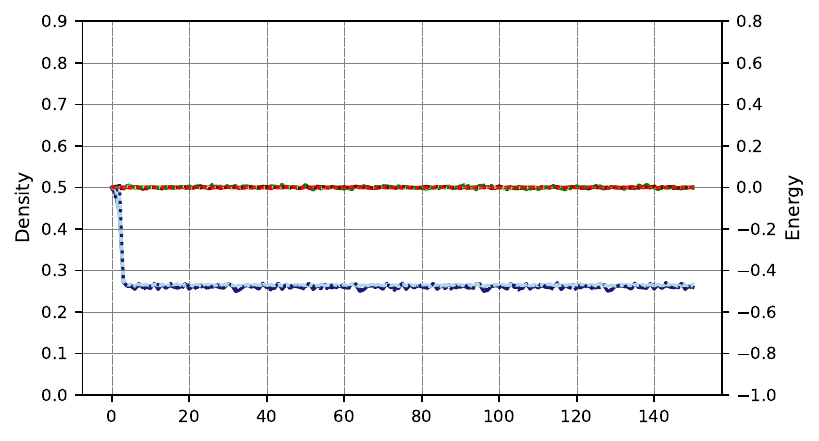}
\includegraphics[width=\textwidth]{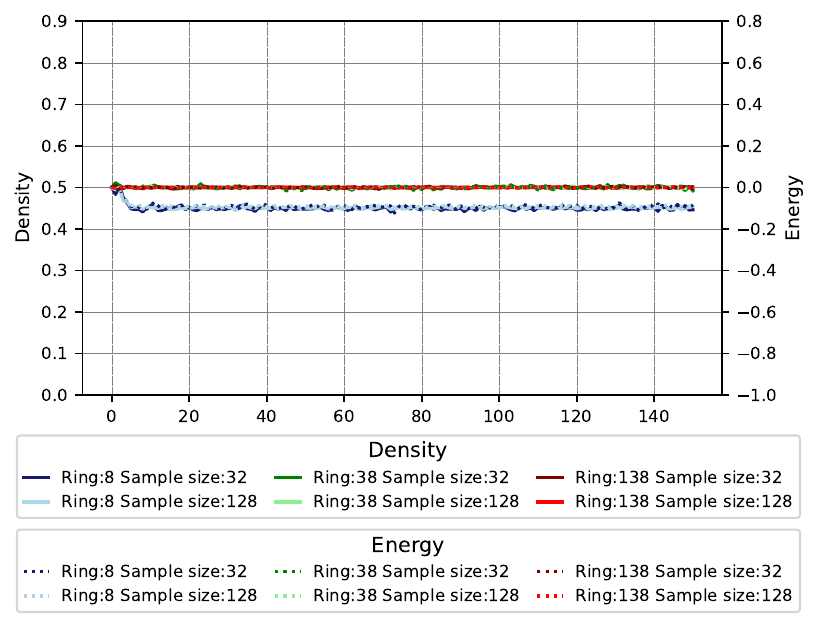}
\caption{Density (left y-axis) and normalized energy (right y-axis) for a ring of sizes $n=8$ (blue), $n=38$ (green) and $n=138$ (red) under rule 90 with sequential (top) and block-sequential (bottom) update mode with sample sizes of $s = 32$ and $s=128$ initial configurations, over 150 time steps.}
\label{fig:90 seq 38}
\end{figure}

\begin{figure}[ht]
\includegraphics[width=\textwidth]{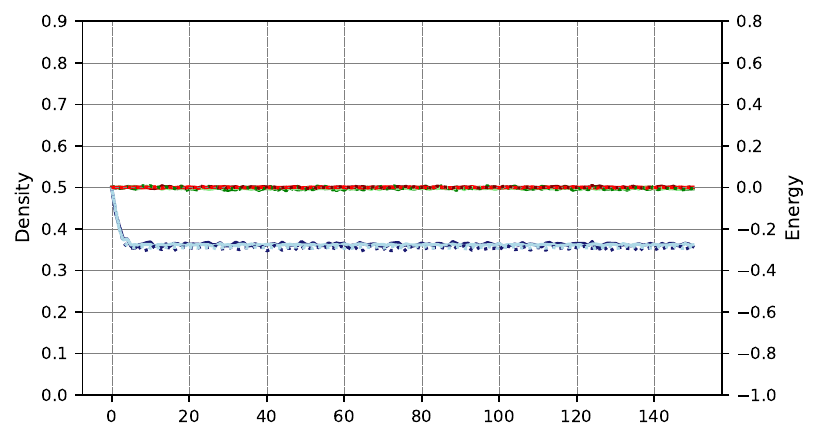}
\includegraphics[width=\textwidth]{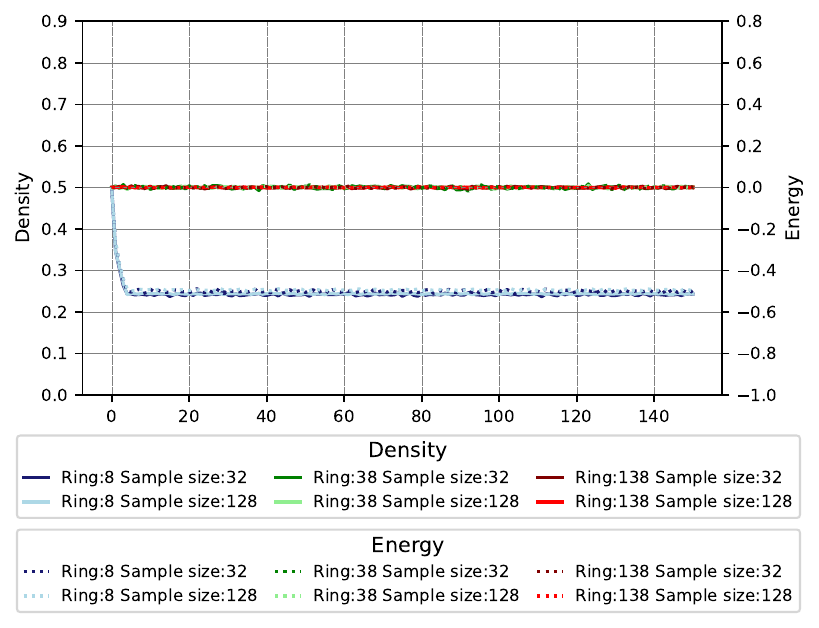}
\caption{Density (left y-axis) and normalized energy (right y-axis) for a ring of sizes $n=8$ (blue), $n=38$ (green) and $n=138$ (red) under rule 90 with block-parallel (top) and local clocks (bottom) update mode with sample sizes of $s = 32$ and $s=128$ initial configurations, over 150 time steps.}
\label{fig:90 lc 38}
\end{figure}

\subsubsection*{Rule 150}
In the case of Rule 150, even more notably than with rule 90, we can see that the energy and density are very stable, regardless of the update mode.

\begin{figure}[ht]
\includegraphics[width=\textwidth]{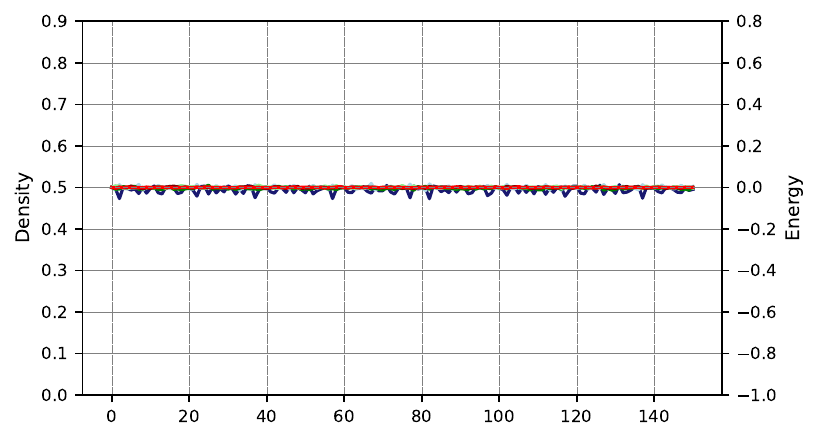}
\includegraphics[width=\textwidth]{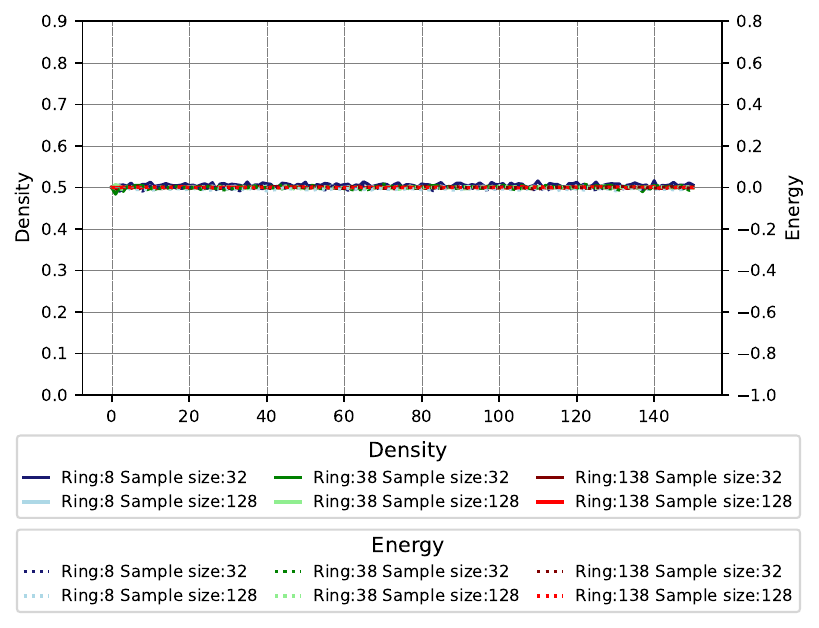}
\caption{Density (left y-axis) and normalized energy (right y-axis) for a ring of sizes $n=8$ (blue), $n=38$ (green) and $n=138$ (red) under rule 150 with sequential (top) and block-sequential (bottom) update mode with sample sizes of $s = 32$ and $s=128$ initial configurations, over 150 time steps.}
\label{fig:150 seq 38}
\end{figure}

\begin{figure}[ht]
\includegraphics[width=\textwidth]{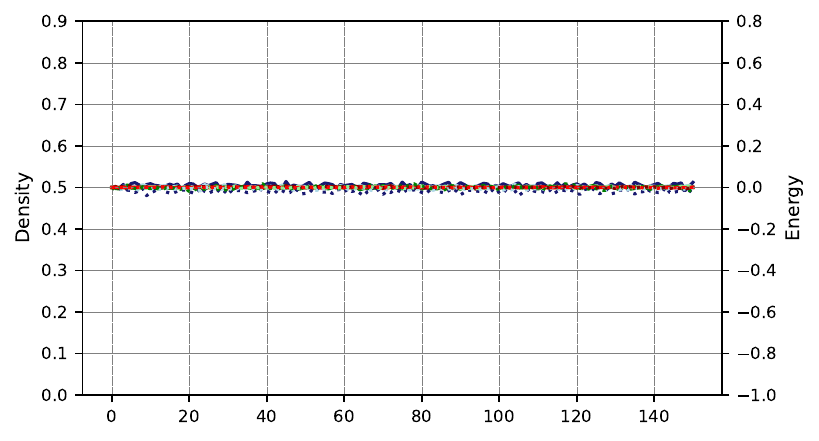}
\includegraphics[width=\textwidth]{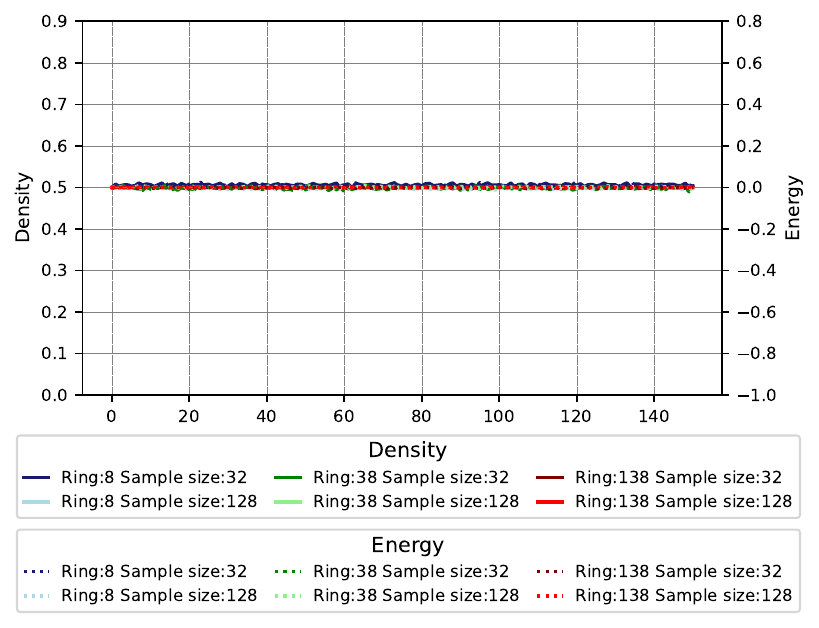}
\caption{Density (left y-axis) and normalized energy (right y-axis) for a ring of sizes $n=8$ (blue), $n=38$ (green) and $n=138$ (red) under rule 150 with block-parallel (top) and local clocks (bottom) update mode with sample sizes of $s = 32$ and $s=128$ initial configurations, over 150 time steps.}
\label{fig:150 lc 38}
\end{figure}

\FloatBarrier
\subsubsection{All configurations}
In this section we will show the results for the average density and energy for all possible configurations of a ring of size $n=16$. 
After the experiments from the previous section, we decided to use a ring of size $n=16$, because it was the order of magnitude in-between, and the experiments were able to be completed for all initial configuration in a reasonable amount of time.

We calculated the dynamics for $m=32$ block-sequential update modes with 3, 4 and 5 blocks; $m=32$ local clocks update modes with periods 2, 4 and 5. As well as $m=32$ sequential and $m=32$ block-parallel update modes.

As expected, block-parallel and local clocks showed the biggest variation for all rules, which agrees with our hypothesis that $\bp$ and $\lc$ are the update modes that induce the greatest differences within the rules. This result also supports our hypothesis that $\bp$ and $\lc$ rank higher in complexity than $\seq$ and $\bs$.

\begin{figure}[ht]
\includegraphics[width=\textwidth]{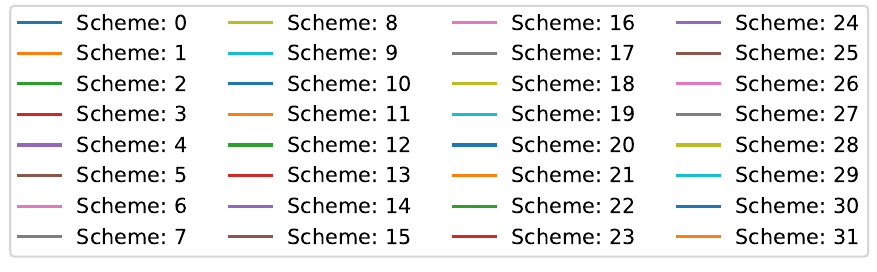}
\caption{Legend for the plots of this section.}
\label{fig:legends}
\end{figure}

Note that the legend shown in Fig~\ref{fig:legends} has the corresponding numeration each of the 32 update modes from each family, details of which can be found in the supplementary material.

\subsubsection*{Rule 110}
\begin{figure}[ht]
\begin{minipage}{.48\textwidth}
\includegraphics{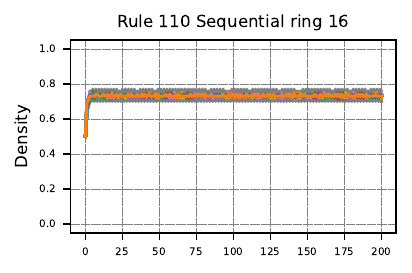}
\includegraphics{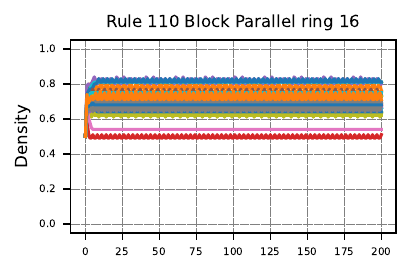}
\end{minipage}
\begin{minipage}{.48\textwidth}
\includegraphics{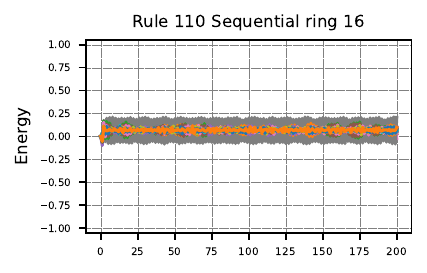}
\includegraphics{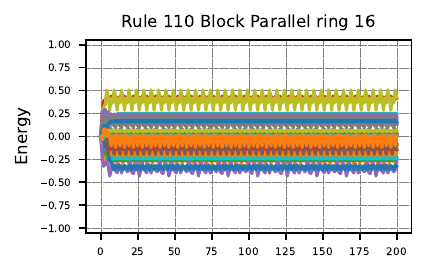}
\end{minipage}
\caption{Density (left) and normalized energy (right) for a ring of size 16 under rule 110 average over all configurations, with different sequential (top) and block-parallel (bottom) update modes, over 200 time steps.}
\label{fig:110 seq 16 all}
\end{figure}

\begin{figure}[ht]
\begin{minipage}{.48\textwidth}
\includegraphics{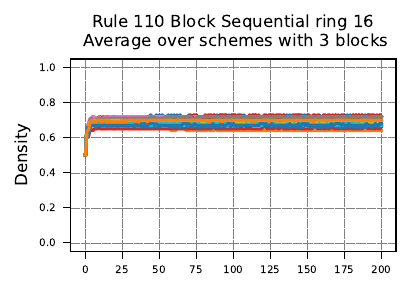}
\includegraphics{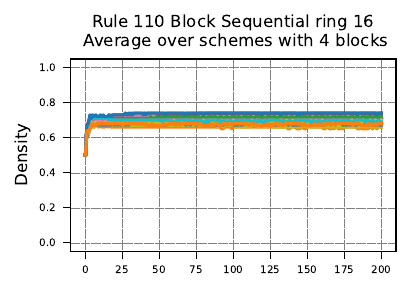}
\includegraphics{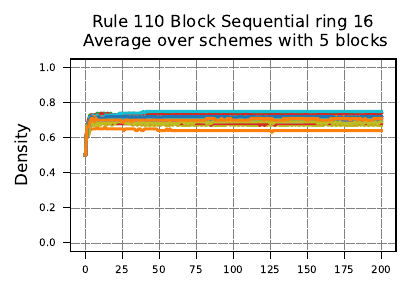}
\end{minipage}
\begin{minipage}{.48\textwidth}
\includegraphics{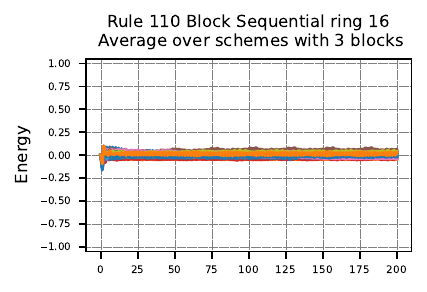}
\includegraphics{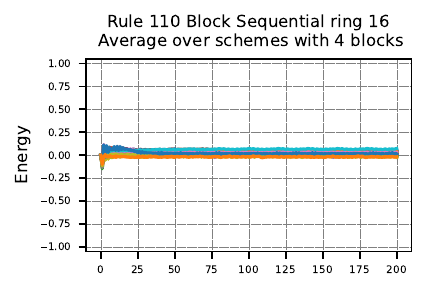}
\includegraphics{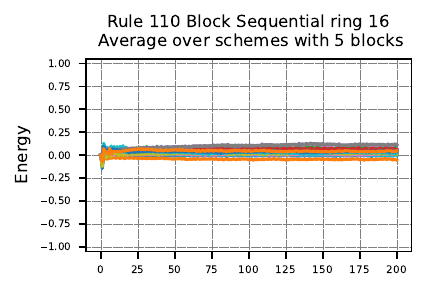}
\end{minipage}
\caption{Density (left) and normalized energy (right) for $n=16$ under $(110,\bs)$ with 3 (top), 4 (middle) and 5 (bottom) blocks, over 200 time steps.}
\label{fig:110 bs 16 all}
\end{figure}

Rule $110$ shows a wide variety of behaviors depending on the update modes. First of all, note that for $(110,\seq)$ (Fig.~\ref{fig:110 seq 16 all}) we see all update modes clustered tightly together both for density and energy. Similarly, for $(110,\bs)$ (Fig.~\ref{fig:110 bs 16 all}), the density stays close together for the different update modes belonging to this category regardless of the different number of blocks. Note that the energy can decrease before returning to hovering around values close to zero. Practically speaking, these two measures behave surprisingly alike when comparing different $\bs$ or $\seq$ update modes, which leads us to conclude that they have little influence over the dynamic for this rule.

\begin{figure}[ht]
\begin{minipage}{.48\textwidth}
\includegraphics{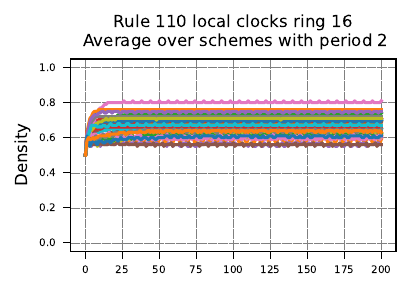}
\includegraphics{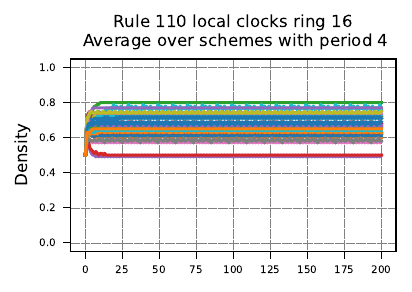}
\includegraphics{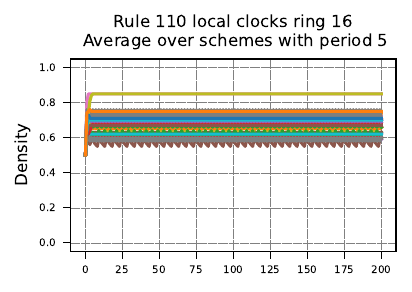}
\end{minipage}
\begin{minipage}{.48\textwidth}
\includegraphics{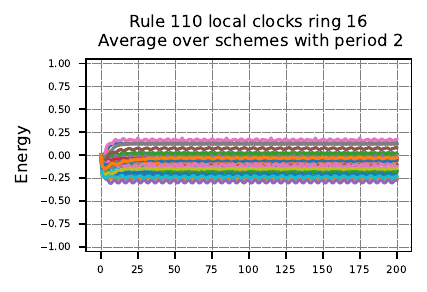}
\includegraphics{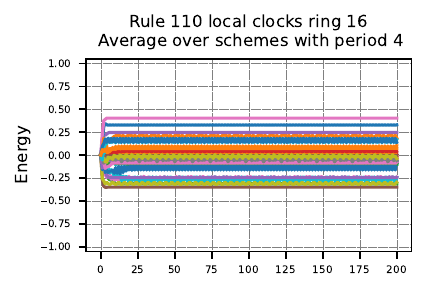}
\includegraphics{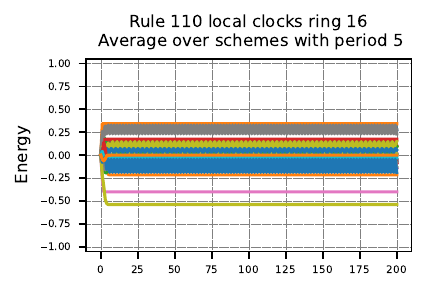}
\end{minipage}
\caption{Density (left) and normalized energy (right) for $n=16$ under $(110,\lc)$ with period 2 (top), 4 (middle) and 5 (bottom), over 200 time steps.}
\label{fig:110 lc 16 all}
\end{figure}

This is very different from what we can observe for $(110,\bp)$ and $(110,\lc)$ (Fig.~\ref{fig:110 lc 16 all} top and bottom, resp.), where the values are much more spread out. We can see that the value of the density does not decrease from $.5$, but for the energy it can lean towards the positive as much as towards the negative. The range of values around which both density and energy is much wider and depends directly on which member of the $\bp$ or $\lc$ family of update modes is being used. Moreover, note that said range can increase with the value of the period assigned to the $\lc$ update mode, especially for the energy.

\begin{figure}[ht]
\begin{minipage}{.48\textwidth}
\includegraphics[width=\textwidth]{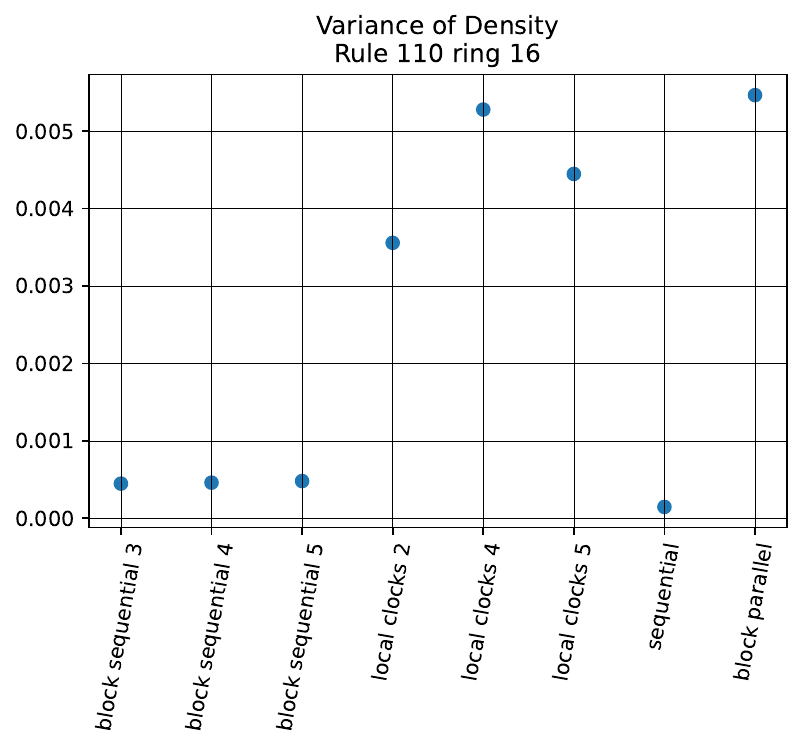}
\end{minipage}
\begin{minipage}{.48\textwidth}
\includegraphics[width=\textwidth]{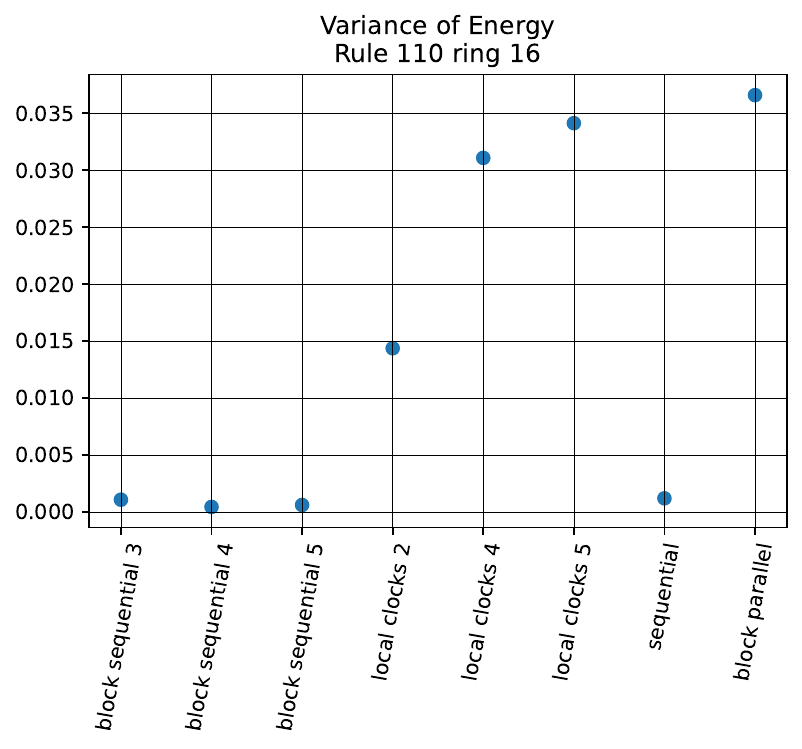}
\end{minipage}
\caption{Variance of Density (left) and normalized energy (right) for $n=16$ for rule $110$. }
\label{fig:var 110 16 all}
\end{figure}

Note that in all cases, the values of density and energy ``stabilize'' very quickly, which is supported by the values of variance shown on Figures \ref{fig:var 110 16 all}. Indeed, said Figure additionally shows that there is a slight increase of variance between update modes of the same class for local clocks and block-parallel when compared to sequential and block-sequential.


\FloatBarrier
\subsubsection*{Rule 54}
Rule $(54,\seq)$ shows a completely stable behavior, identical to $(150,\seq)$, but for update modes under all other categories its behavior is more similar to ones seen under rule $110$, as we can see on Fig.~\ref{fig:54 seq 16 all} using $(54,\bp)$ as an example. Note that it has a much more wider range of values around which both density and energy stabilize compared to rule $(110,\lc)$.

\FloatBarrier
\subsubsection*{All configurations of ring size 16}
\begin{figure}[ht]
\begin{minipage}{.48\textwidth}
\includegraphics{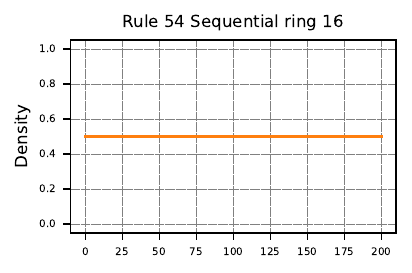}
\includegraphics{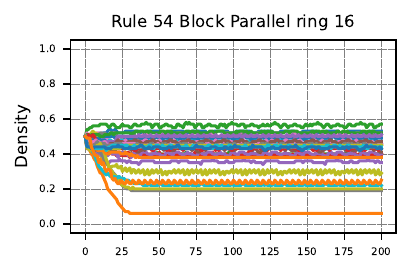}
\end{minipage}
\begin{minipage}{.48\textwidth}
\includegraphics{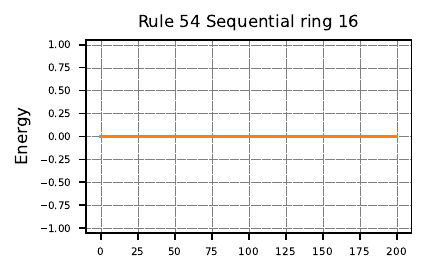}
\includegraphics{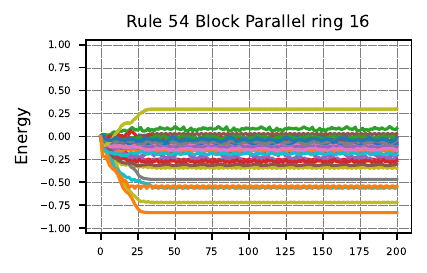}
\end{minipage}
\caption{Density (left) and normalized energy (right) for a ring of size 16 under rule $(54,\seq)$ (top) and $(54,\bp)$ (bottom) average over all configurations, with different update modes, over 200 time steps.}
\label{fig:54 seq 16 all}
\end{figure}

\begin{figure}[ht]
\begin{minipage}{.48\textwidth}
\includegraphics{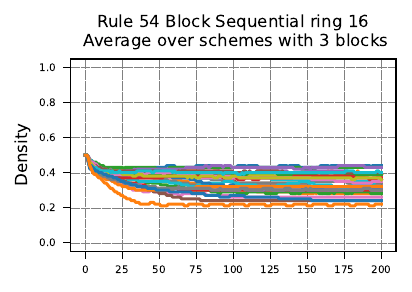}
\includegraphics{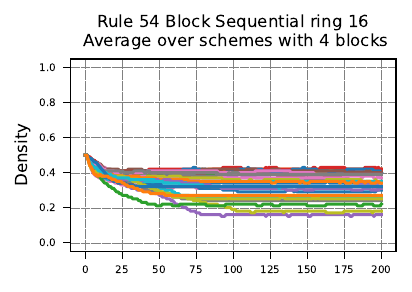}
\includegraphics{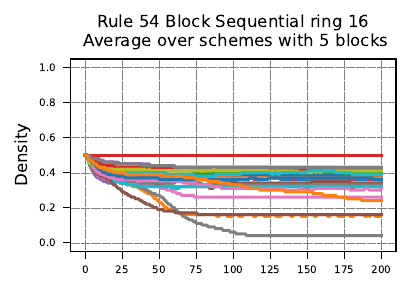}
\end{minipage}
\begin{minipage}{.48\textwidth}
\includegraphics{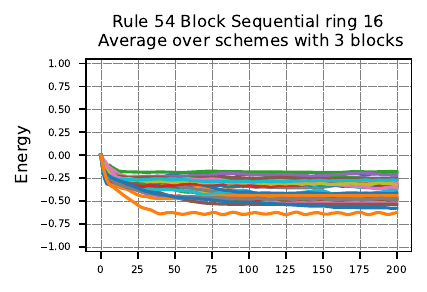}
\includegraphics{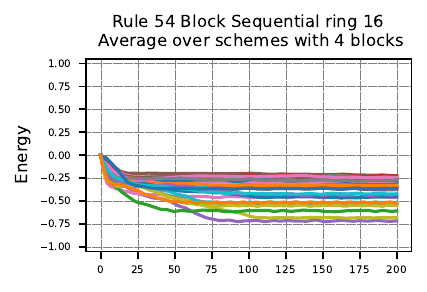}
\includegraphics{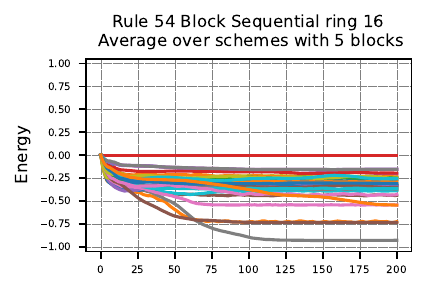}
\end{minipage}
\caption{Average over all configurations of Density (left) and normalized energy (right) for $n=16$ under $(54,\bs)$  with 3 (top), 4 (middle) and 5 (bottom) blocks, over 200 time steps.}
\label{fig:54 bs 16 all}
\end{figure}
\begin{figure}[ht]
\begin{minipage}{.48\textwidth}
\includegraphics{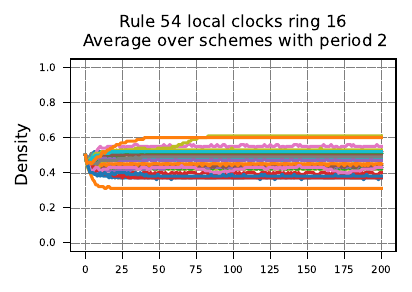}
\includegraphics{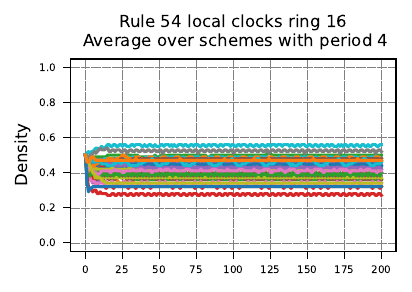}
\includegraphics{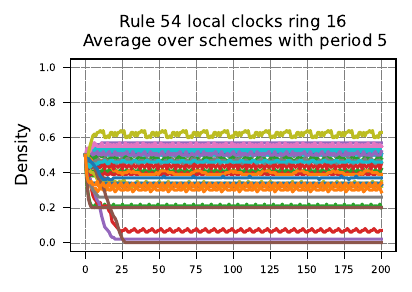}
\end{minipage}
\begin{minipage}{.48\textwidth}
\includegraphics{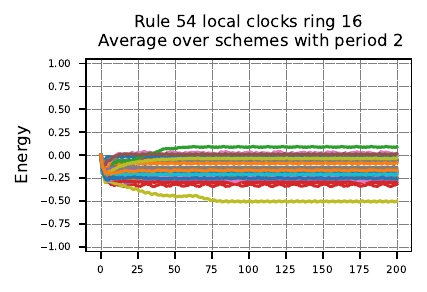}
\includegraphics{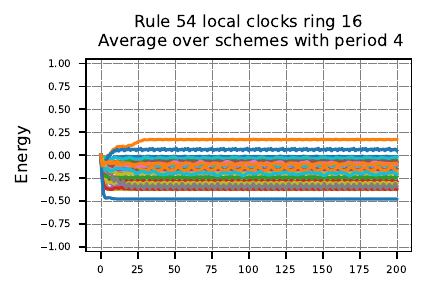}
\includegraphics{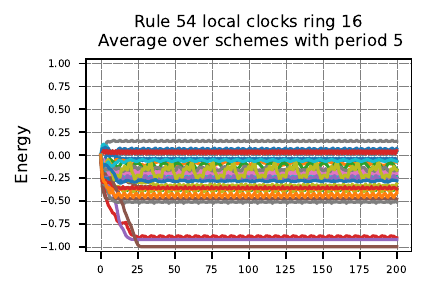}
\end{minipage}
\caption{Density (left) and normalized energy (right) for $n=16$ under $(54,\lc)$ with period 2 (top), 4 (middle) and 5 (bottom), over 200 time steps.}
\label{fig:54 lc 16 all}
\end{figure}

\begin{figure}[ht]
\begin{minipage}{.48\textwidth}
\includegraphics[width=\textwidth]{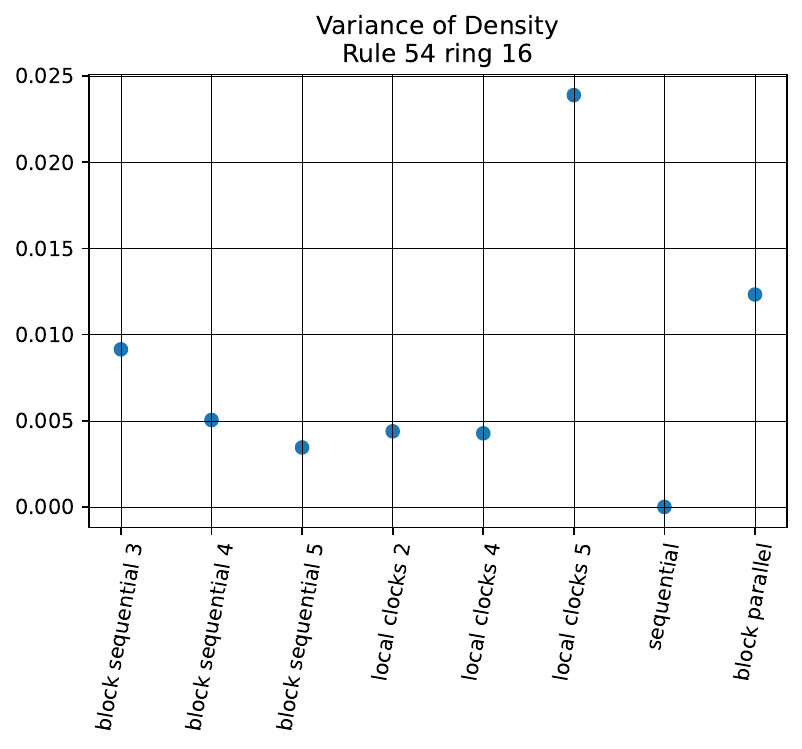}
\end{minipage}
\begin{minipage}{.48\textwidth}
\includegraphics[width=\textwidth]{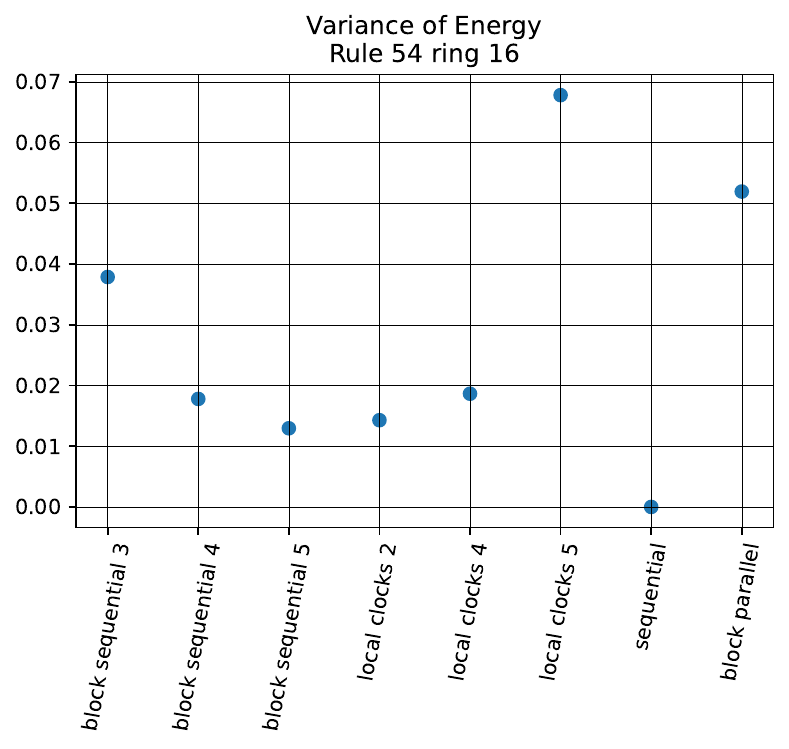}
\end{minipage}
\caption{Variance of Density (left) and normalized energy (right) for $n=16$ for rule $54$. }
\label{fig:var 54 16 all}
\end{figure}

As mentioned in the Results, Figs. \ref{fig:54 seq 16 all} (bottom), \ref{fig:54 bs 16 all}  and \ref{fig:54 lc 16 all} show that indeed these rule $54$ under these update modes behaves similarly to how rule $110$ did. Furthermore, note that local clocks with maximum period 5 and block-parallel have the highest variance when comparing different update modes of that class, as seen of Fig \ref{fig:var 54 16 all}

\FloatBarrier
\subsubsection*{Rules 150 and 90}
\begin{figure}[ht]
\begin{minipage}{.48\textwidth}
\includegraphics{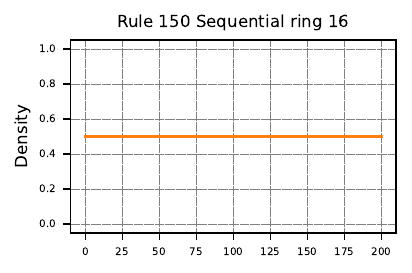}
\includegraphics{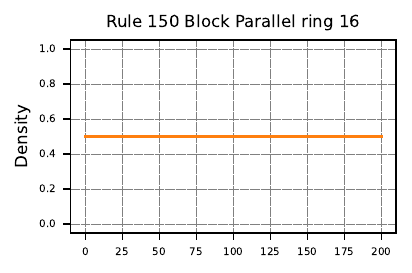}
\end{minipage}
\begin{minipage}{.48\textwidth}
\includegraphics{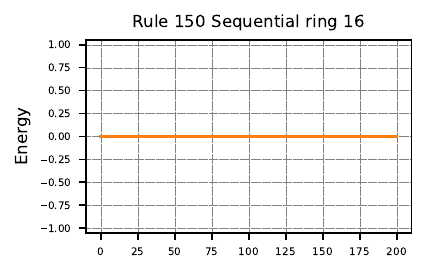}
\includegraphics{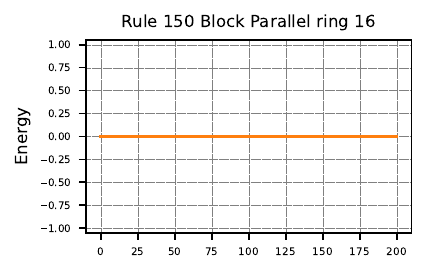}
\end{minipage}
\caption{Density (left) and normalized energy (right) for a ring of size 16 under rule $(150,\seq)$ (top) and $(150,\bp)$ (bottom) average over all configurations, with different update modes, over 200 time steps.}
\label{fig:150 seq 16 all}
\end{figure}

\begin{figure}[ht]
\begin{minipage}{.48\textwidth}
\includegraphics{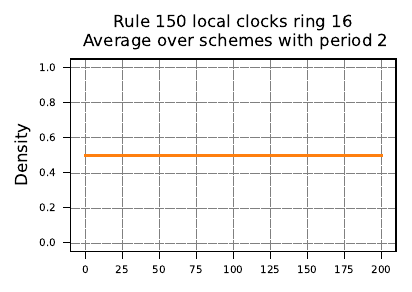}
\includegraphics{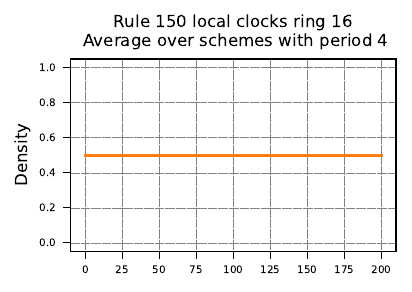}
\includegraphics{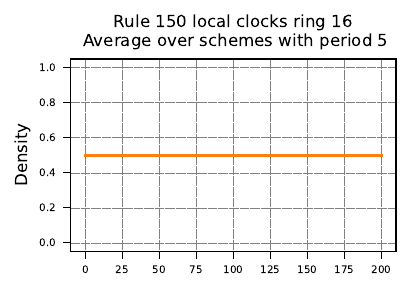}
\end{minipage}
\begin{minipage}{.48\textwidth}
\includegraphics{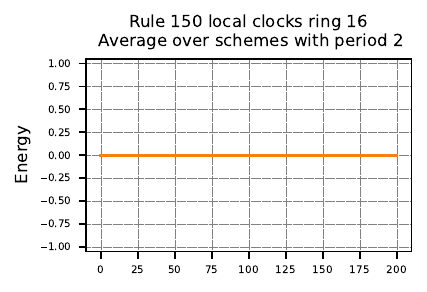}
\includegraphics{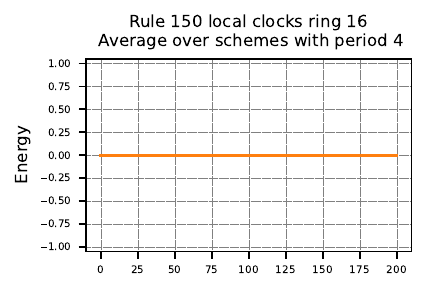}
\includegraphics{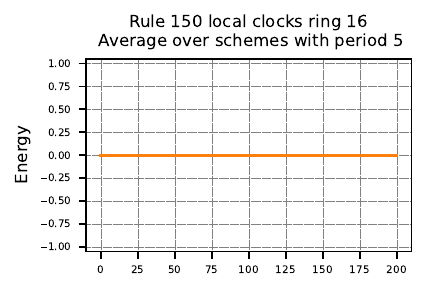}
\end{minipage}
\caption{Density (left) and normalized energy (right) for $n=16$ under $(150,\lc)$ with period 2 (top), 4 (middle) and 5 (bottom), over 200 time steps.}
\label{fig:150 lc 16 all}
\end{figure}

\begin{figure}[ht]
\begin{minipage}{.48\textwidth}
\includegraphics[width=\textwidth]{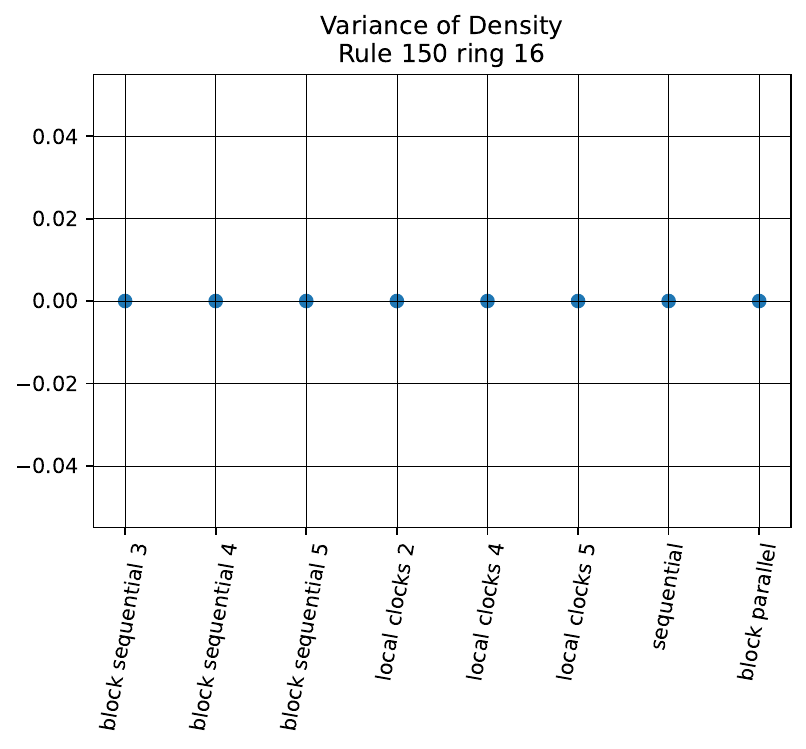}
\end{minipage}
\begin{minipage}{.48\textwidth}
\includegraphics[width=\textwidth]{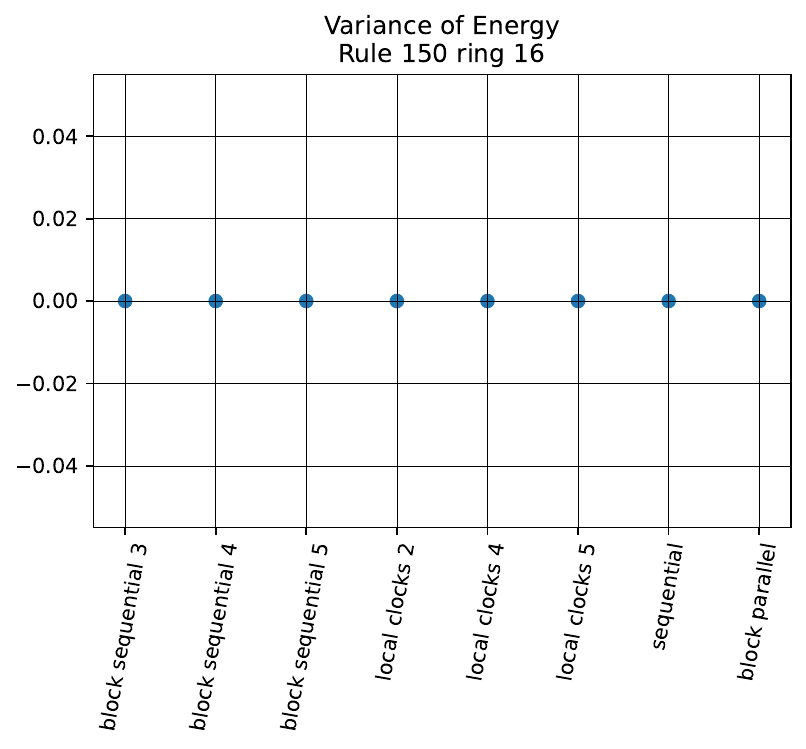}
\end{minipage}
\caption{Variance of Density (left) and normalized energy (right) for $n=16$ for rule $150$. }
\label{fig:var 150 16 all}
\end{figure}

Unlike rule $110$, rule $150$ is completely stable for all update modes, the average density and normalized energy change very little regardless of the update mode, which is shown on  Figs.~\ref{fig:150 seq 16 all} through to~\ref{fig:150 lc 16 all}. Additionally, Fig.~\ref{fig:var 150 16 all} presents the variance which shows that that the variance is indeed 0 for all update modes in each class.

\begin{figure}[ht]
\begin{minipage}{.48\textwidth}
\includegraphics{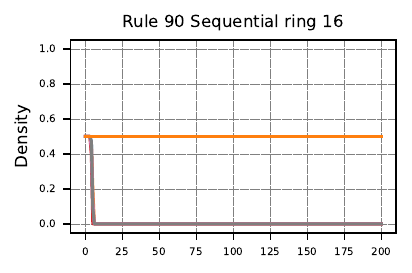}
\includegraphics{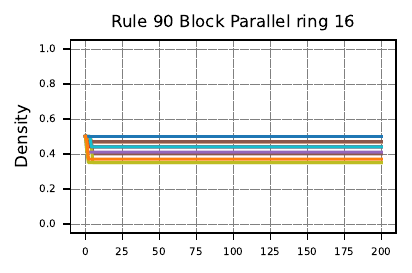}
\end{minipage}
\begin{minipage}{.48\textwidth}
\includegraphics{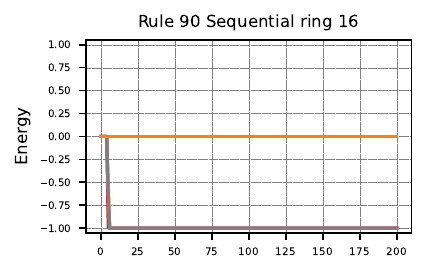}
\includegraphics{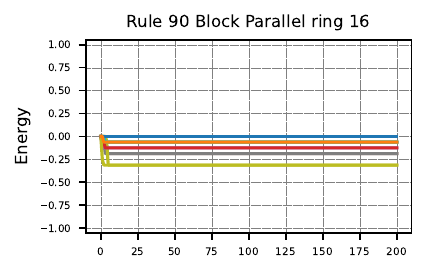}
\end{minipage}
\caption{Density (left) and normalized energy (right) for a ring of size 16 under rule $(90,\seq)$ (top) and $(90,\bp)$ (bottom) average over all configurations, with different update modes, over 200 time steps.}
\label{fig:90 seq 16 all}
\end{figure}

\begin{figure}[ht]
\begin{minipage}{.48\textwidth}
\includegraphics{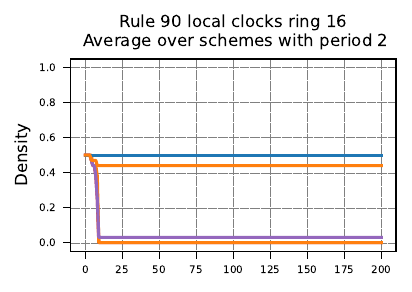}
\includegraphics{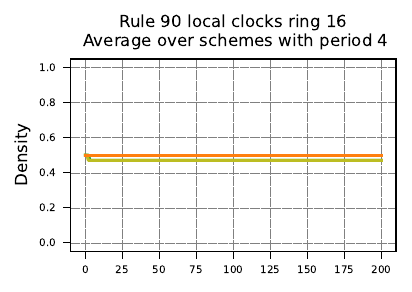}
\includegraphics{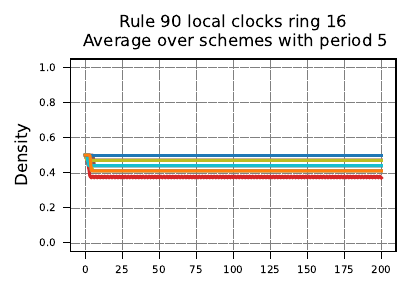}
\end{minipage}
\begin{minipage}{.48\textwidth}
\includegraphics{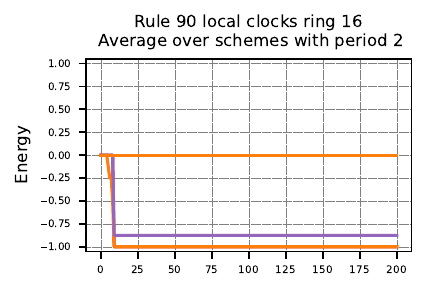}
\includegraphics{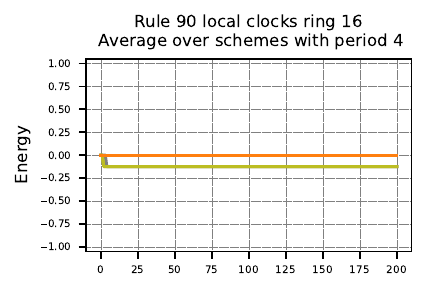}
\includegraphics{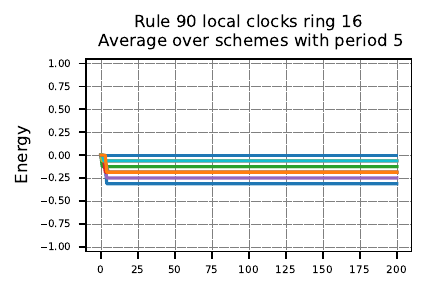}
\end{minipage}
\caption{Density (left) and normalized energy (right) for $n=16$ under $(90,\lc)$ with period 2 (top), 4 (middle) and 5 (bottom), over 200 time steps.}
\label{fig:90 lc 16 all}
\end{figure}

Similarly, rule $90$ (which also belongs to class III) produces the same type of plots; although it is they stabilize at different points than rule $150$. They are shown on Figs.~\ref{fig:90 seq 16 all} through to~\ref{fig:90 lc 16 all}. 

The fact that both rules that belong to class III have such stable average density and energy can be explained by the fact that there is no clear pattern that emerges from this rule. As a consequence we have a larger variety of possible configurations at each time step; thus the values of density and energy for rule 150 tend to be to 0.5 and 0, respectively.

\begin{figure}[ht]
\begin{minipage}{.48\textwidth}
\includegraphics{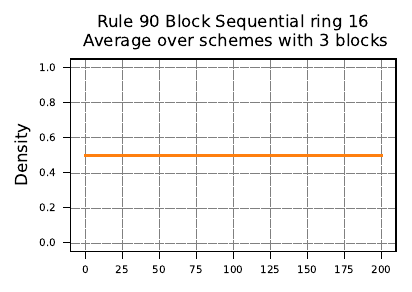}
\includegraphics{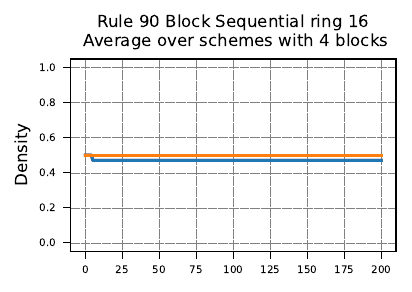}
\includegraphics{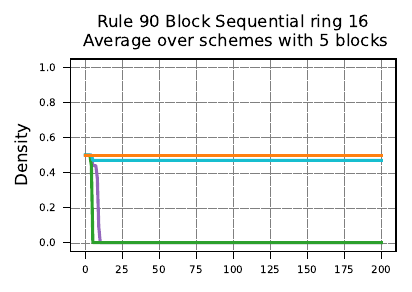}
\end{minipage}
\begin{minipage}{.48\textwidth}
\includegraphics{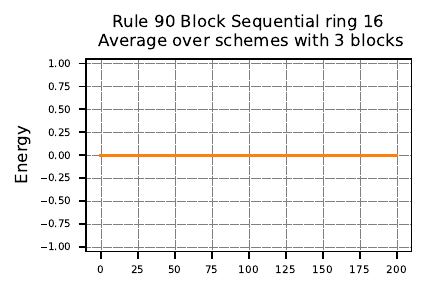}
\includegraphics{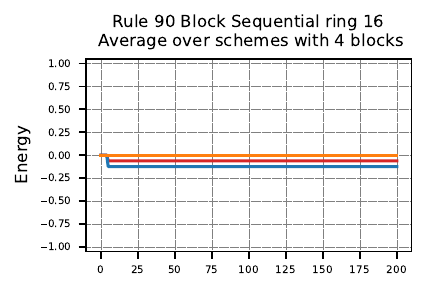}
\includegraphics{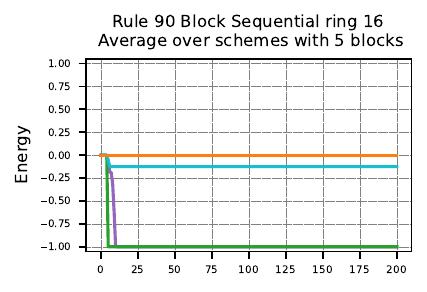}
\end{minipage}
\caption{Density (left) and normalized energy (right) for $n=16$ under $(90,\bs)$ with 3 (top), 4 (middle) and 5 (bottom) blocks, over 200 time steps.}
\label{fig:90 bs 16 all}
\end{figure}

\begin{figure}[ht]
\begin{minipage}{.48\textwidth}
\includegraphics[width=\textwidth]{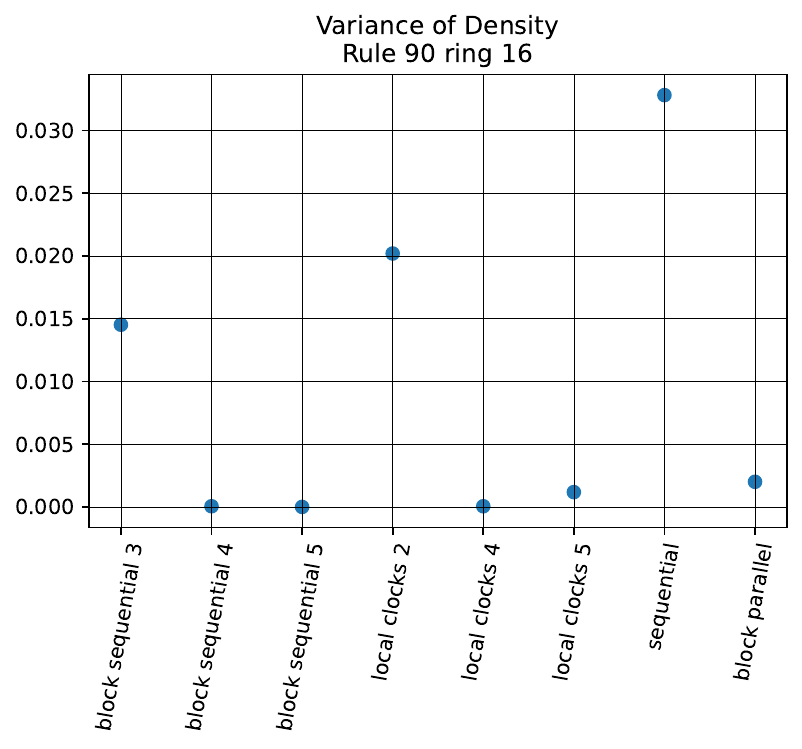}
\end{minipage}
\begin{minipage}{.48\textwidth}
\includegraphics[width=\textwidth]{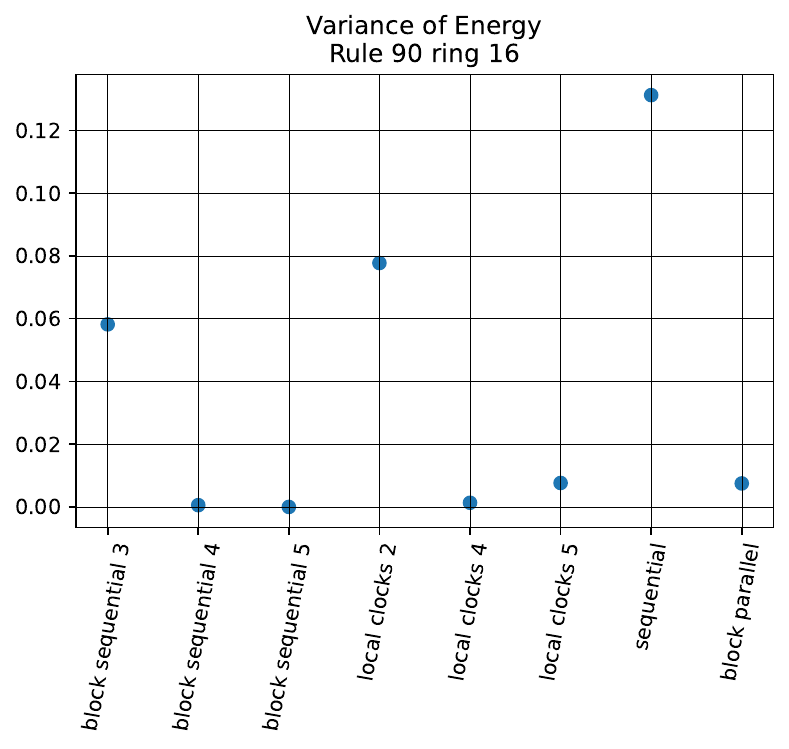}
\end{minipage}
\caption{Variance of Density (left) and normalized energy (right) for $n=16$ for rule $90$. }
\label{fig:var 90 16 all}
\end{figure}

\begin{figure}[ht]
\begin{minipage}{.48\textwidth}
\includegraphics{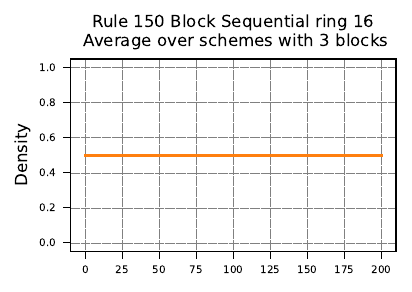}
\includegraphics{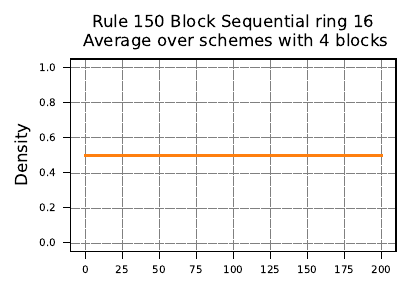}
\includegraphics{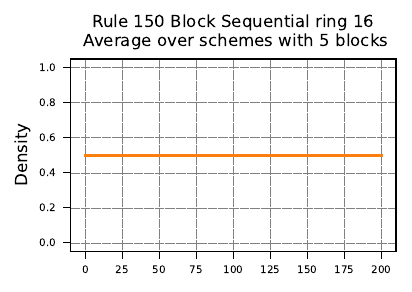}
\end{minipage}
\begin{minipage}{.48\textwidth}
\includegraphics{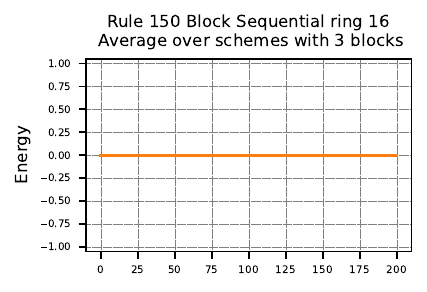}
\includegraphics{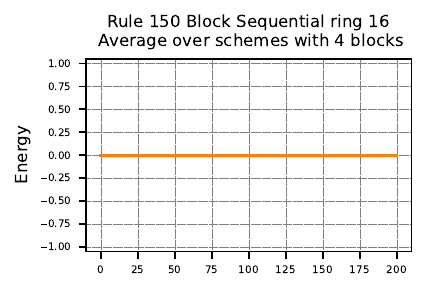}
\includegraphics{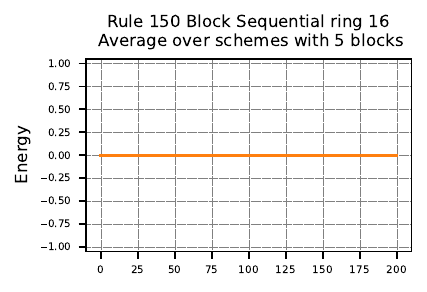}
\end{minipage}
\caption{Density (left) and normalized energy (right) for $n=16$ under $(150,\bs)$ with 3 (top), 4 (middle) and 5 (bottom) blocks, over 200 time steps.}
\label{fig:150 bs 16 all}
\end{figure}

The experiments suggested that for complex rules (class IV) the $\bp$ and $\lc$ update modes have a great influence over the behavior of density and energy, much more noticeable than for $\seq$ and $\bs$. Contrarily, the for the chaotic rules (class III) we saw that the update modes have practically no influence over said parameters, even the two families from which we would have expected it, that is, the $\bp$ and $\lc$ update modes.

\section{Discussion}
\label{sec:persp}

\subsection*{Theoretical Results}
In this paper, we have studied 67 of the 88 Elementary Cellular Automata (ECA) under different update modes, proving that there is a set of rules (15 that always reach fixed points, 10 that always reach cycles of constant length) whose longest limit cycles remain of the same length regardless of the update mode. This shows that some ECA are robust to (a)synchronism.

We have found three other behaviors, summarized in Table~\ref{table:increasing}.
Firstly, we have rules 108 and 156, which present a notorious change in complexity, going from constant (PAR) to superpolynomial (SEQ). Secondly, we have rules 1 and 178, where the increase of the maximum period is more gradual, from constant (PAR), to linear (SEQ) to superpolynomial (BS). And thirdly, we have rules 56, 152 and 184, where the length of the longest limit cycle decreases from linear (PAR) to constat (SEQ) but then returns to linear (BS) and increases once more to superpolynomial (BP)

While we have identified superpolynomial cycle lengths for certain rules under sequential and block-sequential updates modes, in fact, all rules capable of reaching such cycles do so under block-parallel and local clocks update modes. This observation reinforces the hypothesis that, if a hierarchy among periodic update modes exists, BP and LC would occupy a higher rank than the others studied in this paper. 

However, note that there are rules whose complexity remains constant even under BP and LC update modes. Some of these rules are: $12$, $44$, $72$, $76$, $78$,  $132$, $140$, $164$ and $200$, which always reach fixed point and $5$, $13$, $29$, $32$, $36$, $51$, $77$, and $232$ whose limit cycles are always of constant length (more examples can be found in subsection \ref{sec:constante}). Additionally, note that there are also rules whose cycle lengths remain directly proportional to the length of the ring regardless of the update mode, such as rules $15$, $35$, $170$, and most notably rule $18$ that belongs to class III of Wolfram's classification (Subsection \ref{sec:linear} contains more examples).

Finally, note that with the sequential update mode one is able to reach the greatest variety in complexity, depending on the rule.

\subsection*{Experimental Results}
In the experimental part of this work we studied what occurred with the measures of density and energy under different update methods for two chaotic (90 and 150) and two complex (54 and 110) ECA. We selected rules because they serve to represent classes III and IV, and they are very well known and as such will be familiar to most readers. Experimental results suggested that the two complex rules are subject to smaller changes under sequential and block-sequential update modes (Figs~\ref{fig:110 seq 16 all} and~\ref{fig:110 bs 16 all}), while under block-parallel and local clocks update modes the influence of each update mode was evident as seen in Figs~\ref{fig:110 seq 16 all} (bottom), \ref{fig:110 lc 16 all} and~\ref{fig:54 seq 16 all} (bottom). In opposition, the two chaotic rules appeared to be mostly indifferent to the various families of update modes, which is shown in the figures of the corresponding section.

We started by comparing the behavior of said measures with two sample size for three different ring sizes. The resulting plots showed minimal differences across different ring sizes, allowing us to conduct experiments on smaller rings while still testing all possible configurations. This way, the focus would be only on the update modes.

We have studied through measures of density and energy two chaotic rules (90 and 150, both belonging to Class III according to Wolfram's classification), which appear to have little to no change in the average of these measures when the update mode is modified. For rule 90 under most members of each family of update modes, these averages remained at exactly $0.5$ for density and $0$ for energy from start to finish. What is more, for rule 150 it was apparent: under every single one of the update modes that we experimented with the density and energy were always exactly $0.5$ and $0$ respectively.

Under the light of the experiments we can see that there are no favored patterns for rule 150 under any of the analyzed update modes. Which could be explained by the chaotic nature of the rule. 

On the other hand, we saw that the values of density and energy for rules 54 (except under sequential update mode) and 110 find different points of stability for each update mode, and that once those values are reached, the values of density and energy show little variation from one time step to another. Furthermore, local clocks and block-parallel present greater variety of values of stability, which could be attributed to the fact that those update modes allow cells to be updated more than once per step.

\subsection*{Future Work}
The results that we have obtained at this point do not allow us to differentiate between block-parallel and local clocks in terms of complexity. We suspect that local clocks should be at the highest rung, because it is a more general update mode (it includes block-parallel and block-sequential) which leads us to hypothesize that it should allow more freedom and therefore generate a greater complexity according to our definition. Thus, we would like to establish a hierarchy between them eventually.

Naturally, it is of interest to continue the theoretical analysis for rules $25$, $37$, $57$, $58$, $62$, $74$, $154$ (class II), and research what properties (if any) do these rules have that made them incompatible with the strategies developed for other rules in the same class. Conversely, what properties does rule 18 have that allowed us to use methods that did not work for the rest of the rules in class III? Additionally, can we prove the existence of limit cycles of exponential length for some combination of rule and update mode?

Another question is about if we can expand the research about sensitivity under different update modes to one-dimensional cellular automata with different radii, whether some of the approaches found for radius 1 can be of use in those cases. 

We need further experiments to check the behavior of the rest of the rules in classes III and IV, and whether their behavior resembles what we have found for the rules we have presented as examples in this paper. Similarly, for the ones that we were unable to classify belonging to class II: can we expect to find patterns in rules belonging to class II that we did not find for the other two classes? 

For the most part, we know that rule 90 has similar types of graph to rule 150. However, it is worth studying the specific update modes whose density and energy become stable around  values different from $0.5$ and $0$, respectively.

Another possible research line could be the study of higher-dimensional cellular automata under different periodic update schemes. Indeed, an interesting case study could involve applying the methodology outlined in this work (a combined theoretical and applied approach) to totalistic CA rules in two dimensions. On the other hand, one could remain within the realm of ECA and explore other types of update modes that are not periodic. For example in~\cite{goles2020effects} the effects of a memory based update schedule on the dynamics of conjuctive networks were studied. Additionally, in~\cite{gadouleau2024bringing} the authors proposed a framework for update modes based on memory. In this context, it would be interesting to study the effects of these update modes on ECA rules and compare them with the results obtained under periodic update modes.

\paragraph{Acknowledgements}
This work has been partially funded by the HORIZON-MSCA-2022-SE-01 pro\-ject 101131549 ``ACANCOS'' project (IDL, EG, MRW, SS), the ANR-24-CE48-7504 ``ALARICE'' project (SS), the  STIC AmSud 22-STIC-02 ``CAMA'' project (IDL, EG, MRW, SS), ANID FONDECYT Postdoctorado 3220205 (MRW), ANID FONDECYT 1250984 regular (EG) and ANID-MILENIO-NCN2024\_103 (EG).
\bibliographystyle{elsarticle-num}

%\FloatBarrier
\appendix

\subsection{Definition of the Update Modes}
In this section we will show all the definitions for all the update modes used in the experimental section.

\begin{table}
\resizebox{\textwidth}{!}{\centering
\begin{tabular}{|c|c|c|c|}
\hline
\# &Sequential&Block Sequential&Block Parallel\\\hline
0&$(2,3,4,5,7,6,1,0)$&$(\{1\},\{2,7\},\{6,4\},\{5,0,3\})$
&$\{(1,2),(5,3),(7),(6),(4,0)\}$\\\hline
1&$(6,0,2,7,4,5,3,1)$&$(\{2, 5, 4, 3\},\{1\},\{6, 0, 7\})$
&$\{(4, 3), (2), (6, 1, 5, 0, 7)\}$\\\hline
2&$(6,0,7,3,5,4,1,2)$&$(\{6, 5, 1\},\{2\},\{7, 4, 3, 0\})$
&$\{(1, 6), (0, 5, 4), (2, 3, 7)\}$\\\hline
3&$(7,6,5,2,1,3,4,0)$&$(\{6, 7, 2, 0\},\{4, 3\},\{1, 5\})$
&$\{(7, 6, 5, 4, 1, 3), (0), (2)\}$\\\hline
4&$(5,3,2,1,6,7,0,4)$&$(\{3, 1\},\{4, 5, 6, 0\},\{7, 2\})$
&$\{(1), (0, 6, 3), (4, 2, 7, 5)\}$\\\hline
5&$(6,7,0,5,1,2,4,3)$&$(\{7, 2, 6\},\{5, 1, 3, 4, 0\})$
&$\{(7, 6), (4, 3, 5, 1), (2, 0)\}$\\\hline
6&$(5,0,2,6,7,1,3,4)$&$(\{7, 3, 6\},\{1, 2, 5, 4\},\{0\})$
&$\{(2, 0, 3, 6), (1, 7, 5), (4)\}$\\
\hline
\end{tabular}}
\caption{Sequential, Block Sequential and Block Parallel update modes for $n=8$.}
\label{table:seq_n8}
\end{table}

\begin{table}\centering
\begin{tabular}{|c|c|c|}
\hline
\multirow{2}{*}{\#} &\multicolumn{2}{c|}{Definition}\\\cline{2-3}
&$P$&$\Delta$\\\hline
0&$(2, 3, 3, 1, 1, 2, 3, 3)$& $(1, 0, 2, 0, 0, 0, 1, 0)$\\\hline
1&$(3, 2, 3, 2, 5, 2, 2, 2)$& $(2, 0, 0, 1, 3, 1, 0, 1)$\\\hline
2&$(2, 3, 2, 5, 3, 3, 3, 4)$& $(0, 1, 1, 1, 1, 1, 2, 1)$\\\hline
3&$(5, 3, 5, 2, 4, 2, 2, 4)$& $(4, 0, 1, 1, 0, 1, 0, 0)$\\\hline
4&$(4, 3, 3, 2, 5, 2, 2, 2)$& $(2, 1, 1, 0, 3, 1, 1, 0)$\\\hline
5&$(5, 5, 2, 4, 2, 1, 4, 1)$& $(2, 1, 0, 2, 0, 0, 1, 0)$\\
\hline
\end{tabular}
\caption{Local Clocks update modes for $n=8$.}
\label{table:lc_n8}
\end{table}
\FloatBarrier

\begin{table}\centering
\begin{tabular}{|c|c|}
\hline
\# &Definition\\\hline
0&\makecell{
$(30,20,4,14,33,15,34,19,32,29,22,12,6,7,0,35,28,37,27,$\\
$23,36,3,25,26,24,2,9,16,10,11,17,18,21,8,13,5,31,1)$}\\\hline
1&\makecell{
$(33,10,1,0,19,32,13,2,14,18,35,30,15,7,6,8,9,5,21,23,27,$\\
$26,12,34,22,4,20,36,37,17,28,24,29,25,3,11,31,16)$}\\\hline
2&\makecell{
$(30,3,16,27,8,31,28,7,35,33,18,26,11,32,36,1,34,0,5,6,24,$\\
$15,4,20,12,17,14,23,37,21,22,2,29,9,10,19,13,25)$}\\\hline
3&\makecell{
$(36,5,12,4,0,25,27,13,7,21,16,15,35,30,17,6,32,28,2,8,23,$\\
$9,3,34,10,19,29,31,22,24,1,14,33,26,20,18,11,37)$}\\\hline
4&\makecell{
$(24,0,2,16,18,27,28,31,26,20,8,5,13,19,36,1,12,29,6,35,10,$\\
$3,15,21,9,37,32,25,23,22,11,34,7,14,33,4,17,30)$}\\\hline
5&\makecell{
$(9,37,33,20,12,34,36,11,25,24,13,23,32,31,27,19,22,8,5,0,$\\
$21,35,15,4,14,6,18,2,28,10,7,1,17,26,16,29,3,30)$}\\\hline
6&\makecell{
$(35,0,34,19,7,1,2,30,5,12,9,6,23,27,22,24,33,21,20,3,10,32,$\\
$11,25,15,13,17,37,28,36,4,29,14,26,16,31,8,18)$}\\
\hline
\end{tabular}
\caption{Sequential update modes for $n=38$.}
\label{table:seq1_n38}
\end{table}

\begin{table}\centering
\begin{tabular}{|c|c|}
\hline
\# &Definition\\\hline
16&\makecell{
$(5,22,24,10,2,37,4,1,13,29,14,30,27,7,32,16,18,8,25,0,35,$\\$
28,23,3,31,36,26,11,15,9,20,34,12,33,6,21,17,19)$}\\\hline
17&\makecell{
$(7,27,2,17,19,34,36,0,33,22,1,26,12,9,6,28,14,13,21,31,35,$\\$
11,16,20,15,4,3,5,10,29,30,18,23,32,24,37,8,25)$}\\\hline
18&\makecell{
$(9,20,7,11,31,13,28,32,2,16,18,14,21,25,3,34,30,36,29,1,$\\$
33,26,10,35,27,0,6,23,22,5,17,12,8,24,15,37,19,4)$}\\\hline
19&\makecell{
$(27,8,31,12,1,30,24,23,32,19,36,11,5,26,17,37,6,16,0,2,35,$\\$
34,33,21,29,9,10,13,4,22,28,20,7,15,25,18,14,3)$}\\\hline
20&\makecell{
$(22,29,34,6,20,12,8,14,3,31,2,36,4,26,13,28,35,0,27,21,25,$\\$
5,18,7,37,24,33,19,1,16,15,9,30,23,17,32,11,10)$}\\\hline
21&\makecell{
$(6,30,2,7,25,32,0,20,14,12,23,24,27,26,35,28,34,15,22,29,$\\$
18,1,16,33,21,10,11,9,13,17,5,31,36,37,19,4,3,8)$}\\\hline
22&\makecell{
$(20,12,5,24,2,23,28,1,14,21,13,3,35,11,31,25,32,7,27,8,26,$\\$
37,33,10,9,22,18,0,19,4,30,36,16,34,6,17,29,15)$}\\
\hline
\end{tabular}
\caption{Sequential update modes for $n=38$.}
\label{table:seq2_n38}
\end{table}

\begin{table}
\resizebox{\textwidth}{!}{\centering
\begin{tabular}{|c|c|}
\hline
\# &Definition\\\hline
0&\makecell{$(\{5, 14, 9, 6, 16, 31\},\{3, 18, 8\},\{36, 17, 25, 24, 12, 20, 22, 7, 4, 23, 32, 1\},$\\$\{34, 27, 10, 15, 11\},\{30, 26, 33, 37, 2, 29, 19, 35, 28, 13, 0, 21\})$}\\\hline
1&\makecell{$(\{33, 30, 17, 14, 29, 34, 24, 32, 11, 28\},\{2, 7, 20, 6, 22, 19, 37, 13, 10, 4\},$\\$\{8, 31, 9, 12, 26, 27, 25, 1, 23, 5, 0, 16, 18, 21, 36, 3, 35, 15\})$}\\\hline
2&\makecell{$(\{14, 1, 12, 21, 5, 32, 7, 24, 22, 34\},\{13, 36, 11, 33, 0, 10, 35, 26, 16, 37, 19, 17, 4\},$\\$\{30, 28, 8, 20, 18, 23, 15, 29, 9, 27, 6, 3, 25, 31, 2\})$}\\\hline
3&\makecell{$(\{25, 30, 7, 5, 19, 4, 9, 15, 20, 22, 32, 29, 34, 28, 18, 33\},$\\$\{36, 35, 6, 24, 13, 10, 31, 23, 1, 11, 16\},\{21, 2, 26, 14, 17, 0, 12, 8, 37, 27, 3\})$}\\\hline
4&\makecell{$(\{3, 9, 32, 16, 13, 29, 22, 21, 35, 30, 6, 1\},$\\$\{37, 28, 14, 25, 19, 23, 11, 18, 26, 10, 15, 17, 5\},$\\$\{7, 24, 33, 27, 31, 2, 36, 4, 20, 34, 0, 12, 8\})$}\\\hline
5&\makecell{$(\{3, 34, 32, 25, 17, 29, 9, 4, 1, 33, 30, 26, 21, 22\},\{27, 12, 23, 37, 36, 31, 20, 8, 11\},$\\$\{24, 10, 0, 2, 7, 6, 15, 28, 14, 16, 18, 5, 13, 19, 35\})$}\\\hline
6&\makecell{$(\{37, 1, 0, 3, 26, 19, 28, 9, 27, 23, 15\},\{16, 20, 10, 6, 36, 14, 22, 30, 11, 7, 29, 4, 34\},$\\$\{33, 8, 2, 35, 12, 5, 21, 25, 18, 13, 17, 31, 24, 32\})$}\\
\hline
\end{tabular}}
\caption{Block Sequential update modes for $n=38$.}
\label{table:bs1_n38}
\end{table}

\begin{table}
\resizebox{\textwidth}{!}{\centering
\begin{tabular}{|c|c|}
\hline
\# &Definition\\\hline
16&\makecell{$(\{4, 30, 7, 35, 36, 27, 29, 28, 22, 21, 18, 26, 14\},$\\$\{8, 17, 13, 25, 20, 1, 32, 34, 6, 0, 24, 37, 23\},$\\$\{12, 31, 3, 33, 5, 9, 10, 15, 19, 16, 2, 11\})$}\\\hline
17&\makecell{$(\{15, 10, 33, 21, 9, 20, 0, 18, 27, 19, 26\},\{16, 1, 5, 13, 24, 25, 7, 28, 30, 12, 23, 22, 6\},$\\$\{35, 17, 14, 31, 3, 34, 36, 11, 2, 37, 8, 4, 32, 29\})$}\\\hline
18&\makecell{$(\{26, 7, 34, 29, 2, 15, 22, 35, 28, 14, 10, 13, 21\},$\\$\{24, 31, 18, 11, 37, 20, 32, 8, 6, 30, 23, 0, 36\},$\\$\{5, 4, 25, 1, 27, 33, 19, 17, 16, 3, 9, 12\})$}\\\hline
19&\makecell{$(\{17, 34, 13, 36, 19, 12, 11, 28, 21, 37, 2, 8, 30\},$\\$\{25, 20, 5, 24, 31, 1, 14, 33, 35, 4, 27, 29\},$\\$\{0, 32, 9, 3, 10, 22, 18, 7, 23, 26, 6, 16, 15\})$}\\\hline
20&\makecell{$(\{10, 6, 4, 20, 16, 15, 13, 5, 2\},\{17, 33, 18, 26, 22, 36, 8, 11, 3, 14, 29, 34, 0, 37\},$\\$\{31, 35, 21, 9, 1, 30, 25, 32, 12, 19, 24, 7, 23, 28, 27\})$}\\\hline
21&\makecell{$(\{35, 5, 23, 30, 4, 26, 7, 6, 36, 25, 11, 33, 24\},\{16, 22, 0, 32, 8, 28, 37, 2, 3, 21, 12\},$\\$\{13, 10, 1, 17, 34, 27, 20, 15, 31, 29, 19, 14, 18, 9\})$}\\\hline
22&\makecell{$(\{6, 3, 17, 26, 37, 35, 8, 18, 30, 5, 16, 2, 29, 36, 19, 24\},$\\$\{27, 7, 25, 14, 9, 32, 15, 12, 31, 22, 1\},$\\$\{21, 0, 13, 11, 20, 33, 34, 23, 10, 28, 4\})$}\\
\hline
\end{tabular}}
\caption{Block Sequential update modes for $n=38$.}
\label{table:bs2_n38}
\end{table}

\begin{table}
\resizebox{\textwidth}{!}{\centering
\begin{tabular}{|c|c|}
\hline
\# &Definition\\\hline
0&\makecell{$\{(30, 19, 33, 15, 31, 8, 32), (16, 25, 3, 14, 4), (21, 10, 20, 9, 34, 29, 12, 13), $\\$(24, 6, 5, 7, 18, 1, 22, 27, 36, 23), (28, 17, 2, 26, 11, 0, 35, 37)\}$}\\\hline
1&\makecell{$\{(25, 24, 37, 33, 0, 8, 34, 6, 29, 10), $\\$(12, 9, 31, 11, 36, 17, 2, 15, 19, 28, 27, 16), $\\$(4, 3, 20, 32, 14, 35, 30, 22, 5, 7, 18, 13, 21, 1, 26, 23)\}$}\\\hline
2&\makecell{$\{(13, 1, 6, 10, 2, 19, 33, 32, 16, 7, 8, 24), $\\$(25, 9, 3, 26, 22, 18, 29, 27, 12, 34, 20, 28, 5, 23, 35, 21),$\\$ (31, 15, 0, 17, 4, 30, 11, 36, 37, 14)\}$}\\\hline
3&\makecell{$\{(28, 33, 22, 17, 12, 7, 19, 4, 29, 8, 13, 0, 31, 16, 10, 21, 1, 6),$\\$ (34, 27, 11, 24, 20, 9, 25, 3),$\\$ (23, 26, 37, 14, 15, 2, 18, 30, 32, 35, 36, 5)\}$}\\\hline
4&\makecell{$\{(35, 15, 1, 7, 4, 33, 12, 32, 19, 16, 8, 13, 17, 3),$\\$ (10, 6, 29, 31, 26, 24, 0, 25, 27),$\\$ (5, 22, 11, 14, 30, 21, 2, 9, 20, 18, 36, 28, 34, 23, 37)\}$}\\\hline
5&\makecell{$\{(15, 32, 27, 30, 2, 19, 18, 4, 8, 22, 5, 6, 0, 16),$\\$ (1, 36, 23, 10, 14, 35, 25, 3, 13),$\\$ (31, 12, 21, 33, 11, 26, 7, 24, 28, 20, 34, 9, 29, 37, 17)\}$}\\\hline
6&\makecell{$\{(19, 5, 32, 34, 25, 37, 21, 8, 13, 16, 15, 4, 10, 30, 36), $\\$(18, 17, 24, 14, 31, 35, 12, 29, 28, 7, 23),$\\$ (27, 3, 6, 20, 33, 9, 1, 0, 22, 26, 11, 2)\}$}\\
\hline
\end{tabular}}
\caption{Block Parallel update modes for $n=38$.}
\label{table:bp1_n38}
\end{table}

\begin{table}\centering
\begin{tabular}{|c|c|}
\hline
\# &Definition\\\hline
12&\makecell{$\{(6, 19, 12, 25, 17, 9, 33, 0, 20, 7, 27, 32, 28, 16),$\\$ (4, 22, 23, 30, 34, 13, 1, 29, 2, 37, 31, 18, 8, 5, 3, 35, 24),$\\$ (21, 11, 36, 15, 26, 10, 14)\}$}\\\hline
13&\makecell{$\{(36, 5, 15, 7, 10, 12, 14, 16, 18, 13, 1, 37, 20, 11, 8),$\\$ (22, 26, 29, 30, 2, 23, 6, 0, 21, 4, 19),$\\$ (27, 33, 34, 28, 24, 35, 3, 9, 32, 31, 17, 25)\}$}\\\hline
14&\makecell{$\{(18, 5, 19, 10, 0, 4, 2, 23, 16, 1, 36, 22, 27, 25, 26),$\\$ (7, 11, 30, 24, 31, 14, 28, 32, 6, 9, 12, 13, 17),$\\$ (3, 21, 37, 15, 8, 33, 29, 34, 35, 20)\}$}\\\hline
15&\makecell{$\{(28, 21, 20, 11, 33, 34, 12, 37, 27, 18, 26, 32), $\\$(3, 14, 36, 6, 31, 15, 29, 13, 16, 17, 25, 22),$\\$ (5, 30, 2, 35, 4, 19, 23, 24, 7, 0, 8, 9, 1, 10)\}$}\\\hline
16&\makecell{$\{(1, 13, 11, 17, 3, 30, 27, 35, 4, 19, 9, 15),$\\$ (12, 24, 21, 14, 25, 23, 8, 32, 37, 34),$\\$ (0, 18, 5, 22, 36, 29, 6, 16, 31, 2, 28, 26, 20, 7, 10, 33)\}$}\\\hline
17&\makecell{$\{(17, 35, 11, 2, 0, 34, 24, 29, 23, 31, 4, 8, 10),$\\$ (15, 6, 13, 9, 21, 27, 14, 1, 5, 19, 28, 36),$\\$ (18, 37, 3, 32, 22, 26, 20, 25, 16, 33, 7, 12, 30)\}$}\\\hline
18&\makecell{$\{(36, 13, 34, 15, 14, 5, 21, 18, 28, 16, 11, 7, 23, 26),$\\$ (30, 1, 29, 24, 9, 25, 2, 20, 35), $\\$(33, 17, 37, 3, 0, 32, 12, 4, 8, 19, 27, 6, 31, 10, 22)\}$}\\
\hline
\end{tabular}
\caption{Block Parallel update modes for $n=38$.}
\label{table:bp2_n38}
\end{table}

\begin{table}
\resizebox{\textwidth}{!}{\centering
\begin{tabular}{|c|c|}
\hline
\# &Definition\\\hline
23&\makecell{$\{(14, 15, 33, 24, 23, 27, 17, 28, 5, 9, 7, 4, 20, 35, 11, 16),$\\$ (30, 21, 10, 25, 18, 2, 0, 3, 13, 22),$\\$ (31, 12, 1, 19, 8, 26, 34, 36, 6, 32, 37, 29)\}$}\\\hline
24&\makecell{$\{(9, 5, 17, 13, 24, 6, 7, 4, 22, 15, 31, 16, 3, 36, 12, 25, 28),$\\$ (34, 2, 29, 35, 0, 10, 8, 27, 1, 33, 21, 26), $\\$(30, 18, 14, 20, 19, 11, 23, 37, 32)\}$}\\\hline
25&\makecell{$\{(36, 11, 9, 33, 24, 14, 26, 8, 28, 13, 2, 23, 5, 3, 16, 35, 12, 22, 17, 21), $\\$(27, 0, 10, 4, 15, 19, 6, 37),$\\$ (20, 1, 32, 29, 34, 31, 30, 25, 18, 7)\}$}\\\hline
26&\makecell{$\{(36, 29, 12, 26, 27, 0, 10, 21, 9, 35, 18, 24, 30, 32),$\\$ (25, 34, 20, 7, 5, 28, 14, 22, 3, 37, 19, 11, 13, 15), $\\$(1, 23, 8, 2, 17, 6, 4, 33, 31, 16)\}$}\\\hline
27&\makecell{$\{(0, 23, 26, 9, 1, 12, 16, 24, 3, 22, 11, 21), $\\$(8, 31, 4, 20, 27, 13, 17, 18, 19, 35, 15),$\\$ (29, 28, 33, 14, 10, 2, 5, 30, 37, 34, 6, 36, 7, 32, 25)\}$}\\\hline
28&\makecell{$\{(37, 22, 13, 10, 6, 19, 3, 33, 25, 20, 5, 0),$\\$ (32, 23, 31, 1, 34, 2, 7, 36, 26, 29, 27, 12),$\\$ (9, 28, 11, 35, 17, 14, 18, 21, 16, 30, 8, 15, 4, 24)\}$}\\\hline
29&\makecell{$\{(31, 14, 23, 29, 20, 28, 18, 12, 17, 13, 3, 6),$\\$ (2, 25, 19, 16, 37, 21, 27, 34, 8, 32, 22, 24, 4, 33),$\\$ (35, 26, 1, 15, 36, 7, 9, 11, 0, 10, 5, 30)\}$}\\\hline
30&\makecell{$\{(0, 23, 30, 21, 25, 34, 6, 14, 32, 9), $\\$(7, 33, 16, 19, 13, 5, 24, 15, 3, 20, 17, 31, 37, 36),$\\$ (10, 1, 8, 11, 18, 4, 35, 29, 22, 26, 28, 2, 12, 27)\}$}\\\hline
31&\makecell{$\{(13, 9, 21, 35, 36, 6, 32, 19, 23),$\\$ (34, 24, 37, 30, 14, 29, 11, 27, 33, 4, 26, 20), $\\$(15, 3, 8, 16, 0, 2, 1, 17, 7, 31, 5, 12, 10, 18, 28, 22, 25)\}$}\\\hline
\end{tabular}}
\caption{Block Parallel update modes for $n=38$.}
\label{table:bp3_n38}
\end{table}

\begin{table}\centering
\resizebox{\textwidth}{!}{\centering
\begin{tabular}{|c|c|c|}
\hline
\multirow{2}{*}{\#} &\multicolumn{2}{c|}{Definition}\\\cline{2-3}
&$P$&$\Delta$\\\hline
0&\makecell{$(2, 1, 1, 2, 2, 1, 1, 2, 2, 1, 1, 1, 1, 1, 1, 2, 1, 2, 2,$\\$ 2, 1, 2, 1, 1, 1, 2, 1, 2, 2, 2, 1, 1, 2, 2, 1, 2, 1, 2)$}
&\makecell{$(0, 0, 0, 0, 0, 0, 0, 0, 1, 0, 0, 0, 0, 0, 0, 1, 0, 1, 0,$\\$ 1, 0, 1, 0, 0, 0, 0, 0, 1, 1, 0, 0, 0, 1, 1, 0, 1, 0, 1)$}\\\hline
1&\makecell{$(1, 2, 5, 4, 4, 1, 1, 2, 2, 2, 1, 2, 5, 2, 3, 2, 4, 2, 4,$\\$ 5, 2, 3, 5, 1, 1, 1, 3, 3, 4, 5, 4, 2, 1, 1, 3, 5, 1, 3)$}
&\makecell{$(0, 1, 1, 2, 3, 0, 0, 1, 0, 0, 0, 0, 1, 0, 0, 0, 2, 1, 0,$\\$ 2, 1, 1, 4, 0, 0, 0, 0, 1, 3, 1, 2, 0, 0, 0, 0, 1, 0, 2)$}\\\hline
2&\makecell{$(2, 5, 3, 5, 5, 5, 5, 5, 1, 5, 3, 2, 1, 1, 3, 2, 1, 2, 2,$\\$ 4, 5, 4, 2, 3, 3, 3, 3, 1, 1, 2, 2, 3, 5, 4, 3, 4, 5, 3)$}
&\makecell{$(1, 0, 1, 2, 0, 1, 0, 0, 0, 1, 2, 1, 0, 0, 1, 0, 0, 0, 0,$\\$ 1, 2, 0, 0, 0, 0, 0, 0, 0, 0, 1, 1, 2, 3, 3, 1, 1, 3, 0)$}\\\hline
3&\makecell{$(2, 2, 5, 1, 5, 1, 2, 4, 5, 4, 4, 3, 1, 1, 3, 4, 4, 2, 1,$\\$ 3, 1, 3, 2, 1, 1, 5, 1, 3, 3, 3, 3, 1, 3, 3, 4, 3, 4, 3)$}
&\makecell{ $(0, 1, 0, 0, 4, 0, 1, 3, 4, 3, 0, 0, 0, 0, 1, 3, 0, 1, 0,$\\$ 2, 0, 1, 0, 0, 0, 1, 0, 2, 2, 0, 1, 0, 0, 2, 0, 1, 3, 0)$}\\\hline
4&\makecell{$(2, 3, 5, 1, 3, 4, 4, 1, 4, 5, 5, 3, 4, 1, 4, 5, 1, 4, 4,$\\$ 1, 1, 1, 3, 1, 3, 1, 1, 5, 4, 4, 4, 5, 2, 1, 3, 4, 1, 1)$}
&\makecell{ $(0, 1, 1, 0, 0, 0, 3, 0, 2, 1, 1, 0, 0, 0, 1, 0, 0, 2, 0,$\\$ 0, 0, 0, 1, 0, 2, 0, 0, 2, 3, 0, 2, 3, 0, 0, 2, 3, 0, 0)$}\\\hline
5&\makecell{$(4, 5, 2, 3, 2, 4, 3, 1, 1, 2, 5, 3, 1, 4, 4, 3, 2, 4, 3,$\\$ 3, 1, 1, 2, 3, 1, 5, 2, 3, 4, 4, 5, 5, 4, 5, 5, 1, 3, 5)$}
&\makecell{$(3, 3, 0, 0, 1, 3, 2, 0, 0, 0, 3, 2, 0, 0, 1, 2, 0, 2, 0,$\\$ 0, 0, 0, 0, 1, 0, 0, 0, 2, 2, 0, 3, 2, 1, 0, 0, 0, 2, 3)$}\\
\hline
\end{tabular}}
\caption{Local Clocks update modes for $n=38$.}
\label{table:lc1_n38}
\end{table}

\begin{table}\centering
\resizebox{\textwidth}{!}{\centering
\begin{tabular}{|c|c|c|}
\hline
\multirow{2}{*}{\#} &\multicolumn{2}{c|}{Definition}\\\cline{2-3}
&$P$&$\Delta$\\\hline
16&\makecell{$(5, 4, 5, 5, 1, 3, 1, 4, 3, 1, 3, 3, 4, 4, 2, 5, 5, 2, 5, 5,$\\$ 1, 2, 4, 5, 3, 4, 5, 4, 5, 5, 4, 4, 1, 5, 2, 4, 3, 2)$}
&\makecell{$(2, 2, 1, 0, 0, 0, 0, 1, 1, 0, 0, 0, 3, 3, 1, 0, 1, 0, 2, 3,$\\$ 0, 0, 3, 4, 1, 0, 0, 1, 2, 2, 0, 2, 0, 4, 0, 2, 0, 0)$}\\\hline
17&\makecell{$(2, 1, 4, 4, 2, 4, 3, 3, 2, 3, 4, 2, 1, 3, 5, 2, 1, 3, 1, 3,$\\$ 3, 5, 4, 5, 3, 3, 2, 3, 3, 1, 4, 4, 5, 1, 2, 3, 5, 5)$}
&\makecell{$(1, 0, 0, 0, 0, 1, 2, 2, 1, 2, 1, 0, 0, 1, 2, 1, 0, 1, 0, 0,$\\$ 2, 4, 0, 3, 2, 0, 0, 2, 1, 0, 1, 3, 2, 0, 1, 2, 2, 4)$}\\\hline
18&\makecell{$(1, 2, 3, 5, 5, 3, 4, 3, 5, 4, 5, 3, 5, 5, 1, 2, 4, 4, 1, 4,$\\$ 5, 5, 4, 5, 4, 4, 3, 3, 3, 2, 4, 4, 3, 4, 2, 4, 1, 1)$}
&\makecell{$(0, 0, 1, 3, 4, 2, 0, 1, 4, 3, 4, 2, 1, 4, 0, 0, 2, 1, 0, 0,$\\$ 2, 2, 3, 1, 2, 3, 2, 1, 2, 1, 2, 0, 1, 2, 1, 0, 0, 0)$}\\\hline
19&\makecell{$(2, 5, 3, 3, 4, 4, 4, 1, 4, 5, 5, 4, 5, 2, 4, 1, 1, 1, 3, 1,$\\$ 3, 5, 3, 3, 2, 5, 5, 4, 3, 5, 2, 1, 4, 4, 1, 3, 5, 5)$}
&\makecell{$(1, 3, 1, 1, 1, 0, 1, 0, 2, 0, 3, 3, 2, 0, 3, 0, 0, 0, 1, 0,$\\$ 0, 4, 0, 0, 0, 3, 3, 2, 1, 0, 1, 0, 3, 1, 0, 0, 0, 3)$}\\\hline
20&\makecell{$(1, 2, 2, 2, 5, 4, 5, 1, 3, 1, 4, 2, 3, 2, 4, 3, 1, 5, 1, 5,$\\$ 5, 3, 2, 3, 4, 1, 3, 5, 2, 2, 4, 2, 4, 4, 4, 4, 2, 4)$}
&\makecell{$(0, 1, 0, 0, 2, 2, 2, 0, 2, 0, 1, 0, 1, 1, 1, 1, 0, 0, 0, 1,$\\$ 3, 2, 1, 2, 0, 0, 1, 2, 1, 0, 0, 1, 2, 3, 3, 2, 1, 3)$}\\\hline
21&\makecell{$(5, 2, 2, 5, 2, 3, 5, 5, 3, 1, 1, 5, 4, 5, 3, 3, 4, 5, 4, 5,$\\$ 4, 3, 5, 3, 5, 1, 1, 2, 1, 2, 4, 1, 3, 4, 1, 3, 2, 2)$}
&\makecell{$(1, 1, 0, 0, 0, 0, 1, 1, 2, 0, 0, 2, 1, 0, 0, 2, 3, 0, 2, 0,$\\$ 2, 2, 3, 1, 3, 0, 0, 0, 0, 0, 0, 0, 0, 2, 0, 0, 1, 0)$}\\
\hline
\end{tabular}}
\caption{Local Clocks update modes for $n=38$.}
\label{table:lc2_n38}
\end{table}

\FloatBarrier

\begin{table}
\resizebox{\textwidth}{!}{\centering
\begin{tabular}{|c|c|}
\hline
\# &Definition\\\hline
0&\makecell{$(
81,88,63,103,119,79,135,76,107,67,18,66,5,126,96,41,27,51,105,100,54,32,$\\$
2,7,112,131,90,42,98,59,108,4,6,70,3,77,123,101,15,106,0,44,120,69,10,48,$\\$
127,111,11,12,93,21,24,35,16,49,72,121,68,8,31,40,58,116,122,64,26,73,124,$\\$
9,25,53,43,28,46,95,50,52,78,102,97,132,91,82,36,86,133,57,1,110,128,80,56,$\\$
14,22,137,13,74,17,92,84,30,75,47,19,20,134,61,62,113,104,65,39,87,99,117,$\\$
136,125,37,85,89,114,129,115,60,83,118,23,34,29,45,94,55,71,109,130,38,33)$}\\\hline
1&\makecell{$(
60,25,113,69,104,84,45,56,130,87,48,38,124,74,71,6,37,110,31,41,127,120,$\\$
51,24,50,125,27,12,0,76,94,4,97,3,81,116,32,46,119,88,42,100,68,11,7,85,$\\$
80,67,132,108,109,59,133,89,28,2,136,18,21,26,128,86,115,47,14,70,131,33,$\\$
103,61,52,137,13,65,66,57,129,95,73,29,64,102,134,93,39,1,5,106,121,83,92,$\\$
114,35,126,123,98,55,75,34,30,22,53,122,112,17,111,44,54,91,10,72,90,63,15,$\\$ 8, 78, 82, 118, 58, 19, 16, 40, 23, 62, 77, 9, 105, 117, 36, 79, 49, 43, 99, 96, 20, 101, 135, 107)$}\\\hline
2&\makecell{$(
68,136,16,74,61,114,137,111,15,59,101,118,125,79,32,121,134,42,4,70,94,133,$\\$
67,91,95,35,36,21,98,130,17,73,52,55,58,33,53,54,5,92,77,75,43,45,128,47,29,$\\$
2,8,112,65,49,51,86,69,56,20,132,97,122,88,100,82,31,76,108,104,89,135,28,1,$\\$
64,81,30,124,40,34,23,120,85,11,62,90,0,22,72,44,109,6,50,48,78,66,84,106,$\\$
123,107,18,10,24,39,60,127,41,102,26,105,129,14,103,38,110,117,37,119,63,9,$\\$ 19, 93, 115, 83, 25, 13, 113, 7, 99, 87, 116, 131, 71, 96, 126, 27, 46, 12, 3, 80, 57)$}\\
\hline
\end{tabular}}
\caption{Sequential update modes for $n=138$.}
\label{table:seq1_n138}
\end{table}
\begin{table}
\resizebox{\textwidth}{!}{\centering
\begin{tabular}{|c|c|}
\hline
\# &Definition\\\hline
7&\makecell{
$(102,112,47,78,39,19,57,72,28,117,31,106,46,62,68,70,77,5,109,40,132,33,121,$\\$
135,44,34,119,125,22,25,26,89,97,115,99,124,116,63,128,6,75,2,49,94,51,11,$\\$
79,123,122,108,118,48,18,114,45,13,12,82,64,15,32,136,14,16,10,133,130,73,$\\$
21,42,58,23,55,27,76,71,113,29,127,59,38,88,98,4,83,24,66,93,100,37,61,105,$\\$
41,20,9,3,104,107,54,85,69,137,129,36,1,86,35,8,60,91,52,96,0,50,90,92,101,$\\$ 110, 131, 80, 53, 56, 84, 30, 120, 81, 17, 43, 126, 134, 103, 111, 74, 7, 67, 87, 95, 65)$}\\\hline
8&\makecell{
$(93,78,46,111,69,65,36,12,45,43,38,83,99,9,132,53,61,16,8,48,109,96,17,116,$\\$
1,94,3,34,29,126,63,2,21,115,104,56,42,90,125,32,68,102,51,62,89,72,81,60,$\\$
98,40,19,106,25,5,127,110,118,24,58,112,18,23,49,119,47,120,131,11,31,33,$\\$
133,85,123,79,105,55,22,84,41,124,117,0,103,52,71,108,67,27,14,101,122,87,$\\$
57,97,54,114,82,20,95,28,74,135,137,37,80,50,76,4,75,64,35,88,107,92,7,15,$\\$ 39, 91, 134, 121, 66, 6, 100, 26, 128, 44, 113, 129, 30, 13, 86, 130, 73, 59, 136, 10, 70, 77)$}\\\hline
9&\makecell{
$(15,137,26,136,4,98,28,45,131,63,107,80,123,87,44,51,54,111,46,104,11,92,3,$\\$
57,85,95,53,25,37,41,97,6,108,102,76,130,19,18,12,23,34,135,132,24,127,29,$\\$
96,74,88,9,91,75,118,7,86,116,50,40,43,109,100,70,1,115,49,84,114,133,105,$\\$
21,5,128,79,99,20,65,129,39,82,117,101,56,8,30,58,32,103,125,67,17,110,78,$\\$
122,52,38,48,13,72,55,16,113,64,83,121,42,2,124,81,69,90,93,68,89,14,60,61,$\\$ 0, 120, 10, 106, 66, 112, 73, 77, 59, 47, 33, 134, 62, 126, 22, 27, 31, 119, 94, 36, 71, 35)$}\\
\hline
\end{tabular}}
\caption{Sequential update modes for $n=138$.}
\label{table:seq2_n138}
\end{table}

\begin{table}
\resizebox{\textwidth}{!}{\centering
\begin{tabular}{|c|c|}
\hline
\# &Definition\\\hline
14&\makecell{
$(123,78,109,79,130,71,50,93,43,25,133,91,117,28,73,13,122,106,37,84,112,75,$\\$
10,21,76,97,61,14,7,36,5,26,47,63,16,41,20,116,81,72,33,80,31,83,69,68,137,$\\$
125,85,107,59,39,35,115,24,86,66,53,90,103,74,88,54,132,114,3,113,119,98,129,$\\$
60,19,135,42,134,104,128,118,96,30,44,38,57,45,94,100,12,102,108,46,89,18,77,$\\$
52,1,64,99,0,58,124,55,40,32,49,111,34,110,127,105,27,82,23,92,11,62,4,126,48,$\\$ 95, 121, 17, 8, 70, 101, 29, 131, 15, 2, 56, 136, 6, 120, 67, 87, 51, 9, 22, 65)$}\\\hline
15&\makecell{
$(69,137,28,102,114,116,117,99,73,87,65,35,26,24,108,30,106,96,6,8,78,40,111,$\\$
125,133,109,91,129,53,92,19,98,0,42,46,118,34,132,39,67,47,27,75,21,29,51,$\\$
66,122,33,128,112,82,88,44,41,131,119,83,74,103,32,97,14,11,54,63,100,50,79,$\\$
86,49,1,77,59,93,101,64,134,113,55,120,56,62,121,36,17,31,84,105,20,85,110,$\\$
12,123,76,126,25,94,16,43,13,135,4,38,90,130,71,61,81,22,104,107,18,9,72,5,$\\$ 70, 136, 89, 57, 60, 58, 37, 3, 2, 15, 68, 45, 23, 124, 80, 115, 7, 95, 48, 127, 10, 52)$}\\\hline
16&\makecell{
$(80,93,95,20,38,43,69,56,103,32,127,26,115,116,101,90,124,19,135,12,73,55,107,$\\$
46,89,62,75,102,105,52,65,11,59,54,88,34,86,70,81,76,94,125,60,129,100,104,$\\$
63,16,51,128,31,64,137,30,18,53,35,7,87,85,133,15,119,71,2,33,5,66,48,97,99,$\\$
123,50,92,84,21,10,67,1,91,9,13,6,57,23,25,14,4,110,68,77,36,45,40,96,114,$\\$
39,98,120,41,27,47,109,113,118,136,17,22,28,72,29,0,111,134,79,126,58,42,49,$\\$ 74, 131, 112, 130, 108, 83, 78, 121, 8, 82, 24, 122, 3, 61, 44, 106, 117, 132, 37)$}\\
\hline
\end{tabular}}
\caption{Sequential update modes for $n=138$.}
\label{table:seq3_n138}
\end{table}

\begin{table}
\resizebox{\textwidth}{!}{\centering
\begin{tabular}{|c|c|}
\hline
\# &Definition\\\hline
20&\makecell{$(
51,133,134,137,104,27,87,58,68,73,84,82,31,40,53,111,117,94,89,1,110,88,35,46,$\\$
125,63,75,69,2,43,101,18,37,15,95,115,0,132,8,30,97,41,103,34,55,33,14,11,16,$\\$
13,78,123,23,54,83,80,71,12,5,136,113,86,74,52,17,67,105,85,57,20,42,93,122,$\\$
3,130,72,90,77,19,10,7,96,28,36,126,60,92,21,47,114,121,112,6,29,129,135,56,$\\$
24,108,49,119,59,118,128,102,106,32,70,61,64,9,26,127,44,124,39,99,76,4,116,$\\$ 25, 79, 131, 45, 48, 100, 38, 22, 50, 107, 81, 109, 120, 98, 66, 62, 91, 65)$}\\\hline
21&\makecell{
$(103,88,32,133,1,71,60,17,115,51,9,18,77,11,16,101,136,50,30,116,135,61,126,$\\$
49,53,5,39,59,57,15,21,33,121,70,130,34,120,131,128,56,93,64,24,14,110,106,$\\$
75,12,26,83,8,36,113,2,47,31,27,89,76,123,87,90,125,82,124,85,35,107,10,78,$\\$
28,137,23,22,42,3,7,84,73,99,45,37,74,95,108,105,127,92,46,122,100,25,63,19,$\\$
102,55,54,91,52,79,134,62,43,72,58,13,68,0,104,29,40,111,65,80,66,97,6,117,98,$\\$ 41, 132, 109, 81, 69, 118, 119, 114, 48, 44, 129, 4, 67, 94, 38, 20, 96, 112, 86)$}\\\hline
22&\makecell{
$(90,34,85,80,73,30,18,65,101,126,106,103,78,39,129,70,38,137,52,104,131,0,$\\$
123,15,48,121,118,72,108,114,136,102,130,62,109,35,23,74,3,28,69,14,95,59,$\\$
16,40,110,105,75,45,12,77,47,37,7,100,31,98,10,87,119,64,66,115,63,86,21,9,$\\$
26,94,132,4,13,56,107,46,1,117,68,2,79,97,49,67,128,125,6,135,122,8,20,113,$\\$
22,43,76,53,51,24,134,54,82,17,84,120,50,127,81,32,41,58,19,27,91,111,133,$\\$ 93, 92, 36, 25, 55, 89, 96, 83, 33, 124, 61, 99, 11, 60, 29, 42, 116, 88, 71, 57, 5, 44, 112)$}\\
\hline
\end{tabular}}
\caption{Sequential update modes for $n=138$.}
\label{table:seq4_n138}
\end{table}

\begin{table}
\resizebox{\textwidth}{!}{\centering
\begin{tabular}{|c|c|}
\hline
\# &Definition\\\hline
26&\makecell{
$(77,93,110,90,109,114,106,121,39,137,118,27,30,61,92,35,46,59,112,51,54,57,$\\$
94,62,105,111,83,82,40,91,38,100,16,66,41,14,128,58,81,126,31,69,88,131,20,$\\$
129,89,67,78,49,37,122,103,79,56,8,134,123,65,2,29,86,76,73,102,98,9,85,60,$\\$
19,44,6,22,24,136,53,107,13,71,33,97,34,119,28,47,4,117,99,7,0,135,87,21,75,$\\$
15,17,95,68,70,1,48,116,63,120,72,104,10,26,11,23,52,101,80,64,43,130,74,113,$\\$ 32, 42, 84, 3, 18, 55, 5, 127, 124, 132, 25, 36, 133, 45, 12, 96, 115, 125, 108, 50)$}\\\hline
27&\makecell{
$(18,86,123,103,1,95,21,54,106,75,134,62,129,97,115,81,107,32,36,96,128,91,69,$\\$
22,47,25,65,83,109,113,11,41,61,105,64,46,87,120,45,26,88,92,27,28,0,33,15,$\\$
44,7,71,23,39,3,14,5,136,42,127,40,31,125,57,82,30,50,63,52,35,77,84,56,133,$\\$
132,58,85,119,80,53,38,90,43,124,12,94,116,66,17,89,131,49,24,8,110,137,72,$\\$
29,114,102,117,73,118,130,126,60,2,10,4,98,59,99,19,67,112,74,76,68,48,34,6,$\\$ 111, 9, 78, 55, 101, 104, 79, 37, 122, 121, 13, 51, 108, 93, 70, 135, 16, 20, 100)$}\\\hline
28&\makecell{
$(17,93,125,34,101,85,129,39,123,76,58,104,2,83,137,1,136,41,47,30,7,80,62,59,$\\$
135,100,108,124,112,19,119,27,12,103,97,133,66,77,29,117,3,88,109,8,28,70,24,$\\$
42,98,56,16,38,10,22,31,51,86,90,102,130,69,55,79,126,36,35,11,52,118,72,40,$\\$
71,128,67,20,44,84,107,4,45,114,49,110,33,96,0,54,6,116,127,26,134,122,46,$\\$
81,53,91,105,5,95,50,21,32,111,131,74,82,57,65,99,113,18,92,48,63,115,23,68,$\\$ 13, 121, 75, 15, 78, 14, 120, 60, 106, 89, 73, 25, 43, 64, 132, 9, 94, 87, 37, 61)$}\\
\hline
\end{tabular}}
\caption{Sequential update modes for $n=138$.}
\label{table:seq5_n138}
\end{table}

\begin{table}
\resizebox{\textwidth}{!}{\centering
\begin{tabular}{|c|c|}
\hline
\# &Definition\\\hline
0&\makecell{
$(\{39, 57, 56, 102, 12, 29, 76, 120, 78, 18, 99, 72, 8, 89, 85, 97, 65, 73, 14, 61, 115,$\\$ 31, 1, 80, 59, 55, 35, 82, 54, 21, 126\}, \{104, 83, 88, 74, 84, 87, 107, 134, 114, 58,$\\$ 109, 129, 51, 24, 25, 19, 15, 13, 40, 33, 90, 60, 22, 70, 4, 10, 71\}, \{42, 106, 81,$\\$ 125, 100, 111, 113, 117, 6, 23, 119, 46, 93, 53, 44, 116, 136, 122, 95, 128, 121,$\\$ 123, 34, 45, 30, 132, 20\}, \{2, 68, 41, 36, 133, 48, 37, 50, 49, 79, 105, 27, 11, 130,$\\$ 62, 112, 16, 9, 75, 131, 63, 103, 52, 98, 47, 3, 43, 124, 137, 77\}, \{5, 17, 0,$\\$ 7, 92, 101, 32, 110, 69, 38, 64, 91, 67, 108, 94, 26, 96, 135, 86, 66, 28, 118, 127\})$}\\\hline
1&\makecell{
$(\{12, 30, 109, 95, 89, 36, 45, 97, 29, 47, 64, 68, 60, 23, 43, 136, 121, 104, 49, 58,$\\$ 96, 110, 108, 9, 93, 74, 127, 112, 53, 111, 44, 132, 61, 26, 130, 131, 3, 1\}, \{56, 81,$\\$ 103, 5, 124, 8, 122, 7, 50, 106, 113, 55, 37, 41, 52, 77, 17, 135, 129, 19, 48, 39, $\\$62, 94, 76, 107, 0, 133, 51, 115, 14, 87, 79, 134, 46, 90, 24, 91, 105, 99, 78, 86,$\\$ 25, 57, 2, 27, 116, 34, 114, 83, 123, 15, 119, 42\}, \{40, 128, 11, 84, 10, 72, 33,$\\$ 16, 6, 100, 117, 31, 21, 125, 20, 70, 4, 82, 13, 137, 22, 32, 54, 73, 65, 71, 120,$\\$ 85, 66, 118, 92, 63, 88, 80, 98, 75, 28, 18, 59, 102, 101, 67, 35, 126, 69, 38\})$}\\\hline
2&\makecell{
$(\{42, 97, 93, 11, 21, 19, 39, 103, 50, 118, 41, 120, 90, 26, 0, 121, 58, 123, 2, 23, 69,$\\$ 106, 67, 38, 36, 115, 14, 96, 81, 18, 31, 53, 45, 66, 49, 8, 68, 114, 105, 62, 85, 83,$\\$ 37, 94, 113, 107\}, \{71, 32, 57, 61, 82, 130, 98, 136, 112, 9, 34, 73, 55, 64, 52, 5,$\\$ 91, 104, 92, 127, 44, 35, 43, 102, 48, 79, 74, 16, 126, 117, 46, 60, 30, 99, 84, 12,$\\$ 137, 125, 95, 75, 135, 25, 133, 109, 20, 89, 108, 51, 54, 28, 129, 6, 87, 63, 77, 116,$\\$ 22, 76, 3\}, \{24, 70, 40, 119, 47, 111, 128, 124, 110, 65, 15, 80, 29, 122, 86, 78,$\\$ 4, 101, 100, 17, 88, 27, 7, 59, 56, 132, 134, 10, 33, 72, 1, 131, 13\})$}\\
\hline
\end{tabular}}
\caption{Block Sequential update modes for $n=138$.}
\label{table:bs1_n138}
\end{table}

\begin{table}
\resizebox{\textwidth}{!}{\centering
\begin{tabular}{|c|c|}
\hline
\# &Definition\\\hline
5&\makecell{
$(\{116, 111, 41, 38, 98, 27, 61, 34, 37, 68, 136, 7, 100, 1, 81, 67, 102, 64, 26, 88, 77,$\\$ 94, 103, 32, 56, 45, 49, 33, 121, 70, 9, 85, 5, 92, 83, 20, 132, 21, 24, 15, 31\}, \{78,$\\$ 14, 39, 71, 135, 96, 55, 129, 13, 6, 79, 28, 109, 44, 128, 97, 18, 51, 63, 117, 46,$\\$ 52, 115, 74, 86, 30, 101, 99, 113, 65, 104, 120, 123, 36, 40, 29, 133, 12, 0, 90, 87,$\\$ 11, 114, 59, 17, 134, 127, 126, 130, 10, 75, 54\}, \{62, 84, 57, 3, 50, 72, 42, 35, 76,$\\$ 47, 73, 4, 48, 119, 110, 60, 2, 19, 105, 118, 53, 22, 66, 80, 91, 43, 82, 8, 23, 131,$\\$ 69, 124, 107, 95, 108, 125, 106, 137, 112, 58, 122, 93, 89, 16, 25\})$}\\\hline
6&\makecell{
$(\{124, 90, 119, 106, 51, 47, 53, 102, 58, 135, 18, 133, 37, 28, 74, 15, 82, 79, 60, 45,$\\$ 30, 132, 2, 129, 101, 9, 122, 115, 29, 50, 49, 62, 61, 6, 33, 114, 98, 75, 76, 87, 54,$\\$ 26, 97, 57, 63, 24, 94, 73\}, \{35, 136, 80, 12, 118, 123, 68, 56, 43, 10, 71, 4, 130,$\\$ 86, 91, 17, 31, 131, 83, 41, 77, 85, 66, 134, 99, 104, 92, 137, 64, 23, 67, 36, 27, 103,$\\$ 42, 3, 21, 65, 126, 5, 13, 109, 128\}, \{84, 127, 112, 93, 46, 44, 34, 108, 81, 7, 19,$\\$ 78, 116, 111, 55, 88, 22, 113, 96, 11, 110, 48, 20, 16, 39, 40, 69, 14, 107, 100, 120,$\\$ 125, 89, 0, 117, 25, 70, 59, 95, 72, 38, 121, 105, 1, 52, 32, 8\})$}\\\hline
7&\makecell{
$(\{61, 17, 88, 75, 97, 31, 44, 5, 114, 94, 47, 6, 108, 123, 116, 129, 20, 22, 98, 33, 131,$\\$ 81, 50, 72, 42, 3, 36, 137, 65, 19\}, \{83, 76, 86, 52, 18, 25, 132, 121, 26, 126, 127,$\\$ 92, 135, 85, 68, 4, 124, 96, 105, 56, 136, 106, 62, 115, 48\}, \{100, 84, 70, 110,$\\$ 8, 57, 2, 51, 0, 102, 54, 78, 103, 112, 39, 28, 90, 46, 133, 87, 71, 119, 134, 32, 41\},$\\$ \{130, 69, 49, 9, 29, 12, 43, 27, 53, 59, 120, 38, 113, 77, 7, 63, 40, 45, 80, 107, 91,$\\$ 10, 1, 82, 23, 99, 122, 74, 15\}, \{93, 64, 16, 66, 11, 125, 60, 13, 58, 79, 117,$\\$ 35, 104, 128, 111, 109, 67, 34, 89, 118, 24, 55, 30, 101, 21, 95, 73, 37, 14\})$}\\
\hline
\end{tabular}}
\caption{Block Sequential update modes for $n=138$.}
\label{table:bs2_n138}
\end{table}

\begin{table}
\resizebox{\textwidth}{!}{\centering
\begin{tabular}{|c|c|}
\hline
\# &Definition\\\hline
10&\makecell{
$(\{68, 14, 17, 99, 33, 98, 13, 135, 108, 24, 3, 75, 27, 136, 125, 23, 56, 73, 133, 85,$\\$ 39, 87, 83, 43, 107, 128, 97, 131, 64, 45, 60, 4, 82, 109, 10, 53, 113, 22, 104, 124,$\\$ 8, 62, 55, 12, 89, 0, 122\}, \{35, 44, 48, 41, 103, 36, 58, 2, 106, 71, 115, 114, 95,$\\$ 16, 123, 132, 30, 118, 76, 29, 137, 116, 100, 6, 67, 94, 42, 66, 126, 59, 5, 20, 50,$\\$ 34, 120, 32, 7, 28, 57, 92\}, \{1, 101, 47, 90, 110, 84, 70, 26, 9, 121, 105, 38, 65,$\\$ 61, 78, 74, 63, 91, 102, 96, 117, 11, 31, 119, 72, 127, 40, 81, 54, 79, 93, 25, 69,$\\$ 37, 21, 86, 77, 112, 134, 88, 52, 130, 80, 111, 19, 18, 51, 46, 49, 15, 129\})$}\\\hline
11&\makecell{
$(\{35, 79, 8, 120, 59, 93, 4, 39, 113, 28, 1, 14, 72, 118, 2, 36, 22, 23, 80, 123, 75,$\\$ 47, 11, 88, 71, 81, 69, 128, 63, 124, 73, 117, 94, 78, 25, 130, 67, 13, 114, 111, 64\},$\\$ \{45, 107, 119, 42, 17, 0, 49, 134, 106, 16, 74, 68, 15, 29, 62, 99, 10, 46, 125,$\\$ 108, 96, 58, 135, 87, 60, 104, 127, 33, 110, 126, 137, 3, 20, 77, 5, 9, 91, 27, 121,$\\$ 40, 100, 115, 116, 85, 55, 57, 84, 136, 12, 129, 83\}, \{31, 34, 19, 82, 131, 66, 43, 70,$\\$ 7, 32, 133, 105, 52, 95, 112, 97, 86, 56, 76, 103, 41, 18, 132, 122, 37, 65, 102,$\\$ 92, 98, 21, 51, 54, 89, 38, 44, 53, 50, 101, 109, 26, 24, 61, 48, 90, 6, 30\})$}\\\hline
12&\makecell{
$(\{102, 88, 136, 119, 1, 114, 120, 104, 59, 93, 77, 109, 122, 125, 26, 99, 85, 34, 43,$\\$ 135, 66, 80, 17, 33, 98, 16, 76, 92, 70, 113, 11, 127, 12, 82, 22, 89, 86, 25, 111, 94,$\\$ 46, 69, 20, 75, 49, 100, 74\}, \{27, 5, 39, 65, 31, 95, 123, 105, 116, 55, 41, 118, 61,$\\$ 128, 132, 18, 103, 53, 45, 19, 108, 131, 91, 78, 8, 56, 24, 7, 130, 87, 44, 29, 4,$\\$ 51, 48, 84, 121, 21, 124, 50, 30, 37, 96, 58, 3\}, \{54, 32, 40, 106, 90, 47, 81,$\\$ 0, 67, 52, 6, 101, 57, 79, 126, 2, 64, 110, 42, 72, 60, 28, 97, 133, 71, 134, 9, 73,$\\$ 14, 107, 83, 10, 129, 13, 62, 63, 38, 137, 15, 35, 117, 36, 112, 23, 115, 68\})$}\\
\hline
\end{tabular}}
\caption{Block Sequential update modes for $n=138$.}
\label{table:bs3_n138}
\end{table}

\begin{table}
\resizebox{\textwidth}{!}{\centering
\begin{tabular}{|c|c|}
\hline
\# &Definition\\\hline
15&\makecell{
$(\{7, 62, 6, 121, 14, 9, 123, 113, 11, 115, 15, 96, 17, 118, 90, 117, 38, 130, 93, 91,$\\$ 25, 94, 63, 39, 18, 8, 120, 72, 131, 34, 109, 55, 74, 53, 22, 56, 16, 36, 125, 136, 12,$\\$ 42, 76, 111, 68, 49, 129, 101, 102, 97, 135, 83, 122, 65\}, \{20, 2, 59, 77, 99, 103,$\\$ 112, 134, 107, 0, 33, 126, 116, 26, 41, 51, 89, 28, 4, 133, 82, 127, 31, 57, 95, 27,$\\$ 104, 43, 60, 29, 54, 80, 124, 137, 128, 10, 100\}, \{98, 19, 48, 52, 86, 46, 5, 92, 24,$\\$ 70, 105, 132, 67, 88, 58, 78, 32, 44, 64, 37, 81, 110, 114, 119, 40, 71, 69, 108,$\\$ 106, 61, 21, 45, 13, 47, 85, 3, 75, 73, 84, 50, 79, 30, 66, 1, 87, 35, 23\})$}\\\hline
16&\makecell{
$(\{111, 124, 32, 55, 49, 126, 81, 94, 64, 82, 46, 135, 122, 14, 75, 78, 13, 91, 128, 62,$\\$ 53, 109, 69, 90, 68, 121, 4, 97, 19, 95, 63, 57, 85, 52, 38, 58, 21, 28, 54, 119, 65,$\\$ 39, 117, 115, 76, 107, 73, 36, 134, 47, 59, 108, 31, 129, 84\}, \{118, 9, 130, 10, 25,$\\$ 3, 42, 116, 16, 87, 132, 120, 137, 56, 61, 131, 77, 112, 102, 133, 29, 110, 79, 40,$\\$ 33, 45, 8, 22, 71, 24, 37, 41, 0, 44, 12, 34, 99, 136, 2, 7, 70, 101, 17\}, \{74, 50,$\\$ 23, 113, 26, 20, 127, 88, 43, 27, 103, 1, 106, 60, 11, 51, 125, 123, 5, 72, 104, 89,$\\$ 15, 66, 86, 83, 114, 35, 48, 92, 30, 18, 6, 105, 96, 67, 98, 93, 100, 80\})$}\\\hline
17&\makecell{
$(\{63, 115, 79, 78, 40, 114, 112, 133, 54, 74, 129, 95, 65, 34, 15, 42, 2, 9, 98, 127, 60,$\\$ 16, 96, 100, 22, 94, 39, 56, 66, 124, 92, 20, 91, 7, 90, 4, 86, 45, 28, 11, 48, 19,$\\$ 71\}, \{97, 68, 32, 107, 23, 62, 126, 105, 38, 1, 46, 109, 49, 29, 89, 24, 67, 10, 57,$\\$ 88, 87, 44, 84, 102, 37, 82, 122, 81, 17, 120, 130, 73, 80, 83, 72, 13, 25, 59, 116,$\\$ 36, 55, 3, 58, 26, 6, 117, 121, 136, 77\}, \{51, 69, 33, 27, 137, 110, 101, 14, 41, 76,$\\$ 70, 119, 64, 135, 30, 31, 50, 0, 118, 134, 123, 47, 128, 43, 85, 8, 12, 18, 5, 111,$\\$ 75, 125, 21, 52, 53, 106, 93, 61, 113, 108, 35, 103, 99, 132, 131, 104\})$}\\
\hline
\end{tabular}}
\caption{Block Sequential update modes for $n=138$.}
\label{table:bs4_n138}
\end{table}

\begin{table}
\resizebox{\textwidth}{!}{\centering
\begin{tabular}{|c|c|}
\hline
\# &Definition\\\hline
20&\makecell{
$(\{18, 17, 44, 102, 86, 103, 120, 53, 61, 124, 9, 43, 28, 137, 70, 105, 131, 68, 89, 48,$\\$ 33, 42, 113, 75, 134, 60, 72, 38, 123, 104, 39, 15, 62, 58, 93, 108, 64, 63, 119, 23,$\\$ 101, 57\}, \{85, 22, 100, 30, 127, 12, 79, 71, 47, 20, 6, 92, 112, 0, 109, 80, 82, 78,$\\$ 94, 81, 21, 10, 107, 2, 26, 19, 128, 96, 95, 8, 121, 98, 99, 31, 34, 114, 129, 59, 52,$\\$ 77, 3, 73, 88, 1, 74, 5, 87, 69\}, \{66, 32, 25, 29, 117, 106, 37, 132, 11, 65, 115, 16,$\\$ 41, 97, 24, 135, 133, 67, 116, 130, 40, 35, 49, 50, 122, 118, 136, 27, 36, 111,$\\$ 4, 91, 90, 84, 46, 45, 76, 126, 54, 13, 55, 83, 125, 51, 7, 110, 14, 56\})$}\\\hline
21&\makecell{
$(\{134, 106, 7, 45, 51, 18, 34, 130, 70, 80, 41, 22, 50, 19, 93, 37, 58, 0, 131, 36, 68,$\\$ 15, 83, 64, 56, 89, 119, 90, 57, 118, 14, 47, 82, 27, 72, 132, 109, 53, 114, 66\}, \{100,$\\$ 23, 3, 88, 116, 86, 107, 124, 108, 28, 35, 79, 38, 49, 96, 84, 92, 97, 87, 40,$\\$ 62, 73, 32, 9, 123, 75, 77, 120, 1, 110, 128, 76, 8, 63, 95, 105, 10, 48, 42, 26, 102,$\\$ 13, 20, 33, 135\}, \{11, 67, 16, 6, 61, 94, 4, 46, 126, 71, 24, 65, 129, 115, 59, 52,$\\$ 21, 99, 133, 39, 43, 30, 60, 127, 122, 91, 55, 25, 74, 125, 2, 113, 5, 104, 44, 98,$\\$ 103, 31, 137, 69, 54, 117, 12, 111, 17, 101, 78, 112, 85, 136, 121, 29, 81\})$}\\\hline
22&\makecell{
$(\{136, 13, 61, 10, 60, 118, 101, 113, 7, 109, 24, 76, 90, 27, 87, 35, 133, 126, 56, 49,$\\$ 39, 63, 57, 98, 99, 108, 129, 93, 0, 137, 17, 73, 84, 67, 130, 65, 112, 4, 6, 115,$\\$ 77, 42, 1, 48, 94, 15, 105, 85, 44, 75, 33, 83\}, \{58, 16, 110, 51, 23, 116, 121, 86,$\\$ 47, 95, 80, 117, 78, 5, 66, 72, 96, 26, 104, 19, 124, 28, 50, 103, 11, 81, 122, 22,$\\$ 54, 79, 82, 53, 71, 32, 36, 114, 3, 29, 120, 64, 31\}, \{97, 89, 30, 119, 92, 59, 62,$\\$ 69, 127, 14, 55, 132, 100, 34, 128, 123, 25, 107, 41, 135, 38, 9, 21, 2, 102, 131, 52,$\\$ 37, 20, 45, 68, 91, 70, 40, 125, 106, 88, 8, 46, 43, 111, 74, 134, 18, 12\})$}\\
\hline
\end{tabular}}
\caption{Block Sequential update modes for $n=138$.}
\label{table:bs5_n138}
\end{table}
\begin{table}
\resizebox{\textwidth}{!}{\centering
\begin{tabular}{|c|c|}
\hline
\# &Definition\\\hline
25&\makecell{
$(\{98, 101, 3, 114, 31, 103, 124, 123, 88, 94, 9, 72, 2, 58, 48, 90, 11, 13, 54, 105, 62,$\\$ 137, 42, 87, 89, 49, 61, 26, 57, 1, 130, 116, 44, 10, 107, 12, 135, 118, 18, 47, 120,$\\$ 67, 78, 104, 15, 73, 27, 71\}, \{32, 125, 112, 41, 100, 37, 39, 66, 60, 5, 126, 43,$\\$ 0, 115, 133, 38, 50, 53, 23, 34, 74, 96, 86, 22, 102, 68, 21, 63, 28, 55, 91, 69, 25,$\\$ 97, 33, 79, 81, 76, 70, 51, 128, 121, 46, 77, 64, 136, 14, 56, 95, 84\}, \{45, 16, 59, 30$\\$, 82, 99, 117, 92, 24, 132, 110, 93, 129, 29, 17, 111, 4, 20, 108, 85, 113, 122,$\\$ 19, 52, 7, 8, 75, 134, 35, 40, 119, 80, 65, 83, 127, 36, 6, 109, 131, 106\})$}\\\hline
26&\makecell{
$(\{97, 30, 132, 72, 100, 14, 34, 35, 68, 85, 32, 17, 61, 134, 55, 11, 70, 52, 75, 51, 93,$\\$ 54, 63, 126, 5, 136, 130, 41, 121, 37, 56, 3, 102, 135, 84, 66, 44, 133, 12, 39, 86,$\\$ 128, 137, 4, 23, 69, 62, 129, 118, 87, 20, 110\}, \{77, 31, 49, 24, 103, 71, 125, 79,$\\$ 13, 1, 43, 60, 8, 40, 22, 89, 15, 82, 113, 21, 80, 88, 124, 119, 101, 115, 57, 108, 120,$\\$ 74, 28, 10, 47, 2, 64, 81, 0, 7, 109, 117, 25\}, \{16, 104, 27, 65, 116, 26, 45, 76,$\\$ 38, 127, 106, 50, 48, 46, 67, 94, 92, 99, 95, 91, 29, 105, 58, 123, 9, 111, 42, 98,$\\$ 53, 73, 18, 122, 112, 114, 19, 6, 78, 107, 33, 96, 83, 131, 90, 36, 59\})$}\\\hline
27&\makecell{
$(\{10, 56, 60, 2, 35, 104, 76, 127, 12, 134, 30, 81, 52, 85, 67, 136, 31, 27, 86, 107, 95,$\\$ 132, 89, 53, 17, 135, 92, 84, 115, 120, 39, 65, 109, 77, 103, 128, 102, 45, 131, 25,$\\$ 100, 96, 3, 0, 97, 1, 5, 110, 57, 71\}, \{101, 37, 133, 72, 43, 21, 64, 29, 11, 42,$\\$ 44, 73, 40, 122, 14, 62, 19, 94, 125, 18, 49, 87, 23, 55, 113, 121, 112, 7, 69, 16, 78,$\\$ 90, 20, 106, 63, 93, 58, 108, 15, 47, 116, 80, 6, 130, 61, 54, 33, 117\}, \{88, 24, 48,$\\$ 118, 51, 70, 98, 82, 75, 50, 126, 66, 129, 99, 4, 22, 74, 79, 36, 9, 8, 114, 68,$\\$ 32, 38, 105, 46, 91, 119, 59, 137, 13, 34, 123, 26, 124, 41, 111, 83, 28\})$}\\
\hline
\end{tabular}}
\caption{Block Sequential update modes for $n=138$.}
\label{table:bs6_n138}
\end{table}
\begin{table}
\resizebox{\textwidth}{!}{\centering
\begin{tabular}{|c|c|}
\hline
\# &Definition\\\hline
30&\makecell{
$(\{109, 29, 20, 55, 18, 88, 60, 101, 48, 84, 79, 129, 56, 71, 74, 25, 1, 19, 0, 2, 114,$\\$ 39, 102, 91, 126, 12, 57, 107, 111, 75, 32, 38, 70, 35, 112, 62, 13, 116, 82, 36, 80,$\\$ 64, 9\}, \{133, 106, 137, 46, 17, 45, 124, 93, 73, 83, 135, 130, 34, 131, 123, 132, 50,$\\$ 95, 122, 120, 11, 87, 6, 16, 51, 26, 85, 3, 69, 104, 128, 100, 28, 134, 121, 96, 40,$\\$ 77, 99, 54, 68, 81, 59, 14, 72, 37, 21, 125, 90, 92, 127, 118, 24, 115, 98, 61, 63,$\\$ 65, 10\}, \{136, 108, 94, 7, 52, 66, 97, 30, 58, 23, 113, 103, 5, 110, 67, 41, 76,$\\$ 89, 78, 42, 43, 117, 44, 47, 22, 86, 119, 8, 4, 15, 105, 33, 31, 53, 27, 49\})$}\\\hline
31&\makecell{
$(\{90, 75, 57, 38, 65, 59, 82, 64, 137, 113, 48, 28, 111, 106, 107, 116, 77, 32, 50, 76,$\\$ 105, 88, 36, 101, 85, 70, 74, 62, 47, 7, 99, 21, 8, 14, 66, 30, 92, 53, 87, 68, 37,$\\$ 33, 45, 86, 100, 39\}, \{125, 124, 60, 91, 114, 26, 98, 72, 24, 17, 136, 13, 78, 89,$\\$ 122, 130, 109, 1, 18, 71, 80, 95, 52, 120, 34, 61, 127, 117, 132, 2, 84, 6, 115, 16,$\\$ 118, 42, 23, 133, 25, 69, 123, 93, 49, 104, 3, 46, 83\}, \{22, 131, 51, 97, 135, 12, 56,$\\$ 10, 5, 108, 103, 29, 15, 9, 63, 43, 58, 73, 129, 126, 55, 31, 35, 121, 128, 96, 54,$\\$ 41, 81, 40, 94, 19, 79, 20, 27, 44, 0, 67, 110, 4, 11, 102, 119, 134, 112\})$}\\\hline
\end{tabular}}
\caption{Block Sequential update modes for $n=138$.}
\label{table:bs7_n138}
\end{table}

\begin{table}
\resizebox{\textwidth}{!}{\centering
\begin{tabular}{|c|c|}
\hline
\# &Definition\\\hline
0&\makecell{
$\{(97, 47, 34, 9, 109, 129, 128, 74, 87, 5, 58, 44, 15, 123, 90, 125, 88, 126, 21, 135,$\\$ 65, 114, 1, 26, 130, 89, 32, 45, 28, 38, 2, 66, 22, 120, 36, 39, 111, 13, 93, 24, 122,$\\$ 17, 10, 105, 29), (96, 134, 119, 91, 108, 112, 46, 4, 42, 50, 103, 104, 94, 41, 30, 124,$\\$ 18, 31, 102, 12, 131, 57, 127, 61, 71, 78, 64, 82, 59, 20, 118, 60, 76, 54, 80, 63,$\\$ 37, 73, 136), (69, 8, 53, 137, 25, 52, 92, 79, 0, 85, 11, 16, 33, 43, 101, 99, 14,$\\$ 98, 83, 81, 35, 110, 7, 23, 56, 133, 3, 121, 113, 68, 62, 40, 100, 51, 84, 77, 48,$\\$ 67, 6, 55, 19, 116, 72, 132, 49, 115, 75, 27, 107, 86, 106, 95, 117, 70)\}$}\\\hline
1&\makecell{
$\{(114, 21, 8, 74, 40, 50, 20, 17, 112, 93, 55, 95, 92, 72, 86, 56, 53, 62, 127, 57, 126,$\\$ 100, 19, 85, 67, 27, 78, 4, 43, 119, 115, 121, 46, 76, 87, 3, 81, 117, 2, 134, 11,$\\$ 77, 26, 98, 125, 135, 16, 58, 22, 36), (90, 48, 82, 120, 10, 102, 66, 108, 65, 60, 132,$\\$ 69, 47, 103, 99, 32, 1, 63, 111, 52, 128, 113, 96, 133, 136, 14, 104, 31, 45, 51, 83,$\\$ 106, 15, 44, 6, 54, 107, 0, 123, 49, 41, 5, 116, 13), (131, 105, 89, 71, 84, 42, 94, 61,$\\$ 34, 124, 18, 75, 59, 39, 73, 37, 118, 130, 28, 91, 80, 88, 25, 12, 122,$\\$ 110, 64, 137, 24, 33, 109, 129, 38, 29, 70, 79, 68, 35, 7, 30, 23, 101, 9, 97)\}$}\\\hline
\end{tabular}}
\caption{Block Parallel update modes for $n=138$.}
\label{table:bp1_n138}
\end{table}

\begin{table}
\resizebox{\textwidth}{!}{\centering
\begin{tabular}{|c|c|}
\hline
\# &Definition\\\hline
2&\makecell{
$\{(124, 47, 53, 122, 14, 10, 34, 21, 19, 49, 25, 74, 120, 29, 18, 83, 106, 101, 69, 119,$\\$ 30, 60, 89, 23, 131, 84, 71, 90, 38, 76, 28, 5, 81, 35, 132, 13, 78, 113, 72, 130, 31,$\\$ 44, 16, 9, 85, 50, 92, 62), (126, 70, 42, 45, 57, 129, 79, 102, 63, 46, 77, 108, 104,$\\$ 125, 40, 43, 12, 36, 37, 100, 97, 0, 134, 2, 26, 110, 121, 75, 73, 111, 20, 68, 93,$\\$ 98, 94, 136, 103, 55, 22, 137, 91, 4, 15, 33, 27, 67, 116, 128), (105, 123, 65, 41, 99,$\\$ 24, 52, 88, 3, 135, 7, 86, 17, 8, 95, 61, 59, 48, 107, 1, 11, 112, 115, 56, 6,$\\$ 66, 51, 96, 54, 64, 80, 82, 39, 87, 58, 127, 117, 118, 133, 32, 109, 114)\}$}\\\hline
3&\makecell{
$\{(7, 21, 77, 20, 52, 107, 123, 63, 35, 62, 79, 94, 3, 57, 2, 92, 14, 113, 81, 118, 9,$\\$ 50, 22, 133, 106, 47, 136, 49, 89, 121, 99, 12, 134, 111, 29, 30, 65, 83, 68, 44, 73,$\\$ 18, 98, 58, 26), (119, 48, 39, 28, 45, 110, 25, 95, 53, 42, 56, 127, 72, 131, 120, 105,$\\$ 55, 17, 82, 71, 4, 23, 36, 90, 24, 103, 96, 124, 40, 75, 66, 78, 43, 114, 135, 60,$\\$ 37, 61, 5, 32, 27, 115, 16, 59, 85), (108, 41, 117, 122, 51, 129, 8, 104, 6, 10, 46,$\\$ 74, 15, 128, 112, 101, 0, 76, 34, 67, 125, 33, 70, 19, 38, 13, 11, 137, 126, 97,$\\$ 86, 80, 31, 130, 88, 100, 87, 102, 93, 1, 84, 69, 132, 91, 54, 109, 64, 116)\}$}\\\hline
4&\makecell{
$\{(113, 66, 79, 99, 76, 105, 110, 121, 81, 46, 119, 58, 30, 120, 109, 131, 112, 42, 4,$\\$ 53, 92, 56, 85, 127, 103, 33, 17, 28, 1, 84, 0, 100, 88, 29, 116, 55, 12, 34, 57, 13,$\\$ 23, 40, 106, 63, 26, 96, 9, 15, 115, 51, 43, 73, 69, 75, 37), (21, 107, 136, 7, 74, 101,$\\$ 50, 8, 48, 32, 71, 3, 118, 16, 39, 47, 38, 35, 10, 125, 82, 104, 11, 91, 93, 98, 18,$\\$ 89, 134, 27, 44, 24, 5, 117, 62, 97, 68, 19, 80, 2, 64, 135, 122, 129, 111, 70, 14, 78,$\\$ 108, 137), (41, 72, 45, 36, 52, 60, 49, 114, 6, 126, 77, 54, 124, 132, 61,$\\$ 83, 86, 130, 25, 90, 67, 59, 128, 22, 31, 102, 87, 95, 123, 20, 65, 133, 94)\}$}\\
\hline
\end{tabular}}
\caption{Block Parallel update modes for $n=138$.}
\label{table:bp2_n138}
\end{table}

\begin{table}
\resizebox{\textwidth}{!}{\centering
\begin{tabular}{|c|c|}
\hline
\# &Definition\\\hline
7&\makecell{
$\{(130, 80, 86, 3, 135, 94, 93, 116, 100, 81, 56, 49, 122, 13, 127, 30, 37, 108, 12, 55,$\\$ 126, 128, 102, 74, 136, 113, 103, 29, 31, 79, 83, 91, 114, 9, 47, 70, 42, 120, 101, 50,$\\$ 52, 26), (36, 73, 105, 24, 134, 121, 84, 77, 110, 1, 39, 99, 22, 117, 118, 62, 53, 19,$\\$ 71, 78, 0, 45, 82, 23, 44, 72, 112, 107, 58, 8, 96, 133, 10, 111, 88, 43, 57, 60, 5,$\\$ 33, 2, 87, 92, 119, 106, 6, 54, 34), (27, 15, 97, 61, 64, 7, 63, 131, 85, 115, 67, 125,$\\$ 66, 25, 95, 68, 75, 98, 14, 28, 18, 109, 90, 48, 89, 123, 4, 40, 76, 16,$\\$ 104, 32, 65, 51, 59, 11, 21, 41, 17, 132, 38, 35, 124, 129, 69, 46, 137, 20)\}$}\\\hline
8&\makecell{
$\{(78, 67, 20, 122, 137, 66, 102, 103, 109, 107, 97, 98, 79, 74, 30, 22, 130, 6, 121, 123,$\\$ 51, 83, 69, 8, 115, 127, 42, 93, 63, 59, 94, 46, 0, 108, 91, 73, 68, 118, 129, 53, 15,$\\$ 7), (45, 31, 82, 44, 96, 117, 65, 128, 110, 126, 56, 47, 92, 120, 132, 58, 48, 71, 104,$\\$ 41, 13, 125, 23, 111, 16, 80, 72, 105, 88, 50, 81, 124, 27, 77, 29, 38, 3, 12, 28,$\\$ 33, 55, 21, 133, 87, 134, 76, 84, 90, 19, 49, 9, 64, 100, 40), (101, 26, 34, 24, 2, 136,$\\$ 61, 113, 114, 75, 52, 54, 57, 85, 95, 11, 135, 60, 5, 36, 18, 14, 17, 116,$\\$ 112, 39, 35, 32, 70, 37, 86, 106, 89, 119, 99, 131, 25, 62, 43, 1, 10, 4)\}$}\\\hline
9&\makecell{
$\{(101, 125, 13, 33, 52, 18, 113, 71, 15, 41, 103, 108, 28, 107, 25, 62, 43, 35, 49, 63,$\\$ 7, 93, 118, 61, 128, 48, 24, 11, 80, 23, 54, 112, 78, 124, 50, 89, 91, 36, 81, 55, 110,$\\$ 116, 10, 129, 98, 120, 17, 83), (100, 77, 127, 39, 119, 53, 8, 30, 134, 72, 16, 117,$\\$ 122, 45, 29, 31, 56, 9, 1, 126, 132, 59, 42, 109, 85, 115, 14, 106, 136, 65, 3, 135, 37,$\\$ 79, 2, 57, 133, 34, 82, 4, 47, 58), (102, 32, 40, 130, 123, 27, 67, 70, 86, 105, 92,$\\$ 137, 97, 111, 51, 12, 74, 75, 64, 73, 21, 90, 104, 44, 68, 66, 19, 0, 5,$\\$ 96, 38, 95, 26, 99, 6, 94, 87, 20, 114, 22, 60, 46, 76, 88, 121, 69, 131, 84)\}$}\\
\hline
\end{tabular}}
\caption{Block Parallel update modes for $n=138$.}
\label{table:bp3_n138}
\end{table}

\begin{table}
\resizebox{\textwidth}{!}{\centering
\begin{tabular}{|c|c|}
\hline
\# &Definition\\\hline
12&\makecell{
$\{(46, 102, 119, 105, 60, 135, 30, 19, 104, 121, 17, 21, 41, 107, 54, 118, 133, 59, 93,$\\$ 114, 94, 113, 28, 15, 108, 57, 98, 27, 18, 85, 101, 55, 68, 127, 72, 96, 26, 50, 29,$\\$ 112, 116, 77), (23, 47, 71, 129, 132, 69, 24, 8, 7, 66, 137, 58, 125, 25, 110, 117, 20,$\\$ 35, 38, 53, 126, 115, 111, 84, 91, 33, 82, 39, 90, 42, 36, 49, 22, 31, 86, 92, 73,$\\$ 99, 78, 45, 37, 43, 62, 16, 5, 2, 65, 88), (67, 89, 97, 81, 40, 12, 34, 103, 44, 64,$\\$ 100, 134, 4, 136, 75, 56, 109, 106, 48, 95, 123, 131, 52, 51, 9, 13, 3, 11, 63,$\\$ 61, 0, 70, 32, 74, 87, 6, 10, 1, 120, 124, 130, 79, 80, 76, 122, 14, 128, 83)\}$}\\\hline
13&\makecell{
$\{(83, 7, 70, 34, 91, 109, 110, 18, 101, 116, 12, 36, 63, 47, 56, 114, 6, 118, 79, 125,$\\$ 128, 23, 60, 87, 37, 86, 27, 122, 134, 99, 33, 55, 44, 68, 11, 73, 137, 41, 80, 9,$\\$ 78, 84, 113, 2, 89, 49, 1, 4), (132, 8, 105, 131, 24, 3, 108, 65, 130, 14, 104, 39, 85,$\\$ 117, 98, 19, 5, 97, 0, 121, 66, 133, 54, 69, 30, 29, 15, 102, 127, 71, 93, 43, 31, 82,$\\$ 57, 21, 51, 81, 67, 119, 32, 120, 59, 13, 94, 52, 25, 46), (42, 77, 45, 115, 40, 106,$\\$ 74, 96, 88, 135, 38, 62, 92, 90, 107, 100, 136, 112, 53, 103, 58, 124, 26,$\\$ 50, 16, 28, 76, 22, 126, 17, 75, 10, 48, 129, 72, 35, 64, 111, 61, 20, 123, 95)\}$}\\\hline
14&\makecell{
$\{(26, 97, 77, 134, 69, 114, 43, 137, 125, 39, 94, 54, 80, 38, 30, 93, 73, 68, 59, 49,$\\$ 58, 79, 21, 72, 128, 60, 5, 2, 126, 50, 16, 37, 98, 75, 11, 32, 52, 135, 110, 85, 61,$\\$ 47), (17, 14, 86, 31, 123, 87, 41, 76, 111, 92, 1, 20, 36, 62, 65, 71, 91, 127, 74,$\\$ 112, 90, 81, 28, 102, 44, 57, 109, 64, 132, 45, 12, 99, 96, 107, 9, 130, 100, 8, 124,$\\$ 25, 120, 78, 13, 53, 118, 88, 133, 67, 116, 46, 70, 15, 42, 19), (95, 108, 10, 3, 103,$\\$ 131, 56, 51, 29, 105, 7, 18, 48, 129, 117, 122, 33, 119, 101, 89, 63, 0, 115,$\\$ 136, 83, 84, 66, 34, 40, 35, 104, 55, 24, 106, 23, 113, 6, 82, 4, 121, 27, 22)\}$}\\
\hline
\end{tabular}}
\caption{Block Parallel update modes for $n=138$.}
\label{table:bp4_n138}
\end{table}

\begin{table}
\resizebox{\textwidth}{!}{\centering
\begin{tabular}{|c|c|}
\hline
\# &Definition\\\hline
17&\makecell{
$\{(83, 11, 129, 9, 2, 28, 135, 55, 66, 103, 45, 70, 30, 126, 19, 24, 31, 93, 67, 76, 108,$\\$ 51, 61, 86, 34, 48, 136, 18, 37, 85, 90, 15, 78, 17, 88, 96, 53, 27, 22, 107, 79, 77),$\\$ (58, 81, 52, 118, 54, 95, 69, 59, 94, 104, 134, 29, 62, 84, 97, 25, 64, 121, 111, 110,$\\$ 99, 6, 123, 33, 7, 82, 92, 65, 114, 106, 47, 127, 102, 38, 39, 128, 4, 16, 115, 87,$\\$ 63, 91, 137, 112, 3, 40, 13, 120, 117, 109, 132, 89, 113, 10), (41, 42, 71, 49, 44, 50,$\\$ 57, 72, 74, 43, 56, 26, 32, 1, 125, 14, 131, 23, 8, 122, 130, 98, 100, 133,$\\$ 68, 12, 21, 36, 46, 80, 105, 116, 60, 124, 0, 20, 73, 119, 5, 101, 75, 35)\}$}\\\hline
18&\makecell{
$\{(5, 93, 15, 101, 19, 111, 81, 3, 18, 44, 58, 126, 22, 86, 103, 46, 135, 90, 17, 64, 121,$\\$ 84, 10, 9, 8, 124, 33, 41, 1, 132, 133, 114, 73, 0, 71, 87, 127, 40, 99, 82, 119, 52,$\\$ 70, 113, 106), (107, 136, 97, 85, 26, 95, 112, 54, 49, 55, 36, 128, 28, 92, 102, 50,$\\$ 91, 27, 104, 39, 109, 65, 100, 68, 74, 12, 89, 63, 30, 42, 45, 6, 37, 43, 31, 4,$\\$ 72, 7, 110, 88, 96, 20, 62, 24, 83, 134, 120, 60), (137, 38, 116, 118, 115, 29, 76, 47, 78,$\\$ 105, 131, 35, 14, 108, 129, 117, 13, 48, 77, 98, 66, 67, 79, 75, 56, 94, 34,$\\$ 125, 23, 57, 53, 80, 16, 25, 11, 21, 61, 51, 123, 122, 59, 2, 69, 130, 32)\}$}\\\hline
19&\makecell{
$\{(129, 79, 72, 34, 8, 67, 118, 35, 96, 1, 131, 26, 49, 99, 120, 54, 112, 83, 29, 56, 15,$\\$ 5, 52, 93, 30, 132, 44, 6, 69, 47, 97, 71, 24, 62, 14, 119, 134, 110, 74, 137, 127, 128,$\\$ 121, 11, 66, 0, 61, 86), (64, 84, 18, 60, 85, 51, 39, 109, 73, 31, 95, 65, 46, 50,$\\$ 108, 115, 59, 48, 40, 37, 42, 105, 122, 4, 91, 130, 23, 22, 100, 88, 41, 81, 57, 58,$\\$ 113, 80, 20, 135, 3, 103, 38, 106), (16, 68, 43, 125, 104, 78, 13, 94, 75, 98, 19, 89,$\\$ 101, 55, 45, 2, 136, 82, 53, 117, 111, 87, 124, 77, 116, 63, 107, 33, 90, 28,$\\$ 126, 17, 76, 21, 92, 32, 25, 70, 10, 36, 123, 27, 7, 114, 9, 102, 12, 133)\}$}\\
\hline
\end{tabular}}
\caption{Block Parallel update modes for $n=138$.}
\label{table:bp5_n138}
\end{table}

\begin{table}
\resizebox{\textwidth}{!}{\centering
\begin{tabular}{|c|c|}
\hline
\# &Definition\\\hline
22&\makecell{
$\{(41, 23, 12, 54, 101, 1, 29, 38, 91, 22, 82, 92, 96, 123, 71, 43, 86, 109, 77, 17, 20,$\\$ 127, 47, 69, 61, 44, 13, 120, 73, 103, 16, 64, 48, 63, 6, 55, 67, 106, 79, 4, 104, 68,$\\$ 74, 125, 108), (87, 50, 83, 37, 59, 137, 133, 135, 36, 19, 89, 134, 3, 39, 53, 98, 76,$\\$ 7, 95, 45, 15, 114, 14, 111, 105, 18, 122, 62, 58, 66, 90, 121, 75, 46, 5, 10, 2, 8,$\\$ 30, 31, 102, 49, 132, 118, 93), (81, 56, 94, 107, 21, 129, 78, 100, 99, 11, 130, 126, 51,$\\$ 26, 131, 25, 128, 52, 116, 84, 33, 57, 117, 65, 27, 85, 115, 40, 97, 88,$\\$ 0, 72, 110, 136, 124, 119, 34, 24, 60, 113, 70, 35, 32, 9, 28, 112, 42, 80)\}$}\\\hline
23&\makecell{
$\{(66, 130, 28, 49, 37, 76, 134, 12, 23, 9, 33, 103, 104, 55, 73, 56, 115, 86, 110, 7,$\\$ 8, 122, 40, 105, 36, 111, 14, 118, 96, 131, 83, 132, 106, 10, 99, 53, 137, 32, 95, 75),$\\$ (119, 126, 62, 91, 5, 52, 85, 11, 24, 88, 65, 129, 124, 100, 128, 97, 117, 16, 13, 116,$\\$ 64, 57, 21, 78, 81, 136, 3, 42, 60, 58, 1, 39, 34, 27, 108, 92, 70, 29, 94, 44, 71,$\\$ 135, 18, 46, 123, 87, 74, 107, 133), (59, 82, 38, 35, 26, 98, 31, 51, 90, 72, 6, 45,$\\$ 84, 127, 0, 47, 102, 77, 68, 61, 112, 63, 43, 50, 54, 4, 89, 17, 22, 48, 101,$\\$ 2, 25, 80, 79, 120, 121, 114, 15, 113, 30, 109, 20, 69, 19, 67, 41, 93, 125)\}$}\\\hline
24&\makecell{
$\{(86, 126, 125, 128, 30, 84, 22, 110, 106, 71, 43, 36, 24, 23, 82, 116, 37, 2, 21, 66,$\\$ 109, 58, 132, 45, 8, 18, 54, 94, 78, 44, 28, 72, 41, 135, 39, 1, 62, 34, 68, 3, 98,$\\$ 90, 32, 104, 123, 38, 19, 67, 88, 89, 40, 137, 26, 91), (74, 55, 11, 83, 117, 92, 31,$\\$ 14, 120, 0, 13, 118, 80, 20, 113, 70, 129, 96, 127, 95, 63, 101, 51, 17, 122, 134, 7,$\\$ 119, 52, 57, 35, 12, 15, 115, 87, 48, 136, 131, 102, 75, 100, 46), (47, 61, 9, 65, 73,$\\$ 29, 76, 133, 121, 107, 114, 64, 108, 69, 77, 130, 97, 10, 5, 53, 25, 99, 105,$\\$ 93, 81, 4, 112, 59, 33, 124, 79, 42, 16, 111, 60, 49, 6, 85, 50, 27, 103, 56)\}$}\\
\hline
\end{tabular}%
}
\caption{Block Parallel update modes for $n=138$.}
\label{table:bp7_n138}
\end{table}

\begin{table}\centering
\resizebox{\textwidth}{!}{\centering
%
}
\caption{Local Clocks update modes for $n=138$.}
\label{table:lc1_n138}
\end{table}

\begin{table}\centering
\resizebox{\textwidth}{!}{\centering
%
}
\caption{Local Clocks update modes for $n=138$.}
\label{table:lc2_n138}
\end{table}

\begin{table}\centering
\resizebox{\textwidth}{!}{\centering
%
}
\caption{Local Clocks update modes for $n=138$.}
\label{table:lc8_n138}
\end{table}
\FloatBarrier

\begin{table}\centering
%
\caption{Sequential update modes for $n=16$.}
\label{table:seq16}
\end{table}

\begin{table}\centering
\begin{tabular}{|c|c|}
\hline
\# &Definition\\\hline
0&$\{(6),(13, 9, 0),(5, 1, 4, 10, 3),(2, 7, 8, 11, 12, 14, 15)\}$\\\hline
1&$\{(1),(8, 0, 15),(4, 3, 12, 9, 7),(2, 5, 6, 10, 11, 13, 14)\}$\\\hline
2&$\{(14),(11, 7, 10),(13, 6, 9, 2, 1),(0, 3, 4, 5, 8, 12, 15)\}$\\\hline
3&$\{(8),(4, 10, 1),(0, 12, 3, 15, 5),(2, 6, 7, 9, 11, 13, 14)\}$\\\hline
4&$\{(9),(0, 15, 11),(14, 5, 2, 12, 1),(3, 4, 6, 7, 8, 10, 13)\}$\\\hline
5&$\{(3),(1, 11, 6),(0, 14, 13, 15, 7),(2, 4, 5, 8, 9, 10, 12)\}$\\\hline
6&$\{(10),(1, 9, 7),(14, 15, 3, 13, 6),(0, 2, 4, 5, 8, 11, 12)\}$\\
\hline
\end{tabular}
\caption{Block Parallel update modes for $n=16$.}
\label{table:bp16}
\end{table}

\begin{table}\centering
%
\caption{Block Sequential update modes, with 3 blocks for $n=16$.}
\label{table:bs1_16}
\end{table}

\begin{table}\centering
%
\caption{Block Sequential update modes, with 4 blocks for $n=16$.}
\label{table:bs2_16}
\end{table}

\begin{table}\centering
%
\caption{Block Sequential update modes, with 5 blocks for $n=16$.}
\label{table:bs3_16}
\end{table}

\begin{table}
\resizebox{\textwidth}{!}{\centering
%
}
\caption{Local Clocks update modes, with period 2 for $n=16$.}
\label{table:lc1_16}
\end{table}

\begin{table}
\resizebox{\textwidth}{!}{\centering
%
}
\caption{Local Clocks update modes, with period 4 for $n=16$.}
\label{table:lc2_16}
\end{table}

\begin{table}
\resizebox{\textwidth}{!}{\centering
%
}
\caption{Local Clocks update modes, with period 5 for $n=16$.}
\label{table:lc3_16}
\end{table}

%%       \cite{<label>} ==> [1]

%%




%\begin{thebibliography}{00}

%% For numbered reference style

%\bibitem{lamport94}
%  \textit{\LaTeX: a document preparation system},
%  Addison Wesley, Massachusetts,

%\end{thebibliography}
\end{document}